\documentclass[a4paper,fleqn,usenatbib]{mnras}

\usepackage{newtxtext,newtxmath}

\usepackage[T1]{fontenc}
\usepackage{ae,aecompl}

\usepackage{graphicx}	
\usepackage[compatibility=false]{caption}
\usepackage{subcaption}
\usepackage{amsmath}	
\usepackage{amssymb}	
\usepackage{tabularx,multirow}
\usepackage[autostyle]{csquotes}

 \usepackage{multirow}
 \usepackage[utf8]{inputenc} 

 \usepackage{natbib}
 \usepackage{xspace}
 \usepackage{url}
 \usepackage{indentfirst}
 \usepackage{tabularx}
 \usepackage[usenames, dvipsnames]{color}

 \usepackage{multirow}

\title[Cluster Age Gradients Across Spiral Arms of Three LEGUS Disk Galaxies]{Search For Star Cluster Age Gradients Across Spiral Arms of Three LEGUS Disk Galaxies}

\author[F.Shabani et al.]
{F.~Shabani,$^{1}$\thanks{
E-mail: f.shabani@stud.uni-heidelberg.de} E.K.~Grebel,$^1$ A.~Pasquali,$^1$ E.~D'Onghia,$^{2,3}$ J.S.~Gallagher III,$^2$ 
\newauthor 
A.~Adamo,$^4$ M.~Messa,$^4$ B.G.~Elmegreen,$^5$ C.~Dobbs,$^6$ D.A.~Gouliermis,$^{7,8}$ D.~Calzetti,$^9$ 
\newauthor 
K.~Grasha,$^9$ D.M.~Elmegreen,$^{10}$ M.~Cignoni,$^{11,12,13}$ D.A.~Dale,$^{14}$ A.~Aloisi,$^{15}$
L.J.~Smith,$^{16}$
\newauthor 
M.~Tosi,$^{13}$ D.A.~Thilker,$^{17}$ J.C.~Lee,$^{15,18}$ E.~Sabbi,$^{15}$ H.~Kim,$^{19}$ and A.~Pellerin$^{20}$
\\
$^1$Astronomisches Rechen-Institut, Zentrum f\"ur Astronomie der Universit\"at Heidelberg, M\"onchhofstr.\ 12--14, 69120 Heidelberg, Germany\\
$^2$Dept. of Astronomy, University of Wisconsin- Madison, 475 N. Charter Street, Madison, WI 53076--1582, USA\\
$^3$Center for Computational Astrophysics, Flatiron Institute, 162 Fifth Avenue, New York, NY 10010, USA\\
$^{4}$Dept. of Astronomy, The Oskar Klein Centre, Stockholm University, Stockholm, Sweden\\
$^{5}$IBM Research Division, T.J. Watson Research Center, Yorktown Hts., NY, USA\\
$^{6}$School of Physics and Astronomy, University of Exeter, Exeter, United Kingdom\\
$^{7}$Zentrum f\"ur Astronomie der Universit\"at Heidelberg, Institut f\"ur Theoretische Astrophysik, Albert-Ueberle-Str.\,2, 69120 Heidelberg, Germany\\
$^{8}$Max Planck Institute for Astronomy,  K\"{o}nigstuhl\,17, 69117 Heidelberg, Germany\\
$^{9}$Dept. of Astronomy, University of Massachusetts -- Amherst, Amherst, MA 01003, USA\\
$^{10}$Dept. of Physics and Astronomy, Vassar College, Poughkeepsie, NY, USA\\
$^{11}$Dept. of Physics, University of Pisa, Largo B. Pontecorvo 3, 56127, Pisa, Italy\\
$^{12}$INFN, Largo B. Pontecorvo 3, 56127, Pisa, Italy\\
$^{13}$INAF - Osservatorio Astrofisico e di Scienza dello Spazio, Bologna, Italy\\
$^{14}$Dept. of Physics and Astronomy, University of Wyoming, Laramie, WY, USA\\
$^{15}$Space Telescope Science Institute, Baltimore, MD, USA\\
$^{16}$European Space Agency/Space Telescope Science Institute, Baltimore, MD, USA\\
$^{17}$Dept. of Physics and Astronomy, The Johns Hopkins University, Baltimore, MD, USA\\
$^{18}$Visiting Astronomer, Spitzer Science Center, Caltech. Pasadena, CA, USA\\
$^{19}$Gemini Observatory, Casilla 603, La Serena, Chile\\
$^{20}$Dept. of Physics and Astronomy, State University of New York at Geneseo, Geneseo, NY, USA\\
\\
}

\date{Accepted XXX. Received YYY; in original form ZZZ}

\pubyear{2017}

\begin{document}
\label{firstpage}
\pagerange{\pageref{firstpage}--\pageref{lastpage}}
\maketitle

\begin{abstract}
One of the main theories for explaining the formation of spiral arms in galaxies is the stationary density wave theory. This theory predicts the existence of an age gradient across the arms. We use the stellar cluster catalogues of the galaxies NGC~1566, M51a, and NGC~628 from the Legacy Extragalactic UV Survey (LEGUS) program. In order to test for the possible existence of an age sequence across the spiral arms, we quantified the azimuthal offset between star clusters of different ages in our target galaxies. We found that NGC~1566, a grand--design spiral galaxy with bisymmetric arms and a strong bar, shows a significant age gradient across the spiral arms that appears to be consistent with the prediction of the stationary density wave theory. In contrast, M51a with its two well--defined spiral arms and a weaker bar does not show an age gradient across the arms. In addition, a comparison with non--LEGUS star cluster catalogues for M51a yields similar results. We believe that the spiral structure of M51a is not the result of a stationary density wave with a fixed pattern speed. Instead, tidal interactions could be the dominant mechanism for the formation of spiral arms.   We also found no offset in the azimuthal distribution of star clusters with different ages across the weak spiral arms of NGC~628. 

\end{abstract}




\section{Introduction}
\label{Introduction}
Understanding how spiral patterns form in disk galaxies is a long--standing issue in astrophysics. Two of the most influential theories to explain the formation of spiral structure in disk galaxies are named stationary density wave theory and swing amplification. The stationary density wave theory poses that spiral arms are static density waves \citep{Lindblad, LS64}. In this scenario spiral arms are stationary and long--lived. The swing amplification proposes instead that spiral structure is the local amplification in a differentially rotating disk \citep{G, JT, SC, Sellwood, E11, Elena13}. According to this theory indiviual spiral arms would fade away in one galactic year and should be considered transient features. Numerical experiments suggest that non--linear gravitational effects would make spiral arms fluctuate in density locally but be statistically long--lived and self--perpetuating \citep{Elena13}.

To complicate the picture there is the finding that many galaxies in the nearby universe are grand--design, bisymmetric spirals. These galaxies may show evidence of a galaxy companion, suggesting that the perturbations induced by tidal interactions could induce spiral features in disks by creating localized disturbances that grow by swing amplification \citep{ k, B3, Ga, Elena16, P16}. Some studies have been devoted to explore galaxy models with bar--induced spiral structure \citep{conto} and spiral features explained by a manifold \citep{conto, A}. It is also possible that a combination of these models is needed to describe the observed spiral structure. We refer the interested reader to comprehensive reviews of different theories of spiral structure  in \cite{DB14} and to \cite{Shu} for detailed explanations of the origin of spiral structure in stationary density wave theory.

The longevity of spiral structure can be tested observationally. In fact, in the stationary density wave theory, spiral arms are density waves moving with a single constant angular pattern speed. The angular speed of stars and gas equals the pattern speed at the corotation radius. Inside the corotation radius, material rotates faster than the spiral pattern. When the gas enters the higher--density region of spiral arms, it may experience a shock which may lead to star formation \citep{R69}. Consequently, the stars born in the molecular clouds in spiral arms eventually overtake the arms and move away from the spiral patterns as they age. This drift causes an age gradient across the spiral arms. If spiral arms have a constant angular speed, then we expect to find the youngest star clusters near the arm on the trailing side, and the oldest star clusters further away from the spiral arms inside the corotation radius \citep[e.g.,][]{M09}. Outside the corotation radius, the spiral pattern moves faster than the gas and leads to the opposite age sequence. 

\citet{DP10} carried out numerical simulations of the age distribution of star clusters in four different spiral galaxy models, including a galaxy with a fixed pattern speed, a barred galaxy, a flocculent galaxy, and an interacting galaxy. The results of their simulations show that in a spiral galaxy with a constant pattern speed or in a barred galaxy, a clear age sequence across spiral arms from younger to older stars is expected. In the case of a flocculent spiral galaxy, no age gradient can be observed in their simulation. Also in the case of an interacting galaxy, a lack of an age gradient as a function of azimuthal distance from the spiral arms is predicted. A simulation of an isolated multiple--arm barred
spiral galaxy was performed by \cite{grand}, who explored the location of star particles as a function of age around the spiral arms. Their simulation takes into account radiative cooling and star formation. They found no significant spatial offset between star particles of different ages, suggesting that spiral arms in such a spiral galaxy are not consistent with the long--lived spiral arms predicted by the static or stationary density wave theory. In a recent numerical study, \cite{D17} looked in detail at the spatial distribution of stars with different ages in an isolated grand--design spiral galaxy. They found that star clusters of different ages are all concentrated along the spiral arms without a clear age pattern.

A simple test of the stationary density wave theory consists of looking for a colour gradient from blue to red across spiral arms due to the progression of star formation. It is important to note that this method can be affected by the presence of dust. Several observational studies have tried to test the stationary density wave theory by looking for colour gradients across the spiral arms. In an early study of the ($B-V$) colours and total star formation rates in a sample of spiral galaxies with and without grand design patterns, \cite{Bruce86} found no evidence for an excess of star formation due to the presence of a spiral density wave, and explained the blue spiral arm colours as a result of a greater compression of the gas compared to the old stars, with star formation following the gas. \cite{M09} studied the colour gradients across the spiral arms of 13 SA and SAB galaxies. Ten galaxies in their sample present the expected colour gradient across their spiral arms.

A number of observational studies have used the age of stellar clusters in nearby galaxies as a tool to test the stationary density wave theory.  \citet{S09} studied the spatial distribution of 1580 stellar clusters in the interacting, grand--design spiral M51a from Hubble Space Telescope (HST) $UBVI$ photometry. They found no spatial offset between the azimuthal distribution of cluster samples of different age. Their results indicate that most of the young (age < 10~Myr) and old stellar clusters (age > 30~Myr) are located at the centers of the spiral arms. \cite{k10} also mapped the age of star clusters as a function of their location in M51a using HST data and found no clear pattern in the location of star clusters with respect to their age. Both above studies suggest that spiral arms are not stationary, at least for galaxies in tidal interaction with a companion. In order to study the spatial distribution of star--forming regions, \cite{Sanchez} produced an age map of six nearby grand--design and flocculent spiral galaxies. Only two grand--design spiral galaxies in their sample presented a stellar age sequence across the spiral arms as expected from stationary density wave theory. 

In galaxies where spiral arms are long--lived and stationary as predicted by the static density wave theory, one would expect to find an angular offset among star formation and gas tracers of different age within spiral arms \citep{R69}. The majority of observational studies of the spiral density wave scenario have tried to examine such an angular offset \citep{vogel, Rand}. \cite{Tam8} detected an angular offset between HI (a tracer of the cold dense gas) and 24~$\rm \mu$m emission (a tracer of obscured star formation) in a sample of 14 nearby disk galaxies. An angular offset between CO (a tracer of molecular gas) and H$\rm \alpha$ (a tracer of young stars) was detected for 5 out of 13 spiral galaxies observed by \cite{Egusa}. In another observational work, \cite{Foyle} tested the angular offset between different star formation and gas tracers including HI, $\rm H_{2}$, 24~$\rm \mu $m, UV (a tracer for unobscured young stars) and 3.6~$\rm \mu$m emission (a tracer of the underlying old stellar population) for 12 nearby disk galaxies. They detected no systematic trend between the different tracers. Similarly, \cite{F13} found no significant angular offset between H$\rm \alpha$ and UV emission in NGC~4321. \cite{L13} found a large angular offset between CO and H$\rm \alpha$ in M51a while no significant offsets have been found between HI, 21~cm, and 24~$\rm \mu$m emissions. These searches for offsets are based on the assumption that the different tracers represent a time sequence of the way a moving density wave interacts with gas and triggers star formation. \cite{Elmegreen2014} used the S$^4$G survey \citep{Sheth} and discovered embedded clusters inside the dust lanes of several galaxies with spiral waves, suggesting that star formation can sometimes start quickly.

In a recent observational study, \cite{S17} carried out a detailed investigation of a spiral arm segment in M51a. They measured the radial offset of the star clusters of different ages (< 3~Myr, and 3--10~Myr) and star formation tracers (HII regions and 24~$\rm \mu$m) from their nearest spiral arm. No obvious spatial offset between star clusters younger and older than 3 Myr was found in M51a. They also found no clear trend in the radial offset of HII regions and 24~$\mu$m. Similarly, \cite{chandar17} compared the location of star clusters with different ages (< 6~Myr, 6--30~Myr, 30--100~Myr, 100--400~Myr, and > 400 Myr) with the spiral patterns traced by molecular gas, dust, young and old stars in M51a. They found cold molecular gas and dark dust lanes to be located along the inner edge of the arms while the outer edge is defined by the old stars (traced with 3.6~$\rm \mu$m) and young star clusters. The observed sequence in the spiral arm of M51a is in agreement with the prediction from stationary density wave theory. \cite{chandar17} also measured the spatial offset between molecular gas, young (< 10~Myr) and old star clusters (100--400~Myr) in the inner (2.0--2.5~kpc) and outer (5.0--5.5~kpc) spiral arms in M51a. They found an azimuthal offset between the gas and star clusters in the inner spiral arm zone, which is consistent with the spiral density wave theory. In the outer spiral arms, the lack of such a spatial offset suggests that the outer spiral arms do not have a constant pattern speed and are not static. \cite{chandar17} found no star cluster age gradient along four gas spurs (perpendicular to the spiral arms) in M51a.

In conclusion, there have been numerous observational studies aiming to test the longevity of the spiral structure. In many cases, the conclusions show conflicting results and the nature of spiral arms is still an open question.

The main goal of this study is to test whether spiral arms in disk galaxies are static and long--lived or locally changing in density and locally transient. This work is based on the Legacy ExtraGalactic UV Survey (LEGUS)\footnote{https://legus.stsci.edu} observations obtained with HST \citep{C15}. The paper is organized as follows: The survey and the sample galaxies are described in \S~\ref{The LEGUS Galaxy Samples}. The selection of the star cluster samples is presented in \S~\ref{s3}. We investigate the spatial distribution together with clustering of the selected clusters in \S~\ref{location}. In \S~\ref{Azimutahl distribution}, we describe the results and analysis and how we measure the spatial offset of our star clusters across spiral arms. In \S~\ref{2arms} we discuss whether the two spiral arms of our target galaxies have the same nature. In \S~\ref{chandra}, we use a non--LEGUS star cluster catalogue to measure the spatial offset of star clusters in M51a and we present our conclusions in \S~\ref {Summary}.

\section{The sample galaxies}
 \label{The LEGUS Galaxy Samples}
LEGUS is an HST Cycle 21 Treasury programme that has observed 50 nearby star--forming dwarf and spiral galaxies within 12~Mpc. High-- resolution images of these galaxies were obtained with the UVIS channel of the Wide Field Camera Three (WFC3), supplemented with archival Advanced Camera for Surveys (ACS) imaging when available, in five broad band filters, $NUV\,(F275W)$, $U \,(F336W)$, $B \,(F438W)$, $V \,(F555W)$, and $I \,(F814W)$. The pixel scale of these observations is $ \rm 0.04^{\arcsec} \, pix^{-1}$. A description of the survey, the observations, the image processing, and the data reduction can be found in \cite{C15}. 

Face--on spiral galaxies with prominent spiral structures are interesting candidates to study stationary density wave theory. Therefore, three face--on  spiral galaxies, namely NGC~1566, M51a, and NGC~628 were selected from the LEGUS survey for our study. The morphology, distance, corotation radius, and the pattern speed of each galaxy are listed in Table~\ref{tab:properties of galaxies}. The UVIS and ACS footprints of the pointings (red and yellow boxes, respectively) overlaid on Digitized Sky Survey (DSS) images of the galaxies are shown in Fig.~\ref{fig:galaxies} together with their HST red, green, and blue colour composite mosaics. 
\begin{table*}
  \caption{Fundamental properties of our target galaxies.}
  \label{tab:properties of galaxies}
  \begin{tabular}{lcccccccccccccc}
  \hline
   \hline
  Galaxy & Morphology & D [Mpc]& $\rm M_{\star} \, (M_{\sun})$ & SFR (UV) $\rm(M_{\sun} \, yr^{-1}) $ &  $\rm R_{cr}$ [$\mathrm{kpc}$] & $ \rm \Omega_{p}$ [$\rm km\, s^{-1}\, \rm kpc^{-1}$] & Ref \\
  \hline
  NGC~1566 & SABbc &18& $\rm 2.7\times 10^{10}$ & 2.026&10.6 & 23$\pm$2 &1\\
  M51a & SAc &7.6&$\rm 2.4\times 10^{10}$ & 6.88&5.5 &38$\pm$7 &2 \\
  NGC~628   & SAc &9.9 &$\rm 1.1\times 10^{10}$ & 3.6& 7 &32$\pm$2& 3 \\

  \hline
   \hline
  \\
  \end{tabular}
   \vspace{1ex}

    \raggedright Column 1, 2: Galaxy name and morphological type as listed in the NASA Extragalactic Database (NED)  \\
    \raggedright Column 3: Distance\\
    \raggedright Column 4: Stellar mass obtained from the extinction--corrected B--band luminosity \\
    \raggedright Column 5: Star formation rate calculated from the GALEX far--UV, corrected for dust attenuation \\
    \raggedright Column 6: Co--rotation radius\\
    \raggedright Column 7: Pattern speed \\
    \raggedright Column 8: References for the co--rotation radii and pattern speeds: 1- \cite{A04}, 2- \cite{z4}, 3- \cite{Sakhibov}\\
\end{table*}
\subsection{NGC~1566}
\begin{figure*}
\centering

      \begin{subfigure}[t]{0.50\textwidth}
   \includegraphics[width=1\linewidth]{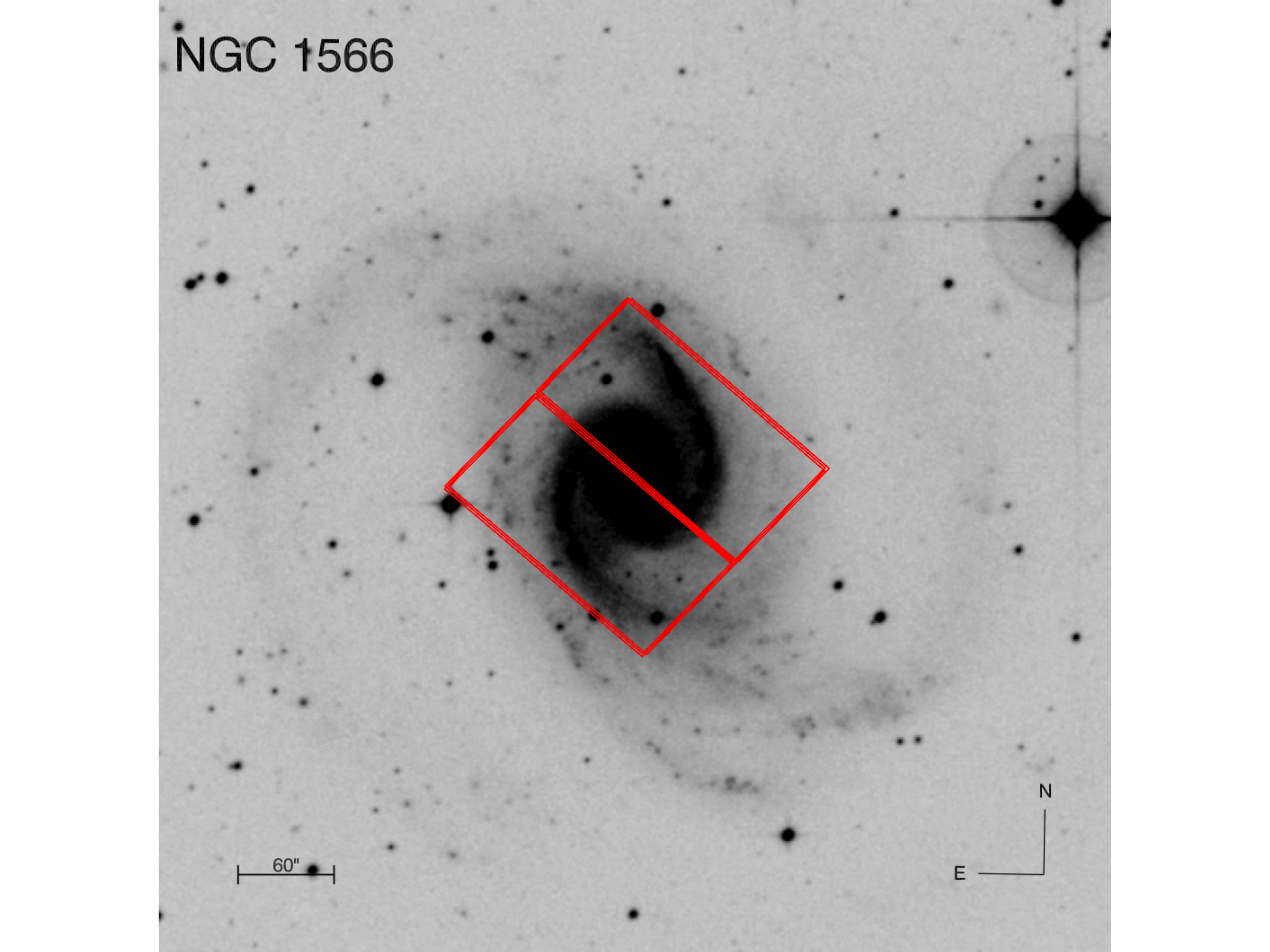}
    \end{subfigure}%
     ~
     ~ \hspace{-1.8cm}
  \begin{subfigure}[t]{0.5\textwidth}
   \includegraphics[width=1\linewidth]{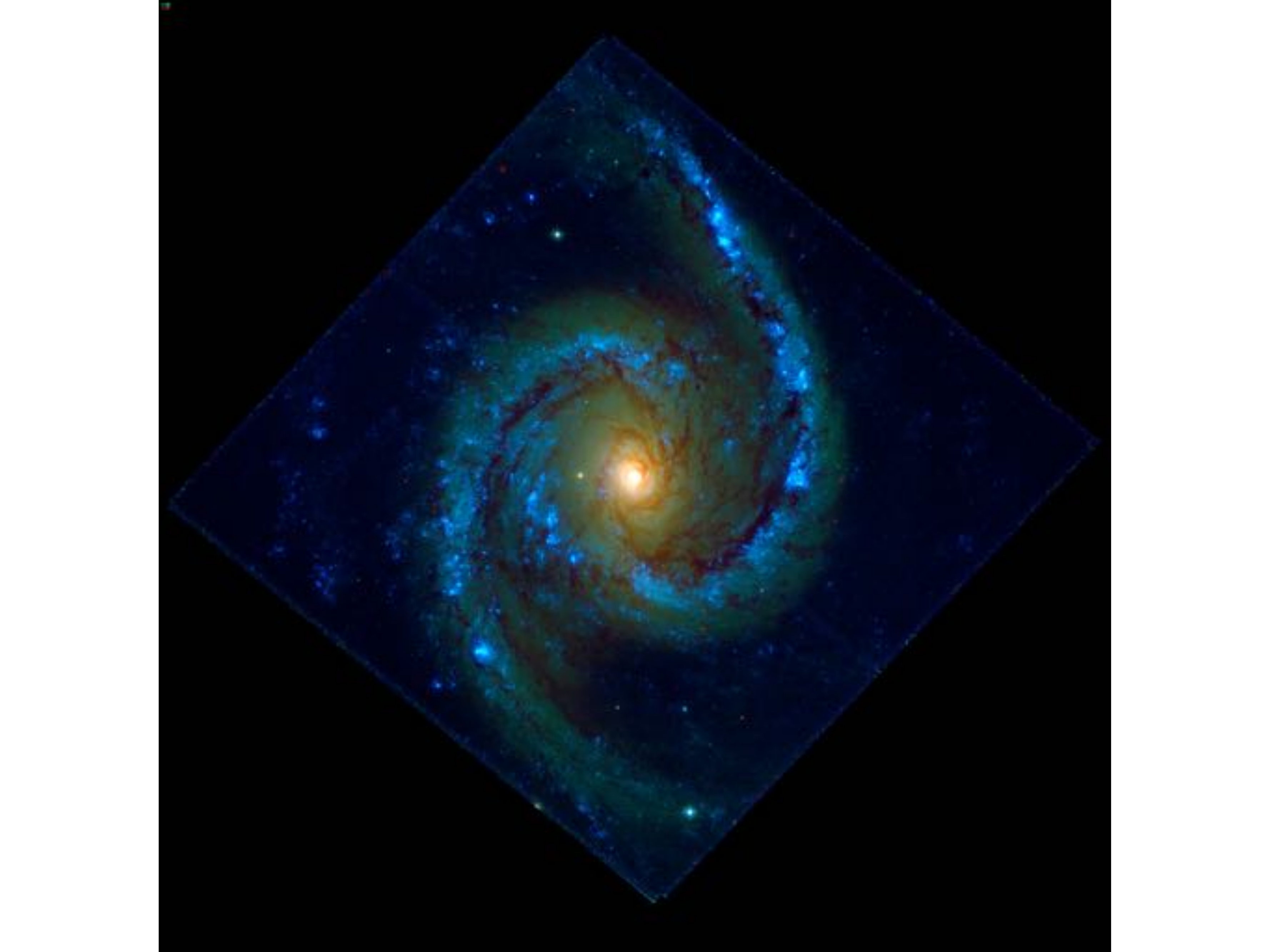}
    \end{subfigure}
    
    \begin{subfigure}[t]{0.50\textwidth}
   \includegraphics[width=1\linewidth]{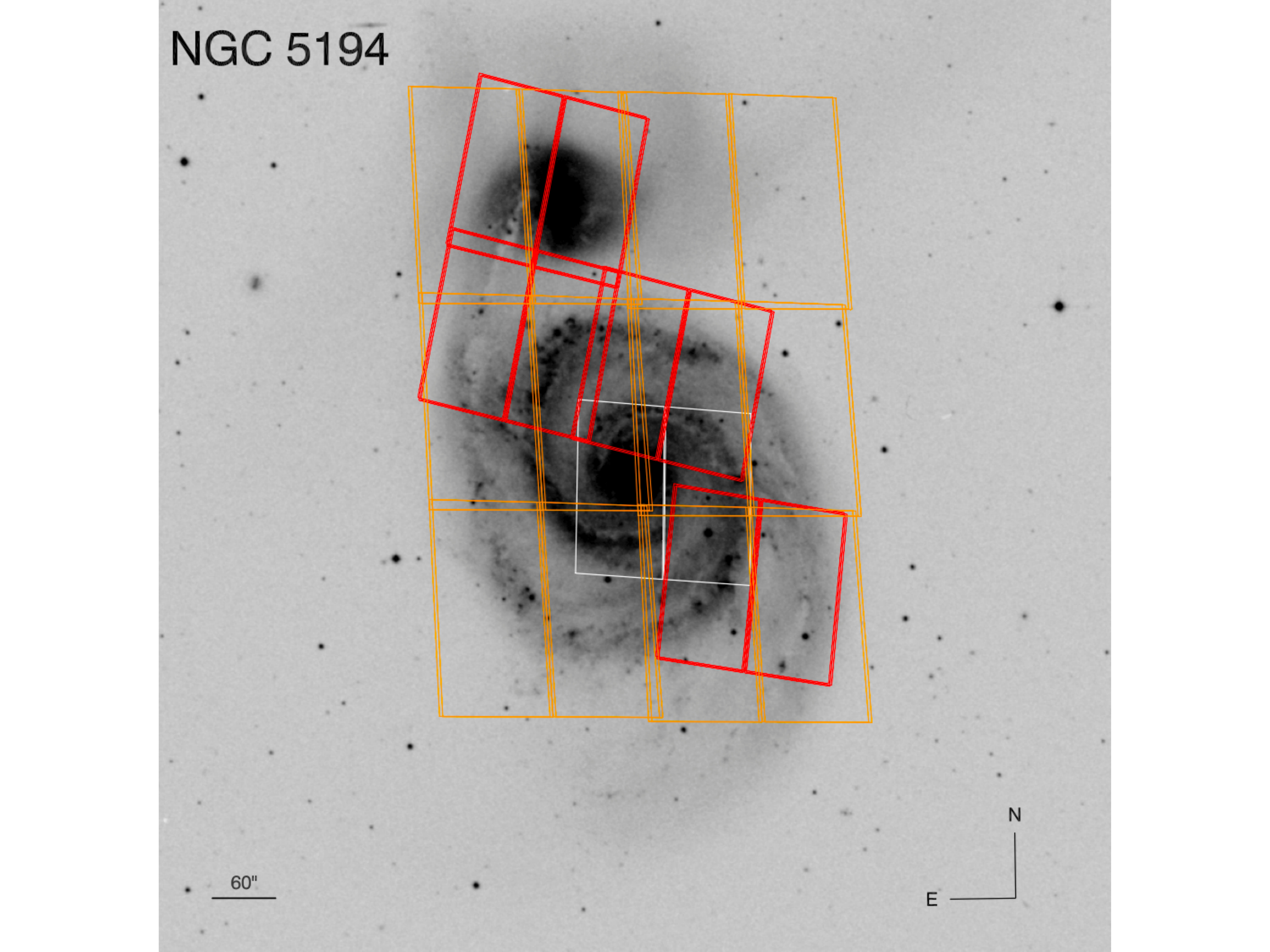}
    \end{subfigure}%
     ~
     ~ \hspace{-1.8cm}
  \begin{subfigure}[t]{0.50\textwidth}
   \includegraphics[width=1\linewidth]{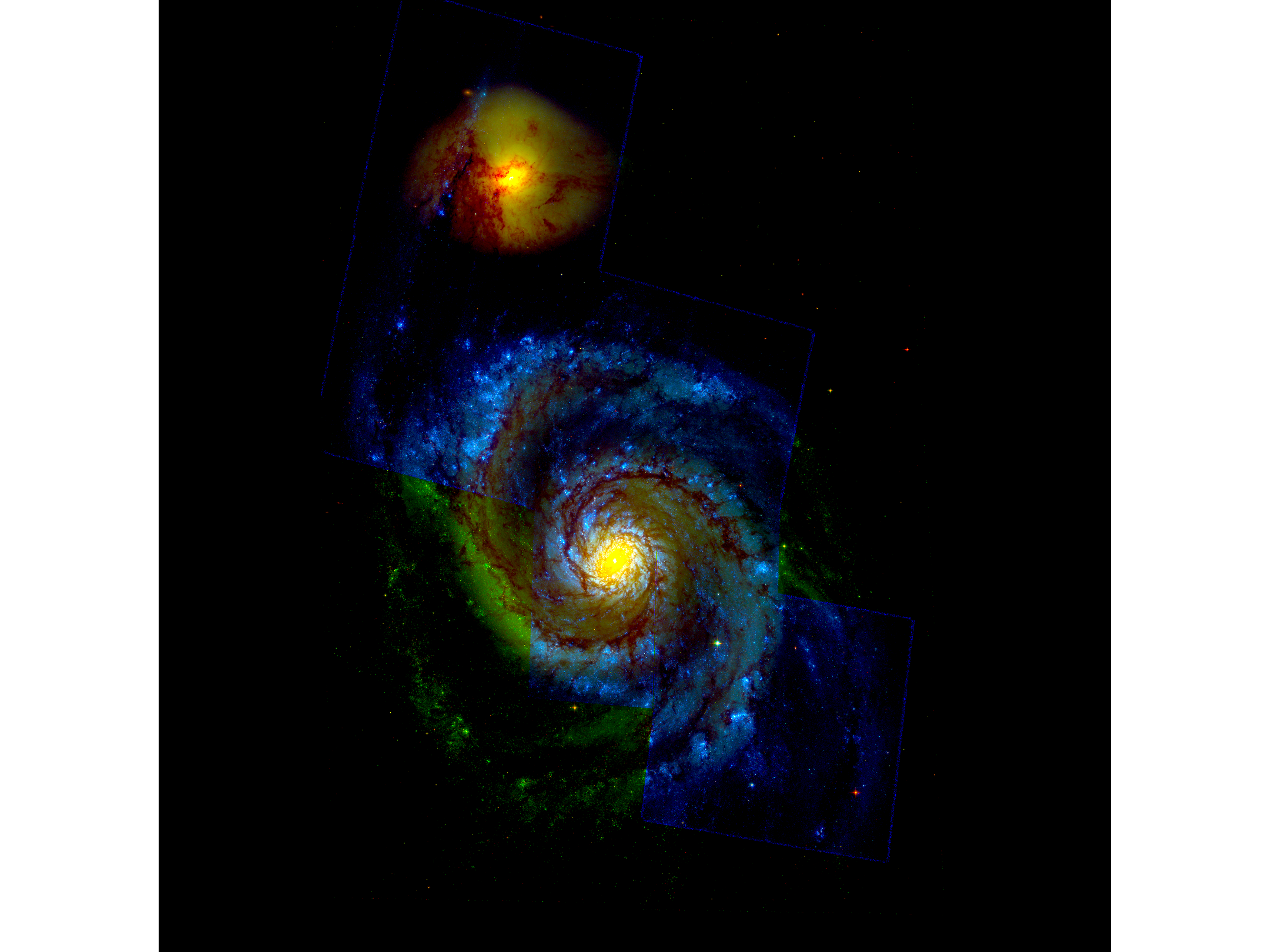}
    \end{subfigure} 
    
    \begin{subfigure}[t]{0.50\textwidth}
   \includegraphics[width=1\linewidth]{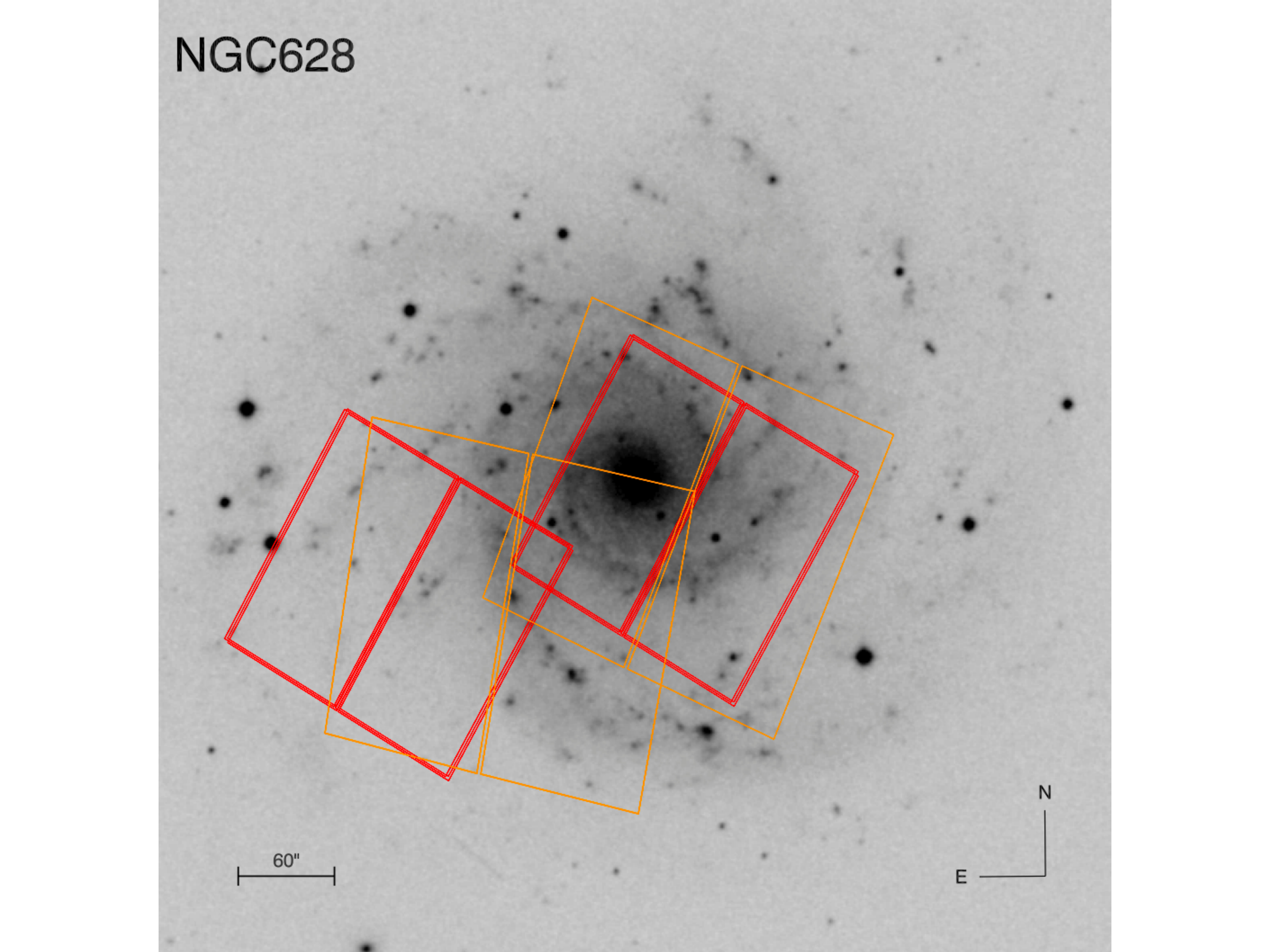}
    \end{subfigure}%
     ~
     ~ \hspace{-1.8cm}
  \begin{subfigure}[t]{0.50\textwidth}
   \includegraphics[width=1\linewidth]{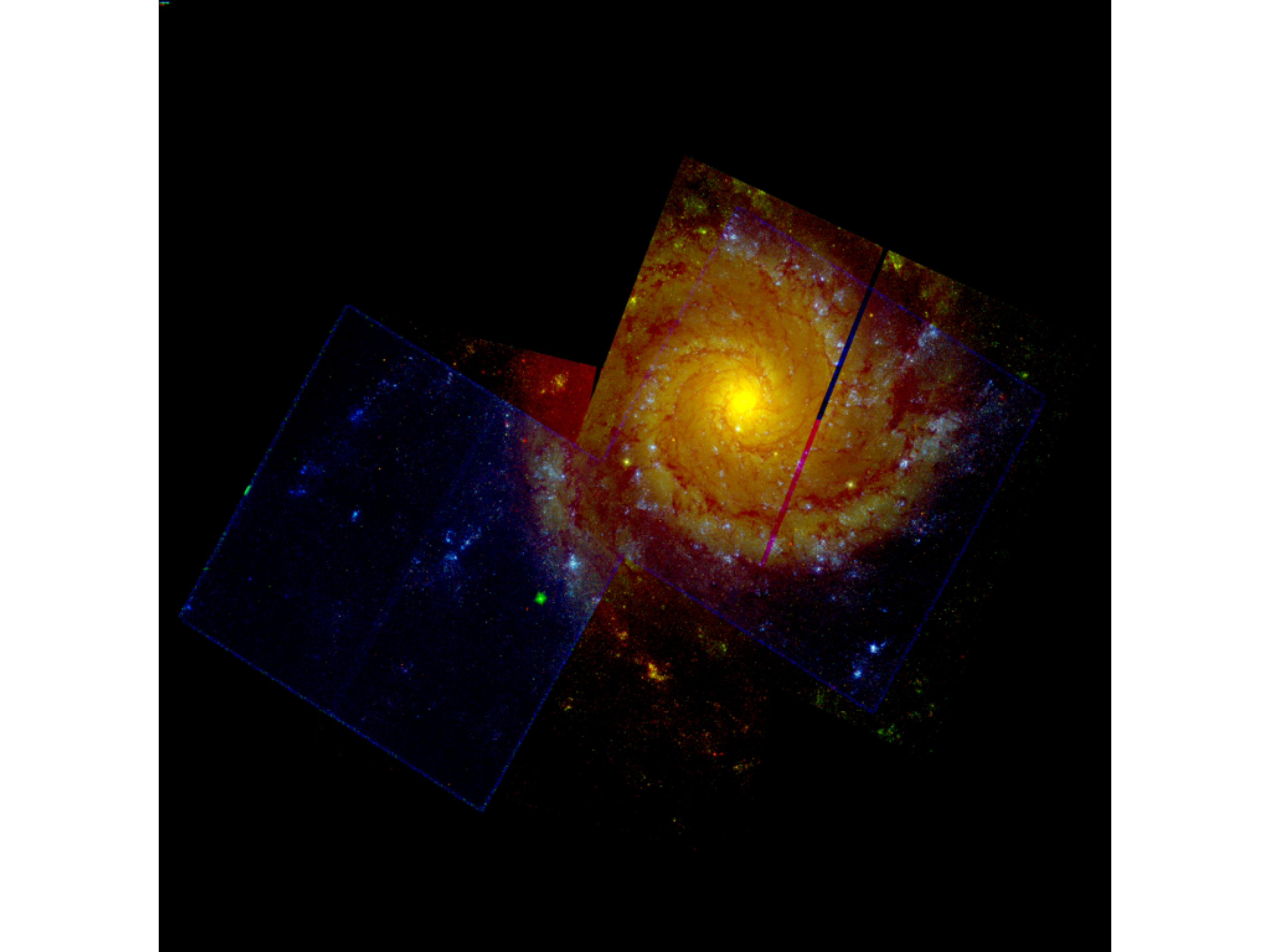}
    \end{subfigure}
\caption{Left: UVIS (red boxes) and ACS (yellow boxes) footprints on DSS images of the galaxies  NGC~1566, M51, and NGC~628 (from top to bottom, respectively). The horizontal bar in the lower left corner denotes the length scale of 60 arcsec. North is up and East to the left. Right: Colour composite images for the same galaxies, constructed from LEGUS imaging in the filters $F275W$ and $F336W$ (blue), $F438W$ and $F555W$ (green), and $F814W$ (red).  The central UVIS pointing (white) of M51a was taken from the observations for proposal 13340 (PI: S. Van Dyk).}
\label{fig:galaxies}
\end{figure*}

NGC~1566, the brightest member of the Dorado group, is a nearly face--on (inclination = $\rm 37.3^{\circ}$) barred grand--design spiral galaxy with strong spiral structures \citep{Debra2}. The distance of NGC~1566 in the literature is uncertain and varies between 5.5 and 21.3~Mpc. In this study, we revised the distance of 13.2~Mpc listed in \cite{C15} and adopted a distance of 18~Mpc \citep{sabbi}. NGC~1566 has been morphologically classified as an SABbc galaxy because of its intermediate--strength bar.  It hosts a low--luminosity active galactic nucleus (AGN) \citep{Combes}. The star formation rate  and stellar mass of NGC~1566 are  $ \rm 2.0 \, M_{\sun }yr^{-1}$and $\rm 2.7 \times 10^{10} \, M_{\sun }$, respectively within the LEGUS field of view \citep{sabbi}.Two sets of spiral arms can be observed in NGC~1566. The inner arms connect with the star--forming ring at 1.7~kpc \citep{S15}, which is covered by the LEGUS field of view (see Fig.~\ref{fig:galaxies}, top panel). The outer arms beyond 100~arcseconds (corresponding to 8 kpc ) are weaker and smoother than the inner arms. 

\subsection{M51a}
M51a (NGC~5194) is a nearby, almost face--on (inclination = $\rm 22 ^{\circ} $) spiral galaxy located at a distance of 7.6 Mpc \citep{Tonry}. It is a grand design spiral galaxy morphologically classified as SAc with strong spiral patterns \citep{Debra2}. M51a is interacting with a companion galaxy, M51b (NGC~5195). M51a has a star formation rate and a stellar mass of  $\rm 6.9 \, M_{\sun } yr^{-1}$and $\rm 2.4 \times 10^{10}M_{\sun }$, respectively \citep{Lee, Both}.
Five UVIS pointings in total were taken through LEGUS observations: 4 pointings cover the center, the north--east, and the south--west regions of M51a, and one covers the companion galaxy M51b.  

\subsection{NGC~628}
 NGC~628 (M74) is the largest galaxy in its group. This nearby galaxy is seen almost face--on ($\rm i = 25.2 ^{\circ}$) and is located at a distance of 9.9 Mpc \citep{Oliver}. It has no bulge \citep{cor} and is classified as a SAc spiral galaxy. Its star formation rate and stellar mass obtained from the extinction--corrected B--band luminosity  are  $ \rm 3.6 \, M_{\sun } yr^{-1}$and $\rm 1.1 \times 10^{10}M_{\sun }$, respectively \citep{Lee, Both}. NGC~628 is a multiple--arm spiral galaxy \citep{Debra} with two well--defined spiral arms. It has weaker spiral patterns than NGC~1566 and M51a \citep{Debra2}. The LEGUS UVIS observations of NGC~628 consist of one central and one east pointing that were combined into a single mosaic for the analysis. 

\section{Stellar cluster samples}
\label{s3}
\subsection{Selection from star cluster catalogues}
In this section, we provide a detailed explanation of the process adopted to select star cluster candidates in our target galaxies. A general description of the standard data reduction of the LEGUS sample can be found in \cite{C15}. A careful and detailed description of the cluster extraction, identification, classification, and photometry is given in \cite{Angela17} and \cite{messa}. Stellar cluster candidates were extracted with SExtractor \citep{Bertin} in the five standard LEGUS filters. The resulting cluster candidate catalogues include sources with a $V$--band concentration index (CI)\footnote{the magnitude difference between apertures of radius 1 pixel and 3 pixels} larger than the CI of star--like sources, which are detected in at least two filters with a photometric error $\leq$ 0.3 mag. The photometry of sources in each filter was corrected for the Galactic foreground extinction \citep{Schlafly}. In order to derive the cluster physical properties such as age, mass, and extinction, the spectral energy distribution (SED) of the clusters was fitted with Yggdrasil stellar population models \citep{Z11}. The uncertainties derived in the physical parameters of the star clusters are on average $\rm 0.1\, \rm dex$ \citep{Angela17}. For some of the LEGUS galaxies, star cluster properties were also estimated based on a Bayesian approach, using the Stochastically Lighting Up Galaxies (SLUG) code \citep{sila}. A detailed and complete explanation of the Bayesian approach can be found in \cite{krumholz}. 

Each source in the stellar cluster catalogue that is brighter than -6 mag  in the $V$--band, and detected in at least four bands, has been morphologically classified via visual inspection by three independent members of the LEGUS team \citep{katie15, Angela17}. The inspected clusters were divided into four morphological classes: Class~1 contains compact, symmetric, and centrally concentrated clusters. Class~2 includes compact clusters with a less symmetric light distribution, Class~3 represents less compact and multi--peak cluster candidates with asymmetric profiles, and Class~4 consists of unwanted objects like single stars, multiple stars, or background sources. Unclassified objects were labeled as Class~0.

In addition, a machine--learning (ML) approach was tested to morphologically classify the stellar clusters in an automated fashion. A forthcoming paper (Grasha et al., in prep.) will present the ML code that was used for cluster classification in the LEGUS survey and the degree of agreement with human classification. An initial comparison between human and ML classification in M51a was already discussed by \citet{messa}. 

For our analysis, we use stellar cluster properties estimated with Yggdrasil deterministic models based on the Padova stellar libraries (see \citet{Z11} for details) with solar metallicity, the Milky Way extinction curve \citep{Cardeli}, and the \cite{Kroupa} stellar initial mass function (IMF).
We also selected clusters based on human visual classification for NGC~628, a combination of human and machine learning classification in NGC~1566, and only machine learning for M51a. Star clusters classified as Class~4 and Class~0 are excluded from our analysis. Among our target galaxies, there is a total number of 1573, 3374, and 1262 star cluster candidates classified as Class 1, 2, and 3 in NGC~1566, M51a, and NGC~628, respectively. 

A detailed description of the properties of the final cluster catalogues of M51a and NGC~628 and their completeness can be found in \citet{messa} and \citet{Angela17}. 

\begin{figure}
\centering
  \begin{subfigure}[t]{0.50\textwidth}
  \includegraphics[width=1\linewidth]{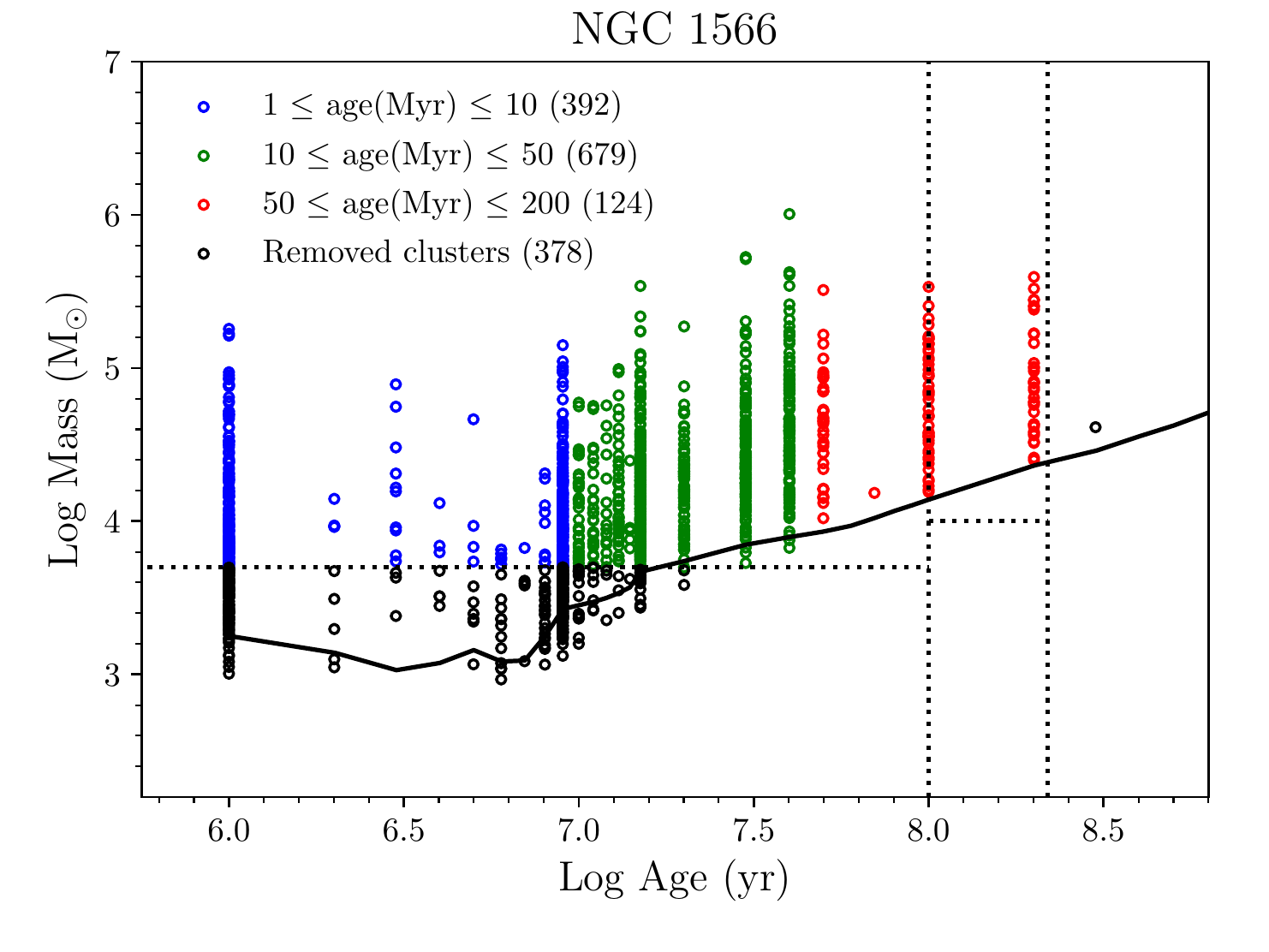}
   \end{subfigure}
     
  \begin{subfigure}[t]{0.50\textwidth}
  \includegraphics[width=1\linewidth]{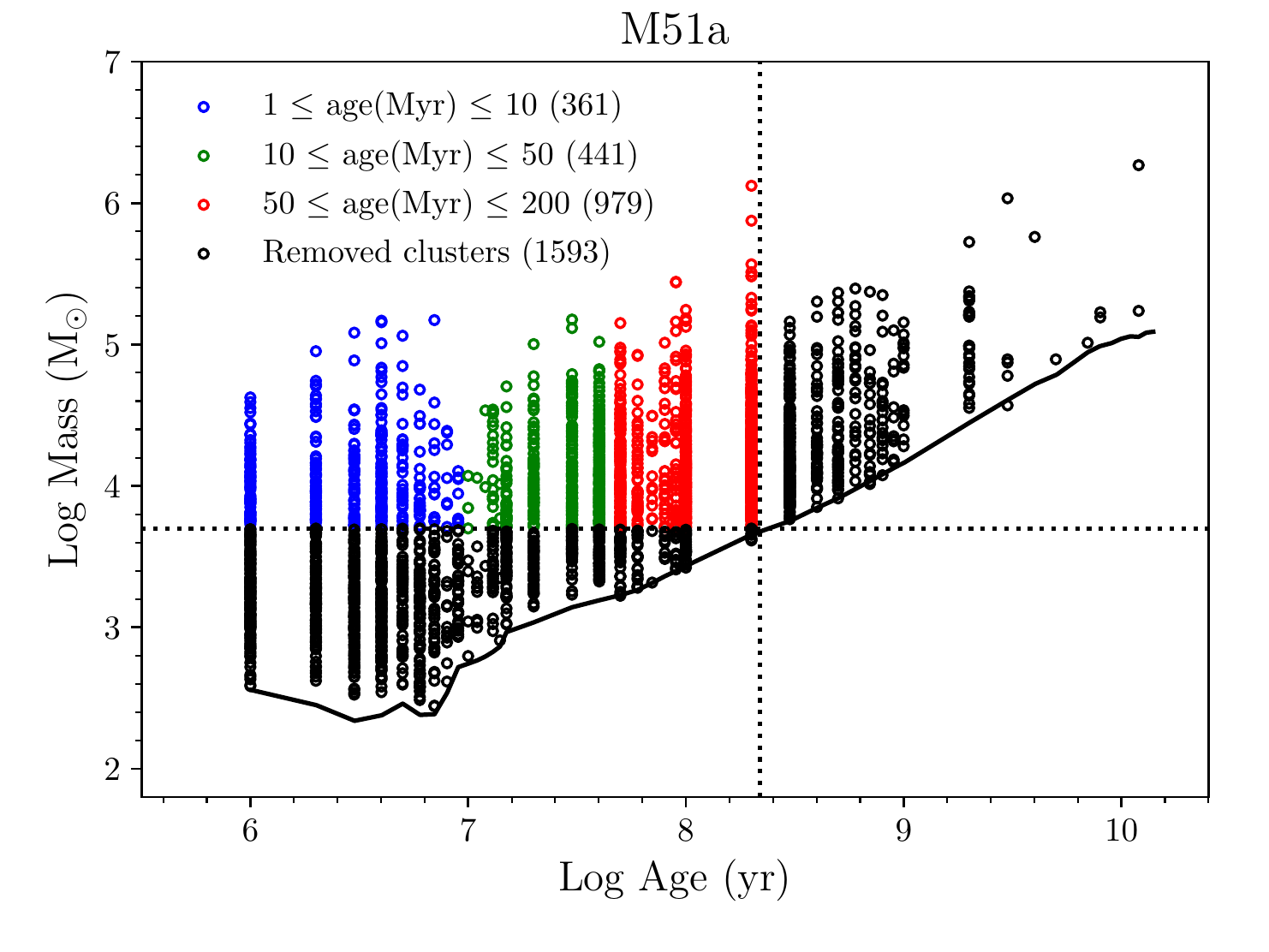}
  \end{subfigure}
    
   \begin{subfigure}[t]{0.50\textwidth}
   \includegraphics[width=1\linewidth]{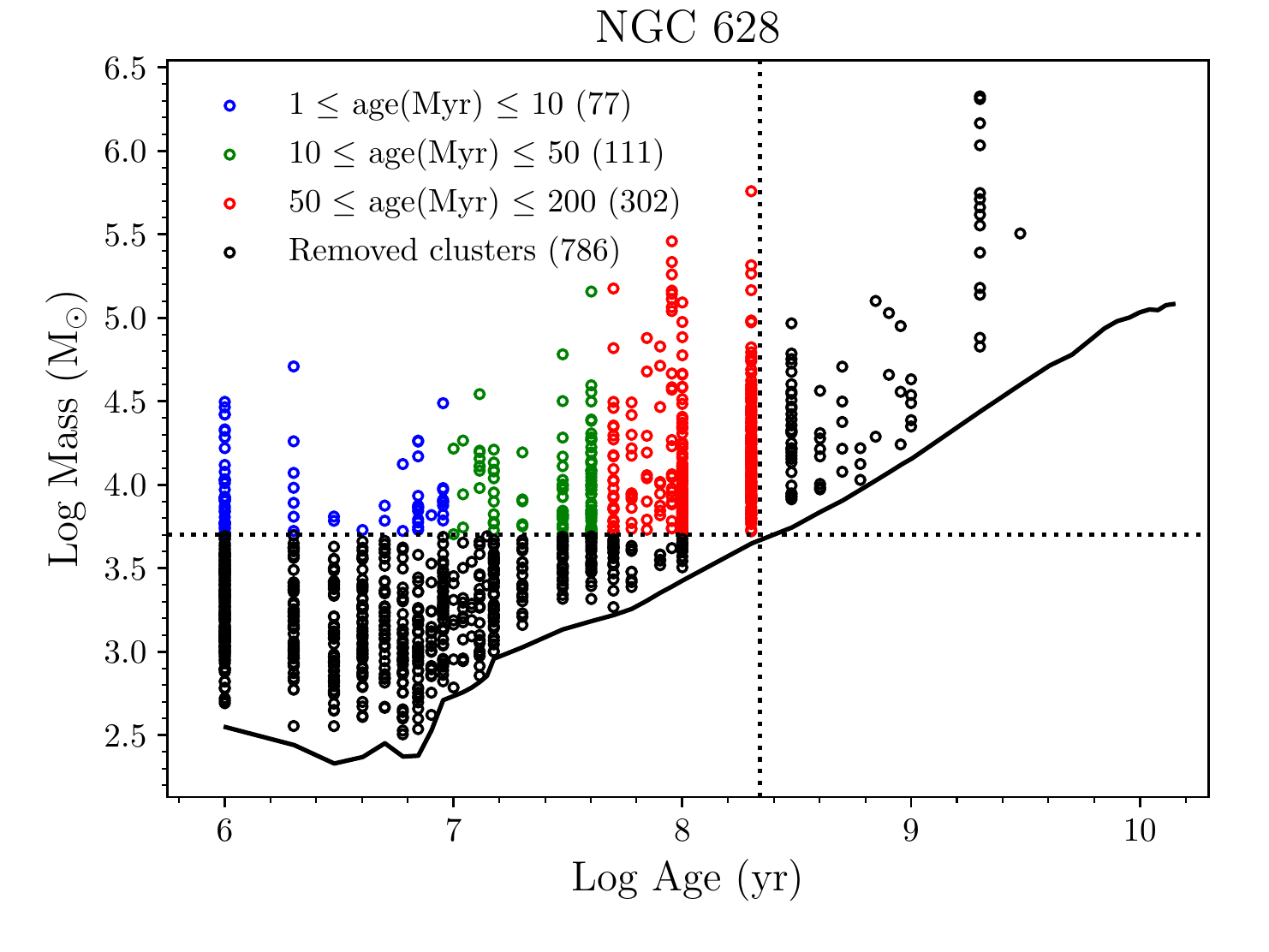}
   \end{subfigure}
    
\caption{Distribution of ages and masses of the star clusters (class 1, 2, and 3) in NGC~1566, M51a, and NGC~628. The colours represent different age bins: blue (the young sample), green (the intermediate--age sample), red (the old sample), and black (excluded star clusters). The number of clusters in each sample is shown in parentheses. The horizontal dotted lines in NGC~1566 show the applied mass cut of 5000 $\rm M_{\sun}$ up to the age of 100~Myr and $\rm 10^{4} \, \rm M_{\sun}$ up to the age of 200~Myr. The applied mass cut of 5000 $\rm M_{\sun}$ up to the age of 200~Myr in M51a and NGC~628 are also by horizontal dotted lines. The solid black line shows the 90\% completeness limit of 23.5 mag in the $V$--band in NGC~1566 and the magnitude cut of $\rm M_{V}$ = -6 mag in M51a, and NGC~628, respectively.}
\label {fig:age_mass}
\end{figure}

\begin{table*}
\centering
\caption{The number of star clusters in the \enquote{young}, \enquote{intermediate--age}, and \enquote{old} samples in our target galaxies.}
\label{tab:cluster sample}
\begin{tabular}{lcccc}

\hline
\hline
Galaxy   & age (Myr) < 10 &  10 $ \leq $ age (Myr) < 50 &  50 $\leq $ age (Myr) $ \leq $ 200 \\ 
\hline
NGC 1566 & 392                                        & 679                                           &124  \\ 
M51a & 361                                       & 441                                            & 979   \\ 
NGC 628 & 77                                       & 111                                            & 302  \\ 

\hline
\hline
\end{tabular}
\end{table*}

\subsection {Selection of star clusters of different ages}
 \label{selection of star clusters of different ages} 
In this study, we use the age of star clusters in our galaxy sample as a tool to find a possible age gradient across the spiral arms predicted by the stationary density wave theory. Therefore, we group star clusters into three different cluster samples according to their ages. 

The estimated physical properties of star clusters based on the Yggdrasil deterministic models are inaccurate for low--mass clusters \citep{krumholz}. A comparison between the deterministic approach based on Yggdrasil models and the Bayesian approach with SLUG  models presented by \cite{krumholz} suggests that the derived cluster properties are uncertain at cluster masses below 5000  $\rm M_{\sun}$. We adopted the same mass cut--off and for NGC~628 and M51a in our analysis. Using the luminosity corresponding to this mass, namely  $\rm M_{V}$ = $-6$ mag ($\rm m_{V}$ = 23.4 and 23.98 mag for NGC~628 and M51a, respectively) results in an age completeness limit of  $\rm \leq 200\, \rm Myr$. In \citet{Angela17} and \citet{messa} the magnitude cut at $\rm M_{V}$ < $-6$ mag is a more conservative limit than the magnitude limit corresponding to 90\% of completeness in the recovery of sources. We have tested our results using different mass cuts as well as by removing any constraint on the limiting mass, and we have not observed any significant change in the age distributions of the clusters as a function of azimuthal distances. Thus, the results presented in \S~\ref{Azimutahl distribution} and \S~\ref{2arms}  are robust against uncertainties in the determination of cluster physical properties.

NGC~1566 is the most distant galaxy within our LEGUS sample. Due to the large distance of this galaxy, the 90\% completeness limit ($\rm m_{V}$ = 23.5 mag) is significantly brighter than $\rm M_{V}$ = $-6$ mag. Therefore, in order to select star clusters in NGC~1566, we used the 90\% completeness limit and a= mass cut of 5000 $\rm M_{\sun}$ for the cluster ages up to 100~Myr and  $\rm 10^{4}  \rm M_{\sun}$ for the 100--200~Myr old star clusters (see Fig.~\ref{fig:age_mass}). Applying these two criteria reduced our cluster samples from 1573 to 1195 clusters for NGC~1566, from 3374 to 1781 clusters for M51a, and from 1262 to 490 for NGC~628.

Then, we selected three cluster samples of different ages for each galaxy as follows:

\begin{description}
  \item[$\bullet$] \enquote{Young} star clusters:  age (Myr) < 10
  \item[$\bullet$] \enquote{Intermediate--age} star clusters: 10 $\rm \leq$  age (Myr) < 50 
  \item[$\bullet$] \enquote{Old} star clusters:  50 $\rm \leq$  age (Myr) $\leq$ 200 
\end{description}

The number of star clusters in the \enquote{young}, \enquote{intermediate--age}, and \enquote{old} samples is shown in Tab.~\ref{tab:cluster sample}. 

Fig.~\ref{fig:age_mass} displays the age--mass diagram of star clusters in NGC~1566, M51a, and NGC~628. The young, the intermediate--age, and the old star cluster samples are shown in blue, green, and red colors, respectively. The excluded star clusters (due to the mass cut) are shown in black. The horizontal and vertical dotted lines show the applied mass cut of $ \rm 5000\, \rm M_{\sun}$  and its corresponding completeness limit at a stellar age of $ 200\, \rm Myr$, respectively. 

\section{Spatial distribution and clustering of star clusters}
\label{location}

\begin{figure*}
\centering
   \begin{subfigure}[t]{0.40\textwidth}
   \includegraphics[width=1\linewidth]{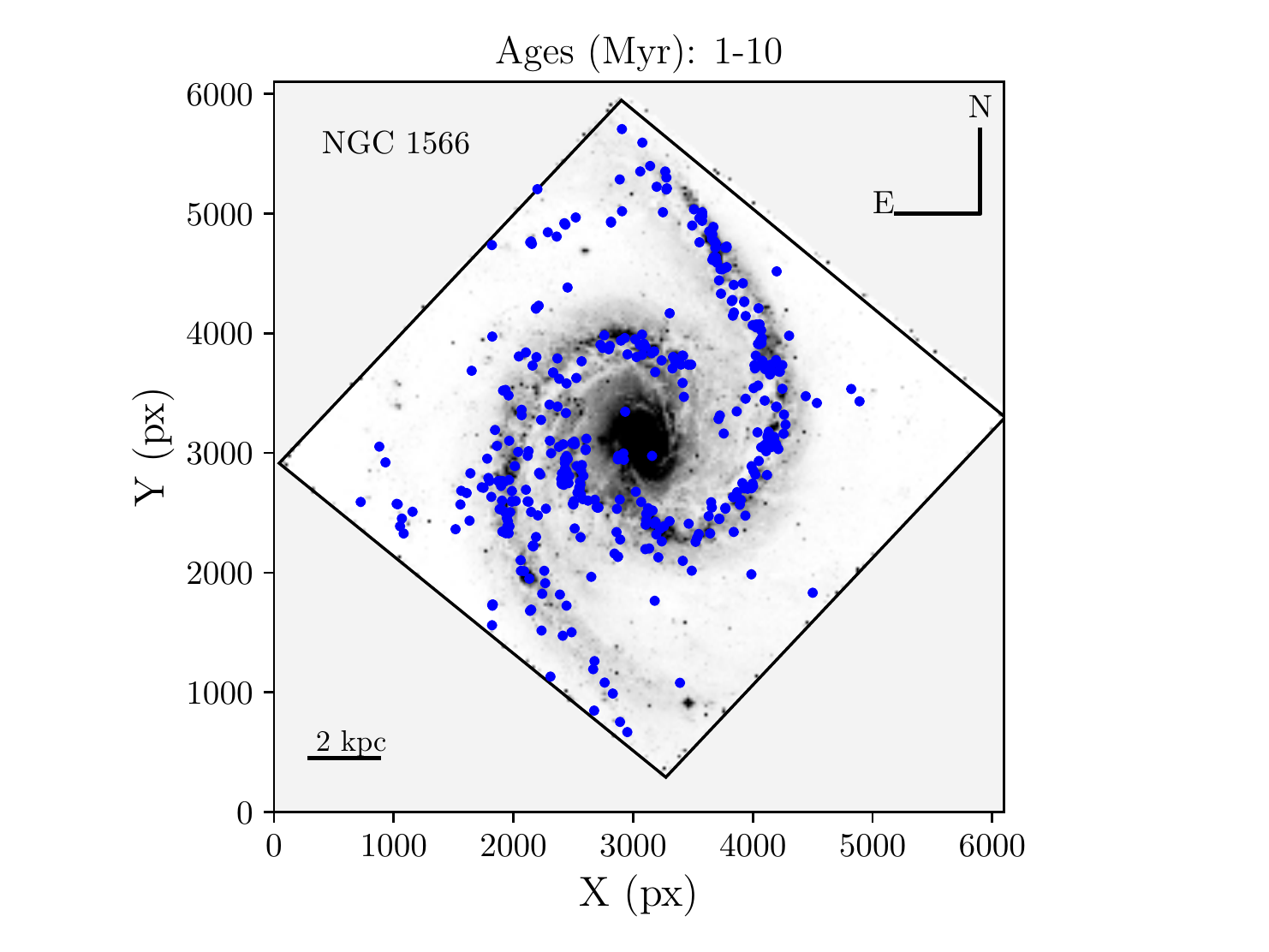}
    \end{subfigure}%
    ~
     ~ \hspace{-1.54cm}
  \begin{subfigure}[t]{0.40\textwidth}
   \includegraphics[width=1\linewidth]{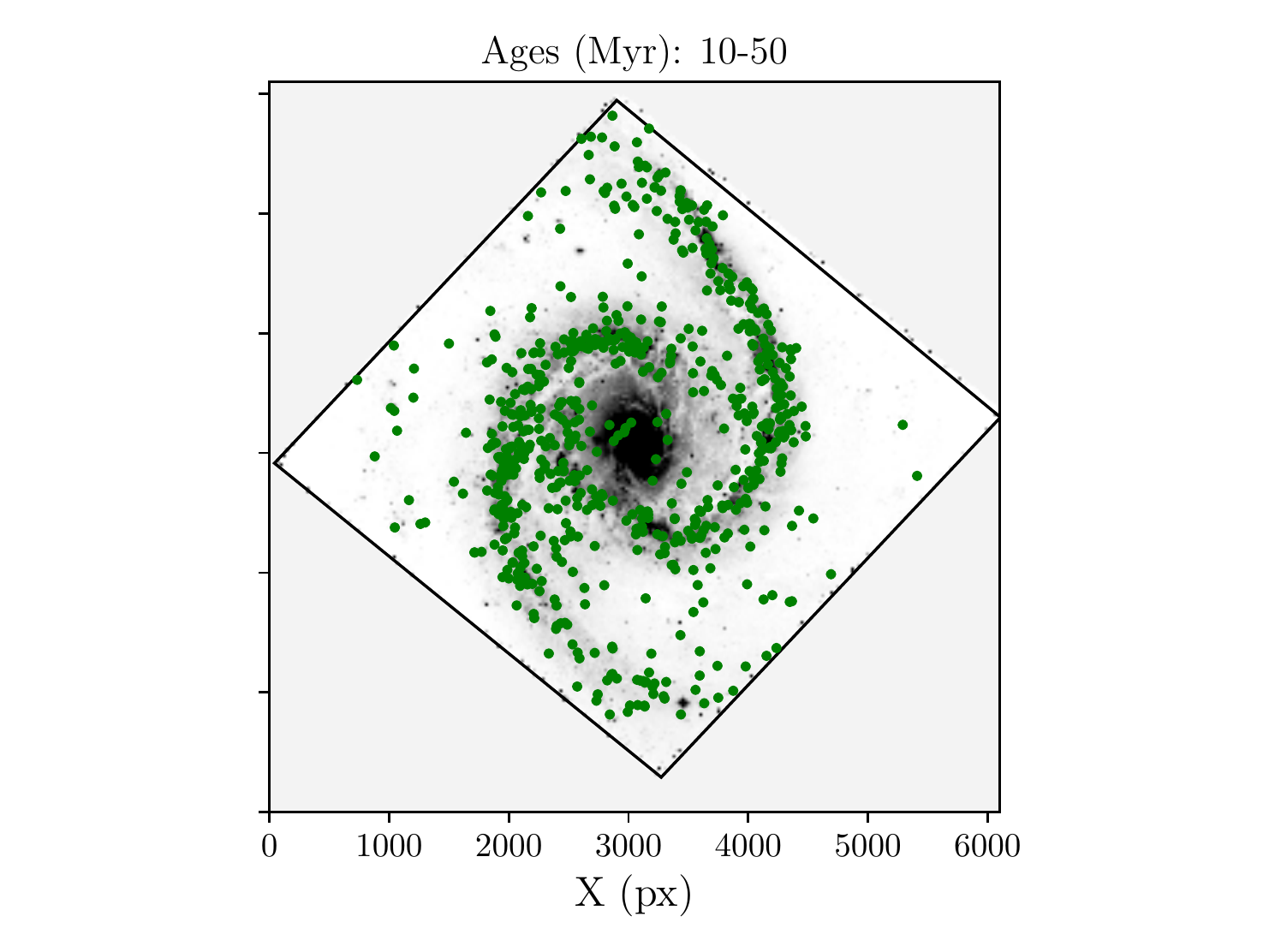}
    \end{subfigure}%
    ~
    ~ \hspace{-1.54cm}
    \begin{subfigure}[t]{0.40\textwidth}
   \includegraphics[width=1\linewidth]{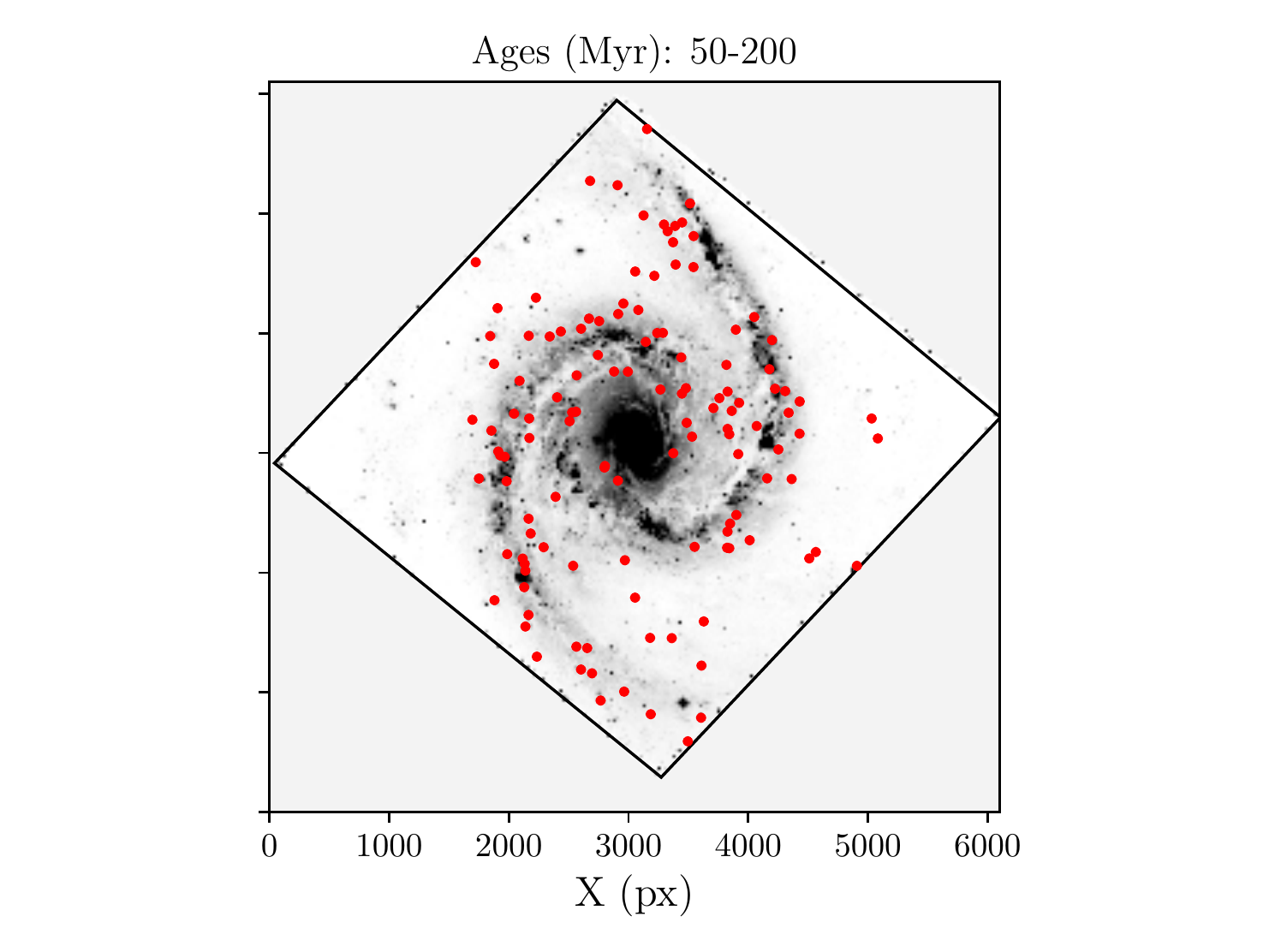}
    \end{subfigure}%
    
    \begin{subfigure}[t]{0.40\textwidth}
   \includegraphics[width=1\linewidth]{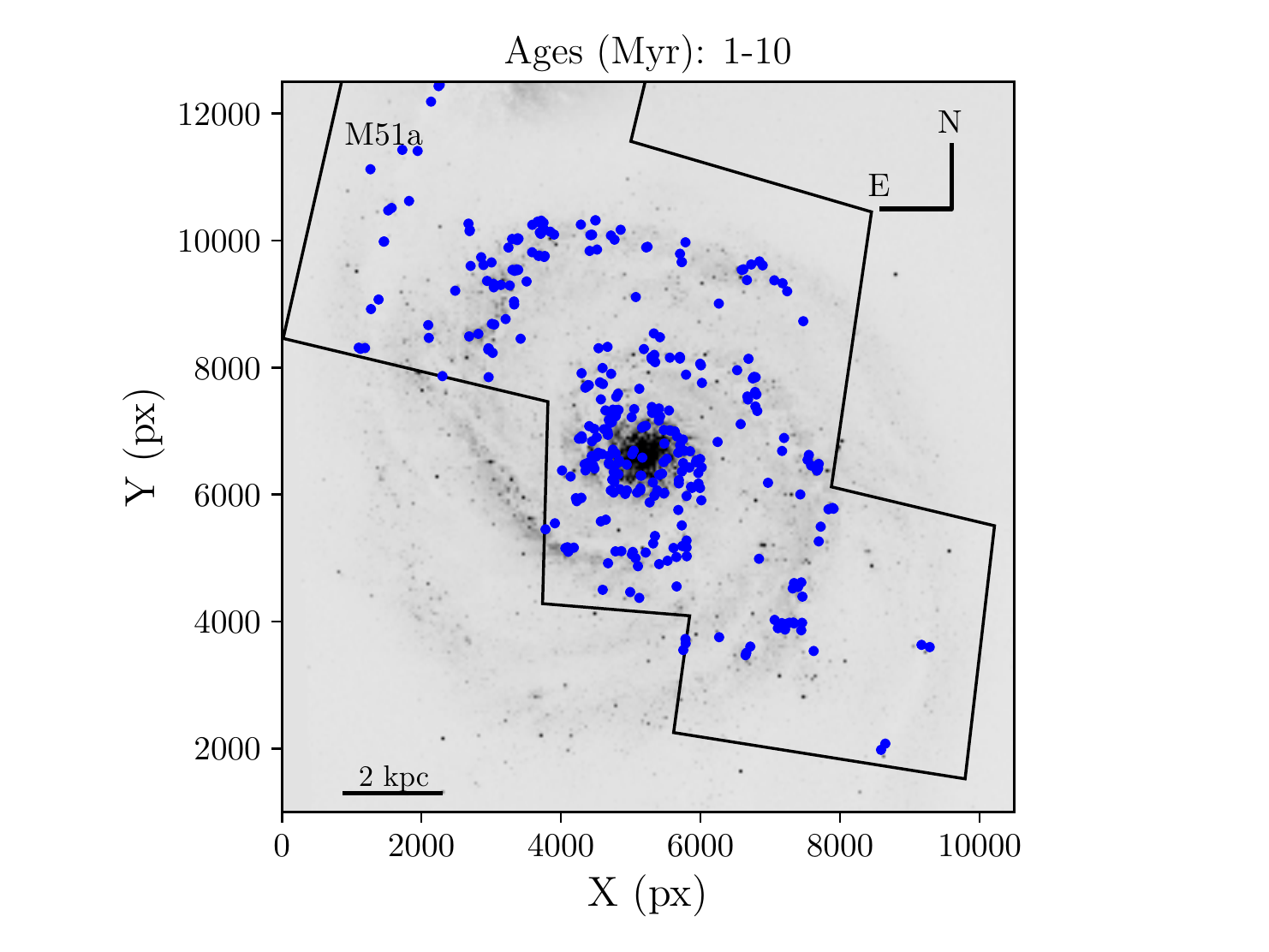}
    \end{subfigure}%
    ~
     ~ \hspace{-1.54cm}
  \begin{subfigure}[t]{0.40\textwidth}
   \includegraphics[width=1\linewidth]{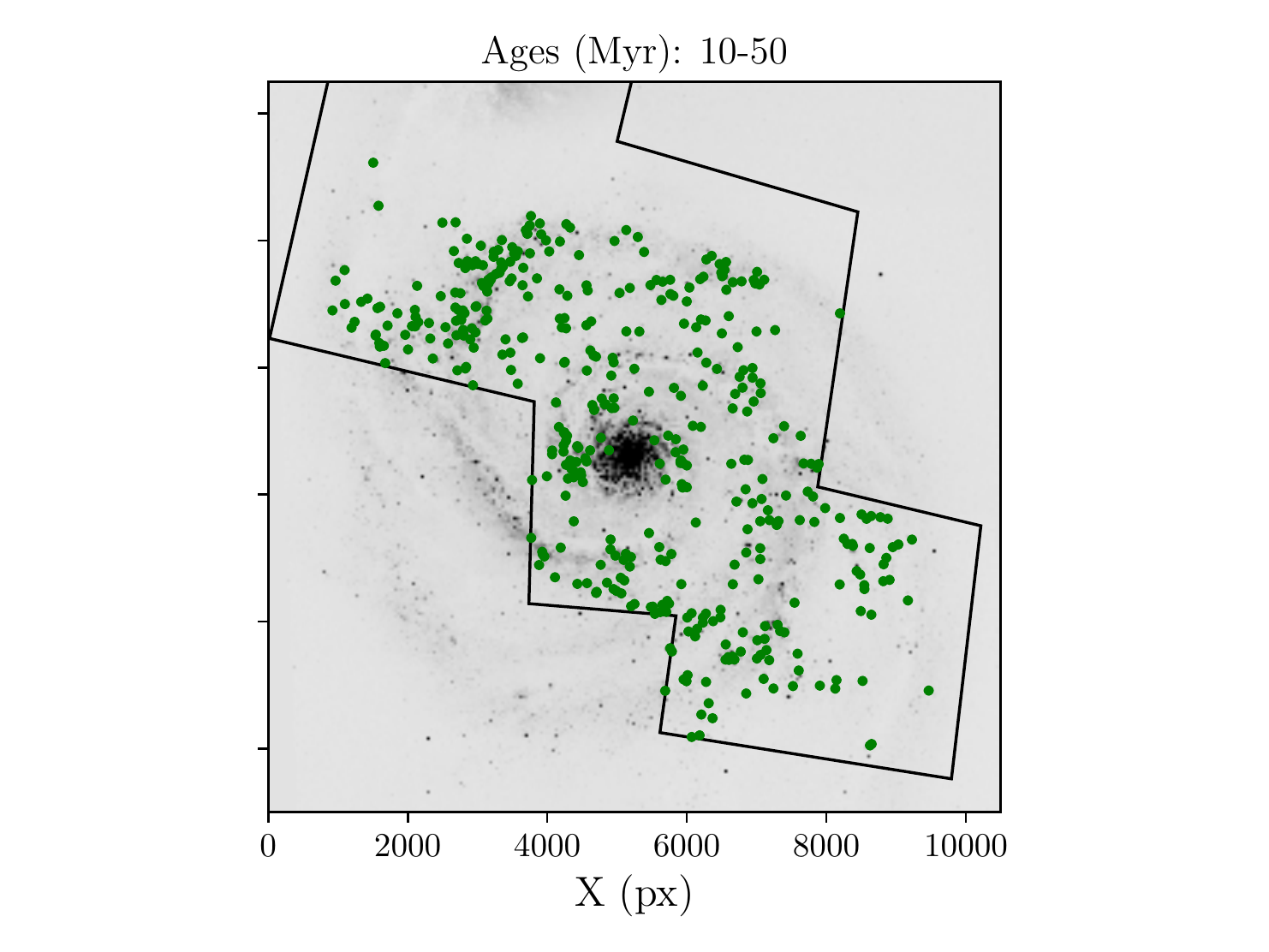}
    \end{subfigure}%
    ~
    ~ \hspace{-1.54cm}
    \begin{subfigure}[t]{0.40\textwidth}
   \includegraphics[width=1\linewidth]{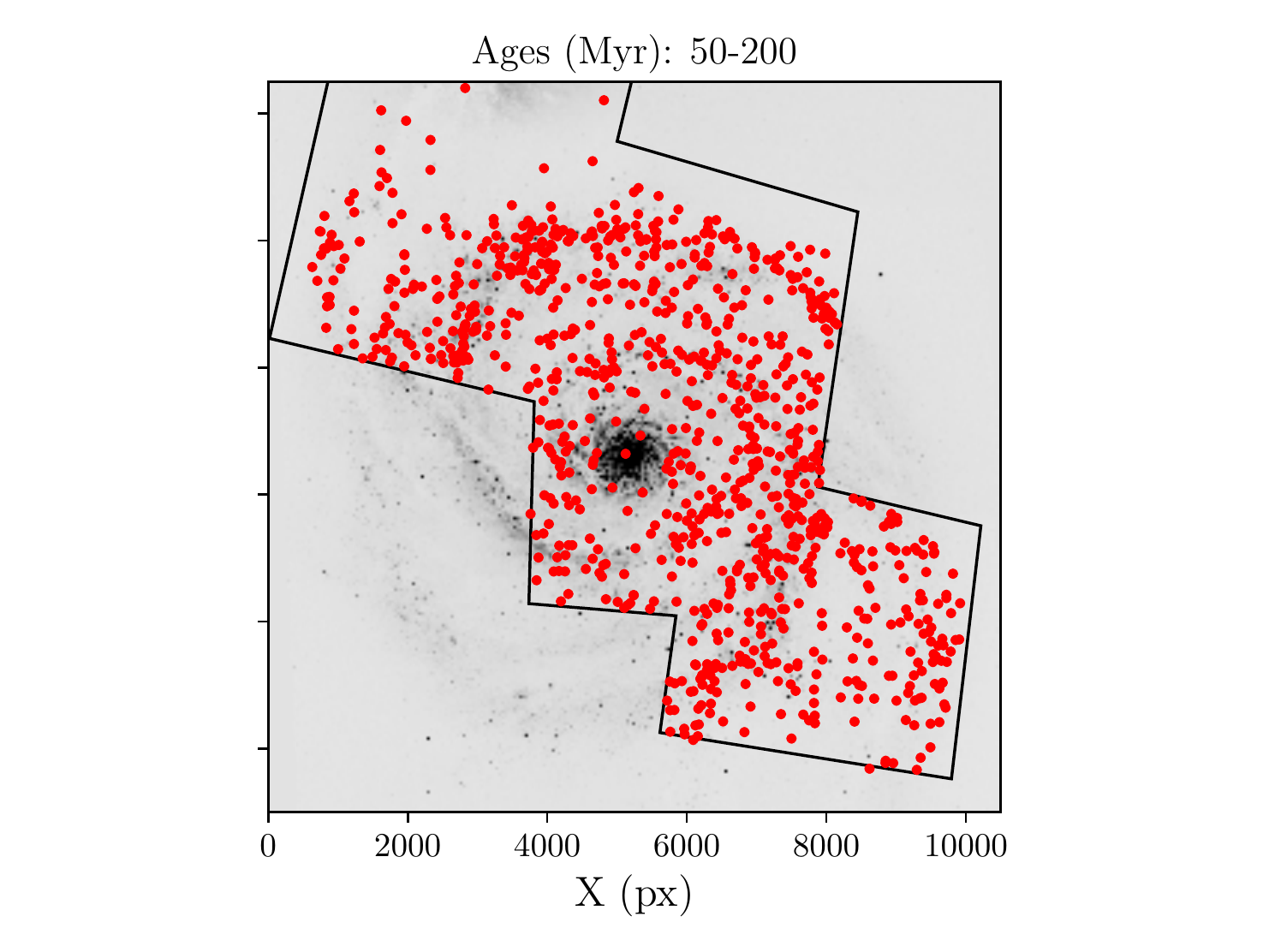}
    \end{subfigure}%
    
    \begin{subfigure}[t]{0.40\textwidth}
   \includegraphics[width=1\linewidth]{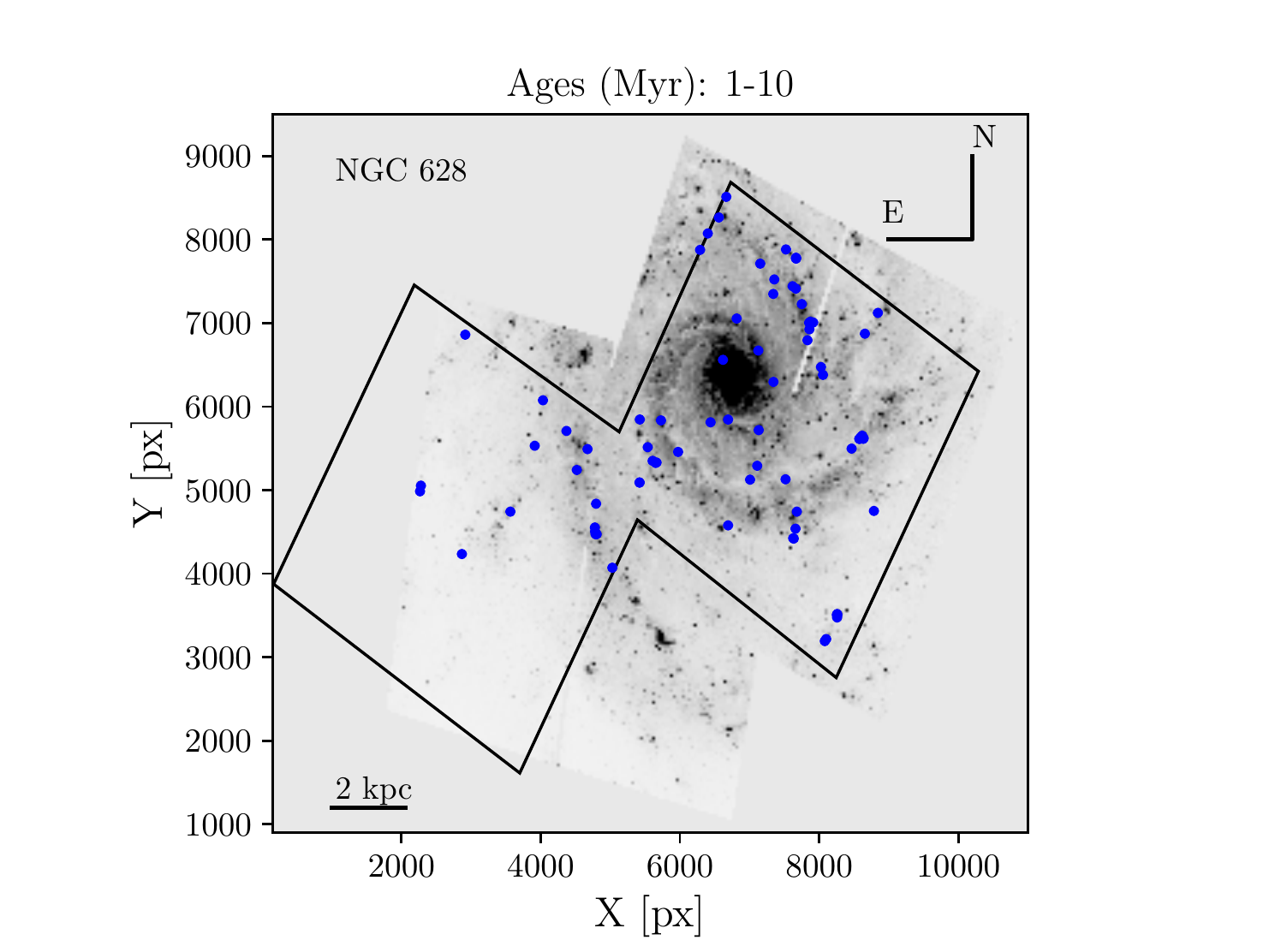}
    \end{subfigure}%
    ~
     ~ \hspace{-1.54cm}
  \begin{subfigure}[t]{0.40\textwidth}
   \includegraphics[width=1\linewidth]{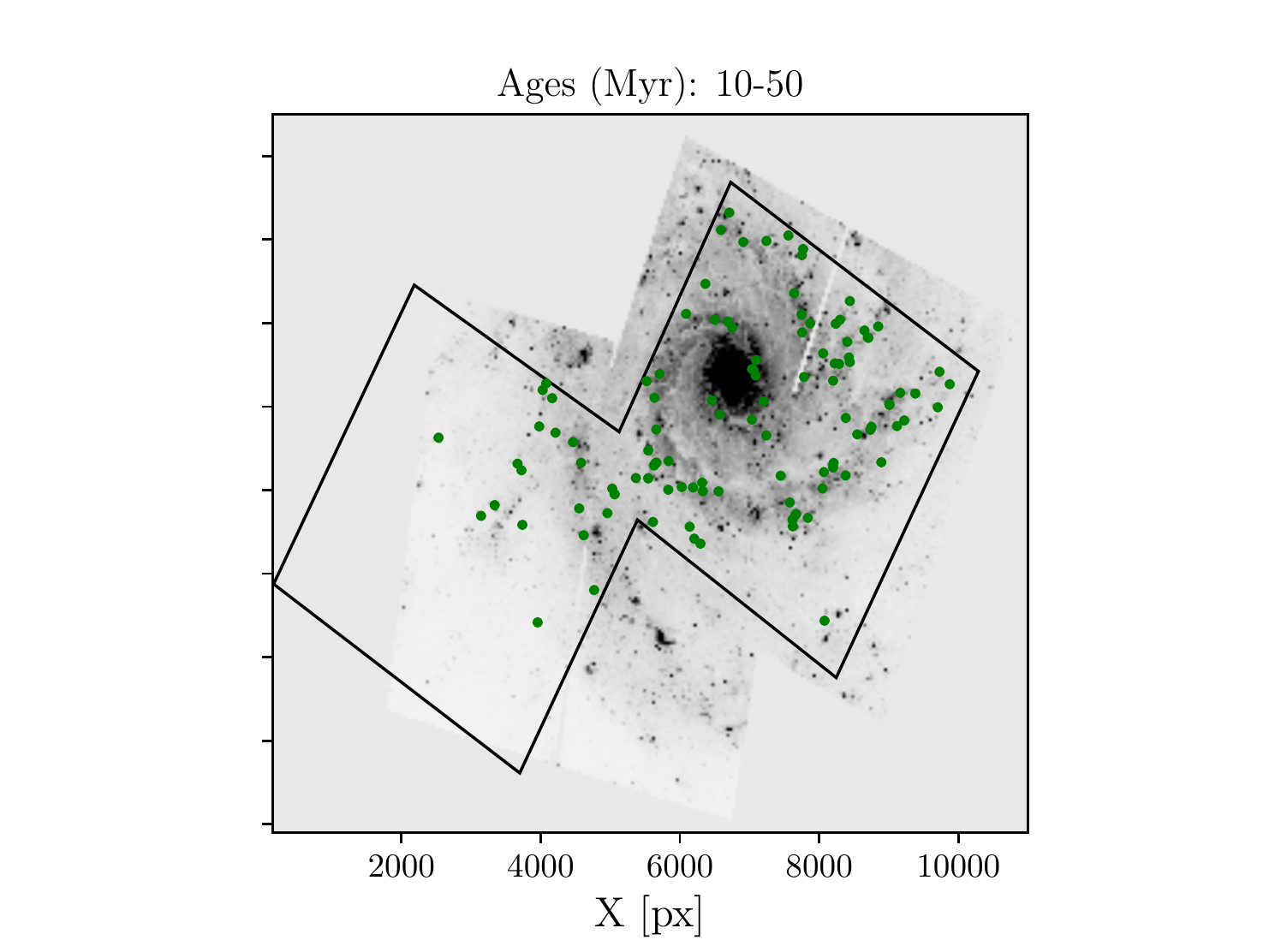}
    \end{subfigure}%
    ~
    ~ \hspace{-1.54cm}
    \begin{subfigure}[t]{0.40\textwidth}
   \includegraphics[width=1\linewidth]{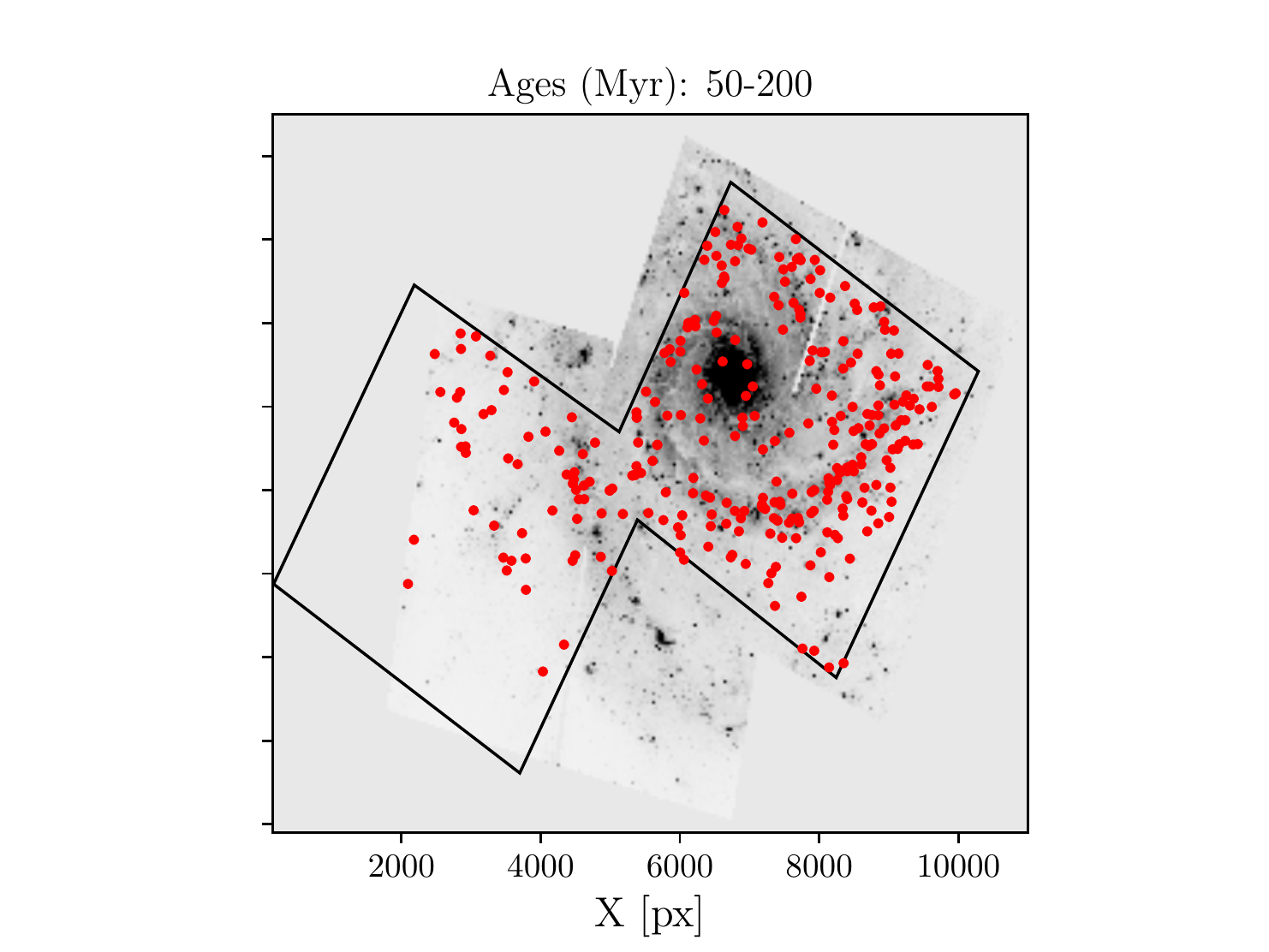}
    \end{subfigure}%

\caption{The spatial distribution of star clusters of different age in the galaxies NGC~1566, M51a, and NGC~628 (from top to bottom) superimposed on the $B$--band images. The blue, green, and red circles show the young (age (Myr) < 10), intermediate--age (10 $ \leq $ age (Myr) < 50), and old star clusters (50 $\leq $ age (Myr) $ \leq $ 200), respectively. The black outlines show the UVIS footprints. The horizontal bar in the lower left corner denotes the length of 2~kpc. North is up and East to the left.}
\label {fig:clusters}
\end{figure*}

In Fig.~\ref{fig:clusters}, we plot the spatial distribution of star clusters of different ages in the galaxies NGC~1566, M51a, and NGC~628. The young, intermediate--age, and old stellar cluster samples are shown in blue, green, and red, respectively. In general, we observe a similar trend in our target galaxies: First, the young and the intermediate--age star clusters mostly populate the spiral arms rather than the interarm regions. This is particularly evident for NGC 1566 and M51a, which show strong and clear spiral structures in young and intermediate--age star clusters. Second, the old star clusters are less clustered and more widely spread compared to the young and intermediate–age star cluster samples.

Our findings are similar to other literature results on the spatial distribution of star clusters of different ages: \cite{D17}, using LEGUS HST data found that in NGC~1566 the 100~Myr old star clusters clearly trace the spiral arms while in NGC~628 star clusters older than 10~Myr show only weak spiral structures. \cite{chandar17}, using other HST data observed that M51a shows weak spiral structure in older star clusters (>100~Myr). 

Clustering of star clusters has been observationally investigated for a number of local star--forming galaxies \citep[e.g.,][]{Efremov,EE}. In a detailed study of clustering of the young stellar population in NGC~6503  based on the LEGUS observations, \cite{d15} found that younger stars were more clustered compared to the older ones. \cite{katie15} investigated the spatial distribution of the star clusters in NGC~628 from the LEGUS sample. Their findings confirmed that the degree of the clustering increases with decreasing age. More recently, \citet{grasha17a} studied the hierarchical clustering of young star clusters in a sample of six LEGUS galaxies. Their results suggested that the youngest star clusters are strongly clustered and the degree of clustering quickly drops for clusters older than 20 Myr and the galactic shear appears to drive the largest sizes of the hierarchy in each galaxy \citet{grasha17b}.

Adopting a similar approach as \cite{katie15}, we use the two--point correlation function to test whether or not the clustering distribution of the clusters in our selected age bins shows the expected age dependence. The two--point correlation function $\rm \omega (\theta)$ is a powerful statistical tool for quantifying the probability of finding two clusters with an angular separation $\rm \theta$ against a random, non--clustered distribution \citep{peebles}. Here we use the Landy--Szalay \citep{LS} estimator, which has little sensitivity to the presence of edges and masks in the data:
\begin{equation}
\omega(\theta) = \frac{r (r-1)}{n (n-1)}\frac{DD}{RR} - \frac{(r-1)}{n}\frac{DR}{RR}+1,
\end{equation}
where $ n$ and $r$ are the total number of data and random points, respectively. $ DD$, $ RR$, and $ DR$ are the total numbers of data--data, random--random, and data--random pair counts with a separation $\rm \theta \pm d\theta$, respectively. We construct a random distribution of star clusters that has the same sky coverage and masked regions (e.g., the ACS chip gap) as the images of each galaxy. 

\begin{figure}
\centering
   \begin{subfigure}[t]{0.50\textwidth}
   \includegraphics[width=1\linewidth]{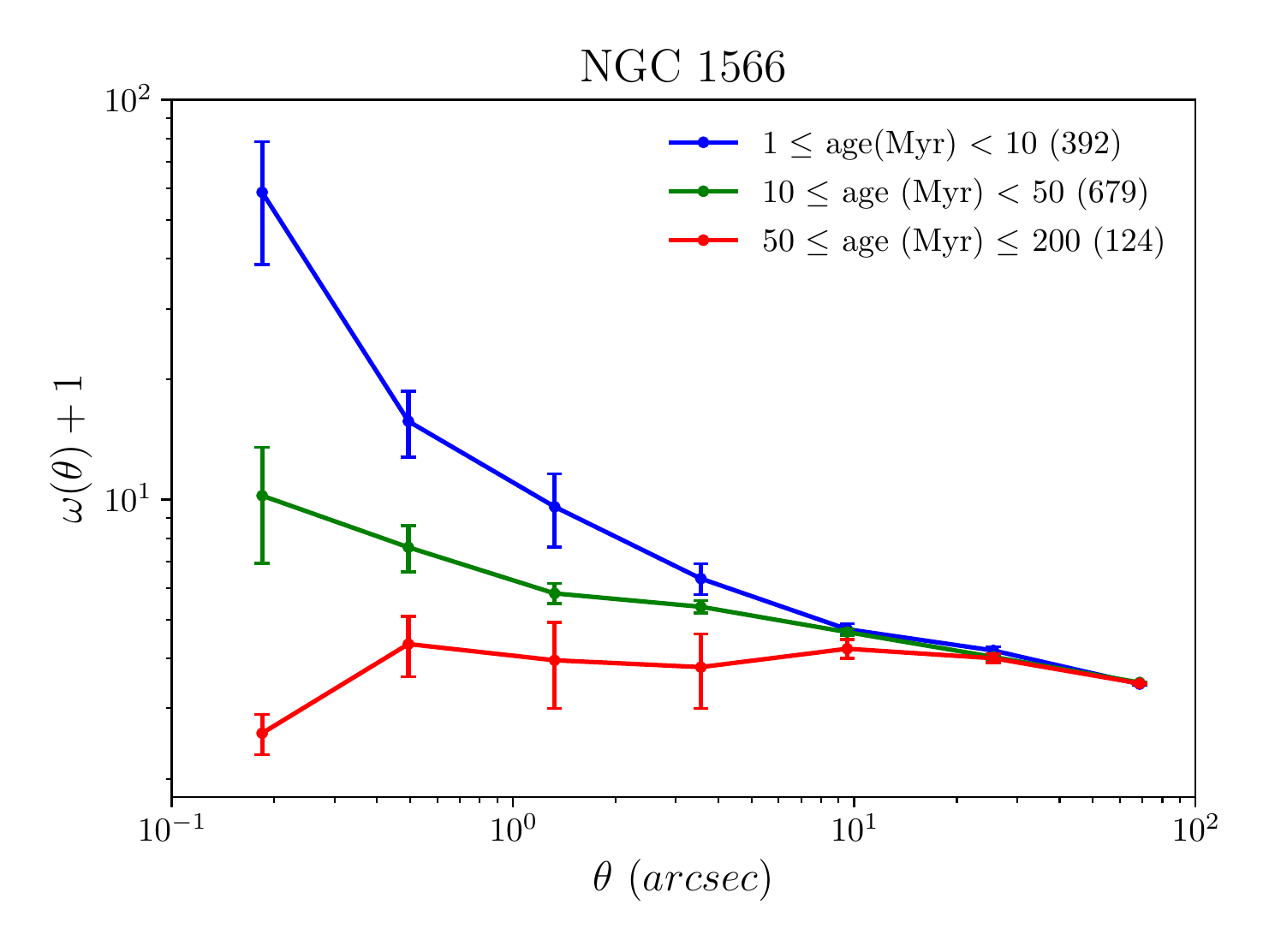}
    \end{subfigure}
   
   \begin{subfigure}[t]{0.50\textwidth}
   \includegraphics[width=1\linewidth]{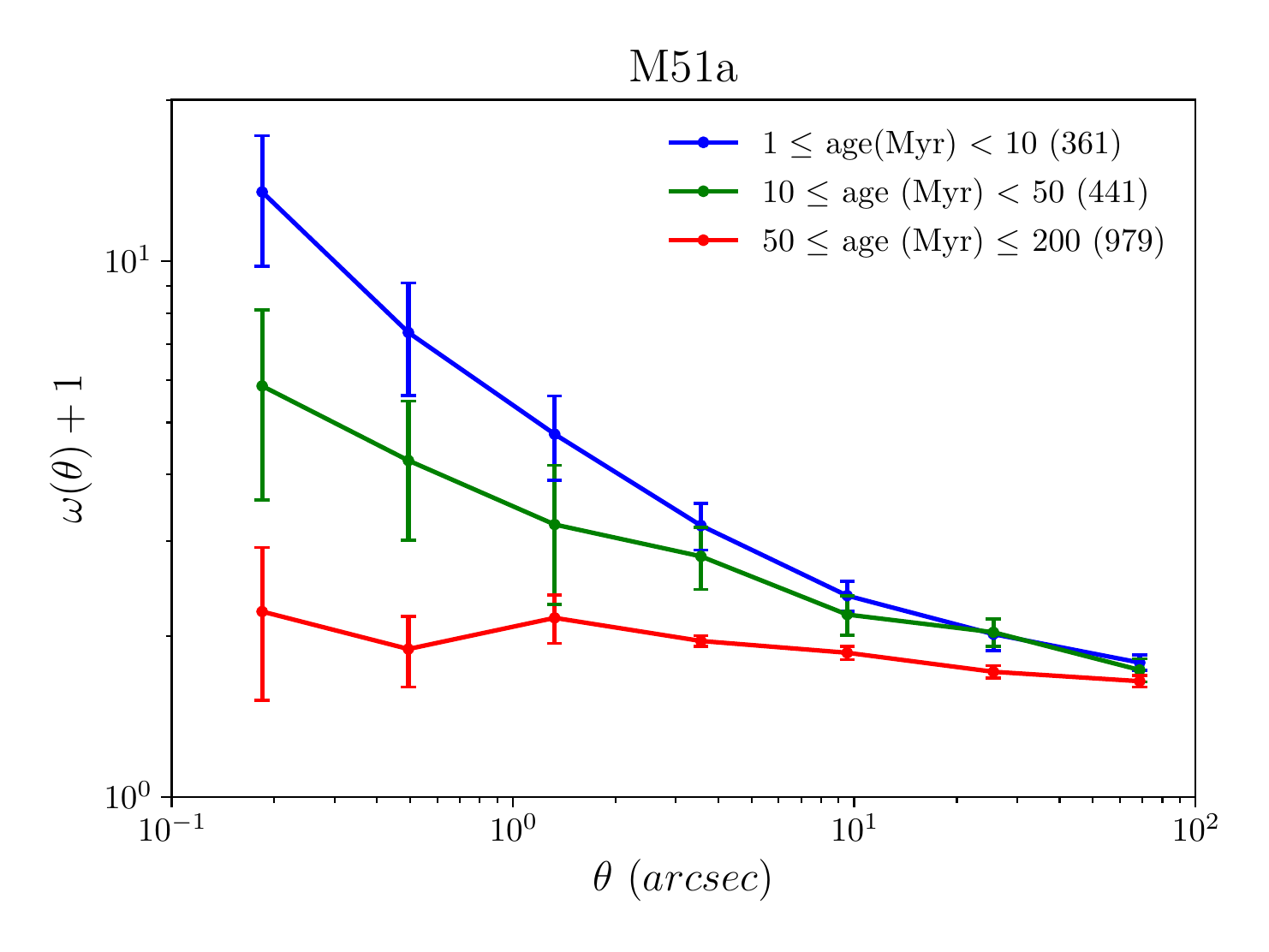}
\end{subfigure}

   \begin{subfigure}[t]{0.50\textwidth}
   \includegraphics[width=1\linewidth]{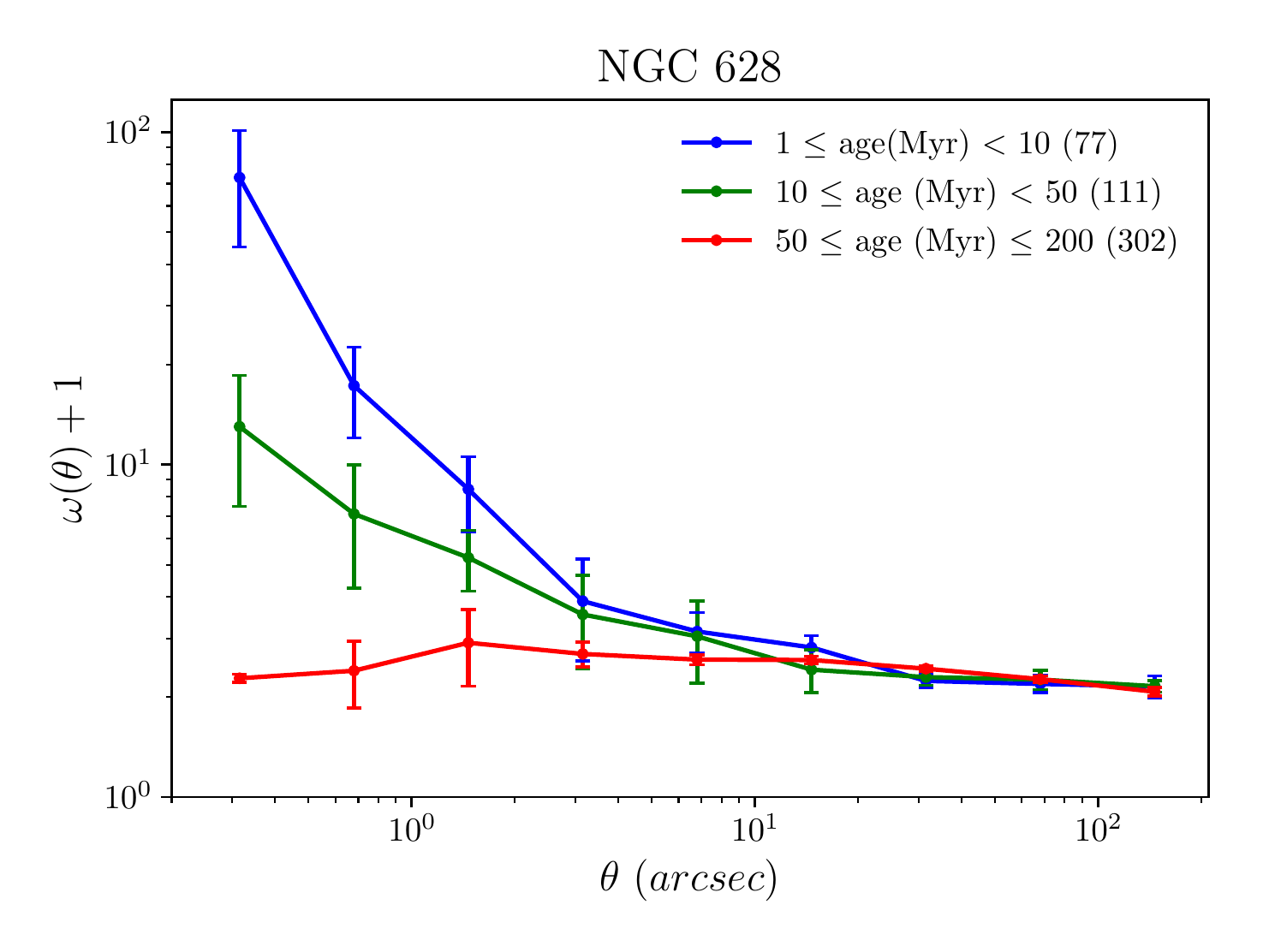}
\end{subfigure}
\caption{ two--point correlation function for the star cluster samples of different ages as a function of angular distance (arcseconds) in NGC~1566, M51a, and NGC~628. The young, intermediate--age, and old star cluster samples are shown in blue, green, and red, respectively. The error bars were computed based on a bootstrapping method. The number of star clusters in each age bin are listed in parentheses.}
\label {fig:two_point}
\end{figure}

Fig.~\ref{fig:two_point} displays the two--point correlation function for the star clusters in different age bins as defined for our galaxy samples. The blue, green, and red colours represent the young, intermediate--age, and old star cluster samples in each galaxy, respectively. The error bars on the two--point correlation function were estimated using a bootstrapping method with 1000 bootstrap resamples. 

The general distribution of the star cluster samples in the target galaxies shows a similar trend: Independent of the presence of spiral arms, young clusters show hierarchical structure, whilst the old star clusters show a non--clustered, smooth distribution.

\section{Are the spiral arms static density waves?} 
\label{Azimutahl distribution}
As discussed in \S~\ref{Introduction}, the stationary density wave theory foresees that the age of stellar clusters inside the corotation radius increases with increasing distance from the spiral arms. In other words, we expect to find a shift in the location of stellar clusters with different ages. 

In order to test whether the distribution of star clusters of different ages in our target galaxies agrees with the expectations from the stationary density wave theory, we need to quantify the azimuthal offset between star clusters of different ages.

\subsection{Spiral arm ridge lines definition}
First of all, we need to locate the spiral arms of our galaxy sample. We wish to define a specific location in each spiral arm so we can measure the relative positions of the star clusters in a uniform way. We use the dust lanes for this purpose because they are narrow and well--defined on optical images. 

As gas flows into the potential minima of a density wave, it gets compressed and forms dark dust lanes in the inner part of the spiral arms, where star formation is then likely to occur \citep{R69}. We have used the $B$--band images for this purpose since most of the emission is due to young OB stars and dark obscuring dust lanes can be better identified in this band.

\begin{figure}
\centering
\begin{subfigure}[t]{0.50\textwidth}
   \includegraphics[width=1\linewidth]{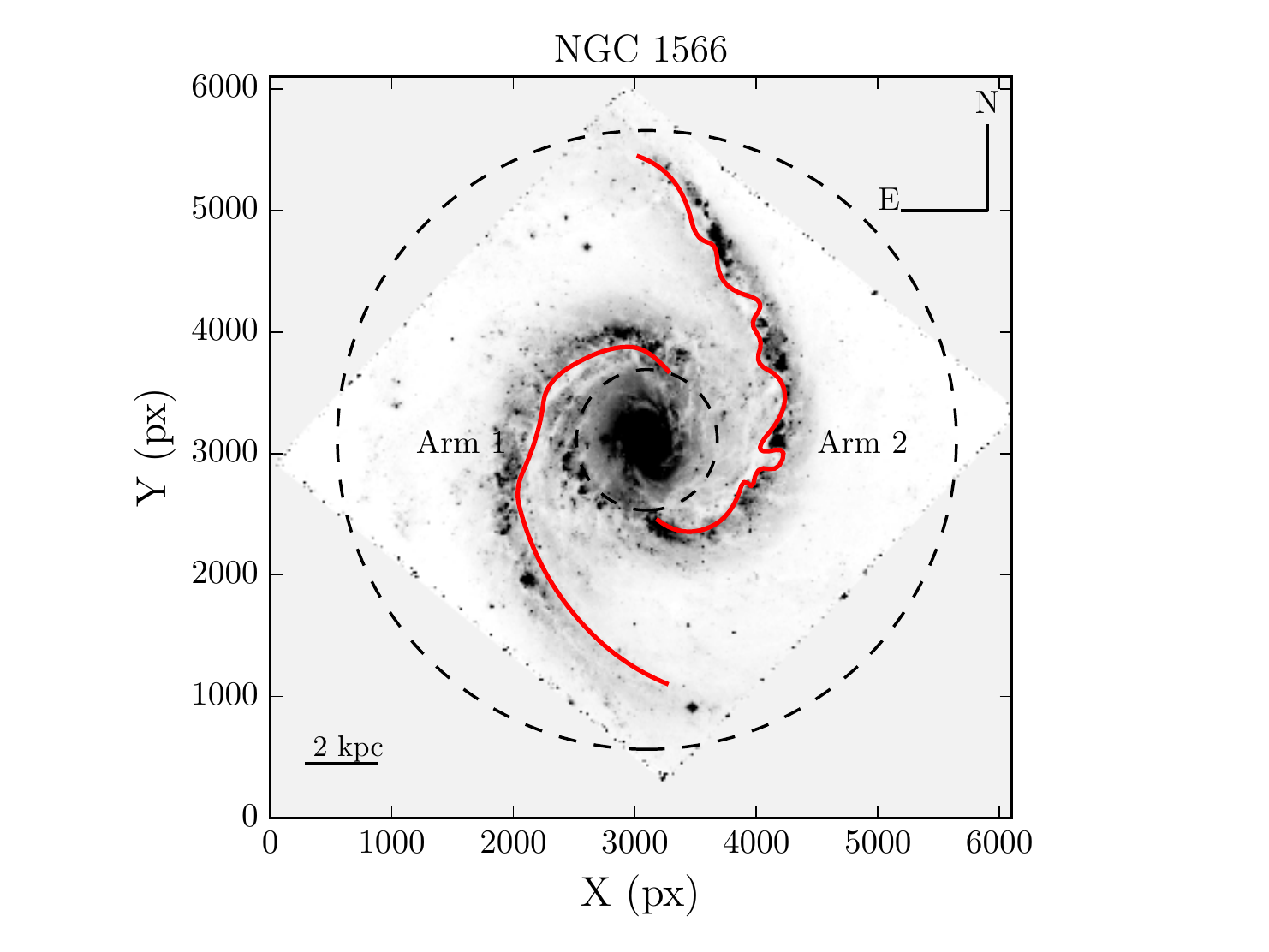}
\end{subfigure}

  \begin{subfigure}[t]{0.50\textwidth}
   \includegraphics[width=1\linewidth]{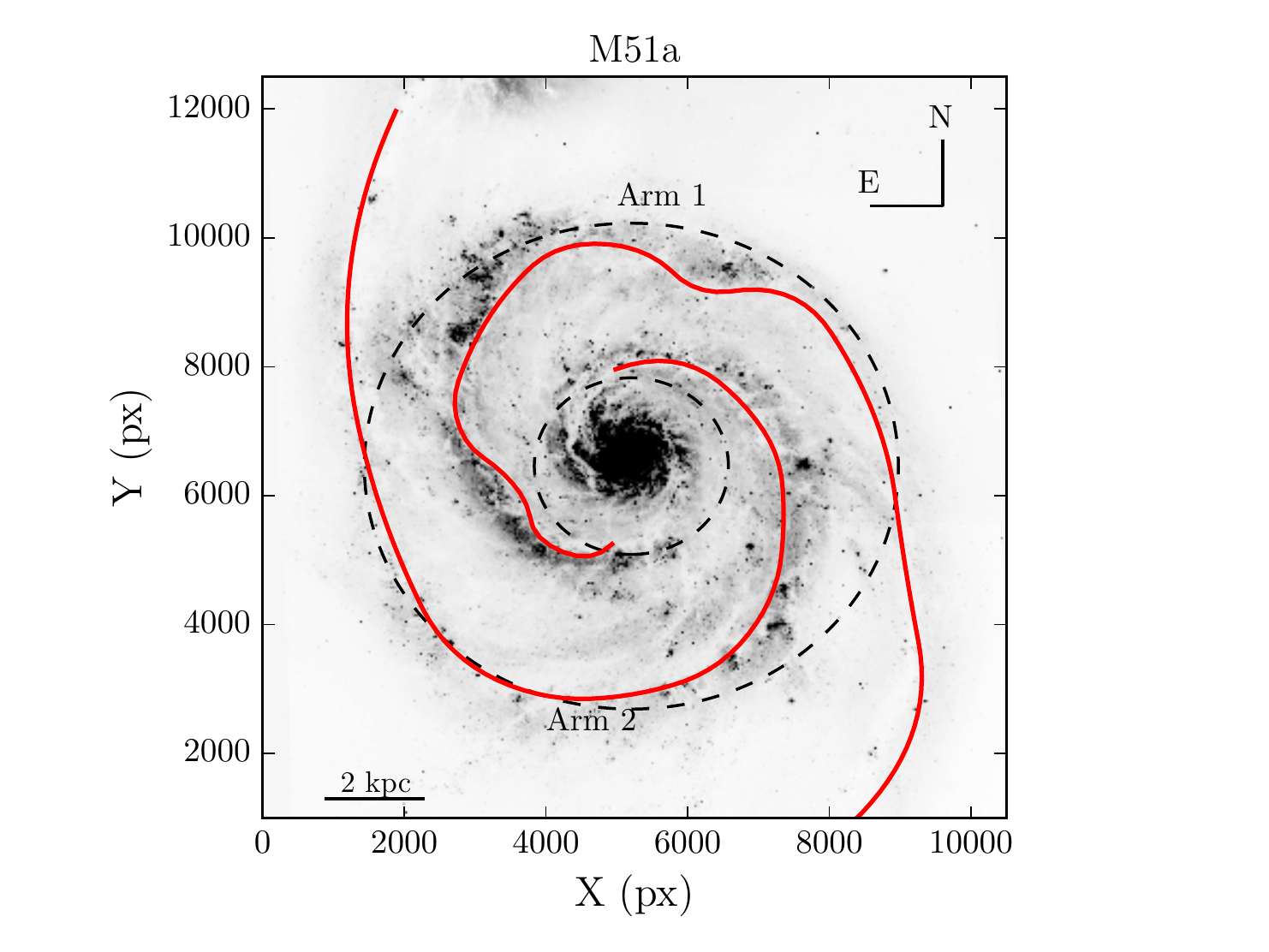}
\end{subfigure}

  \begin{subfigure}[t]{0.50\textwidth}
   \includegraphics[width=1\linewidth]{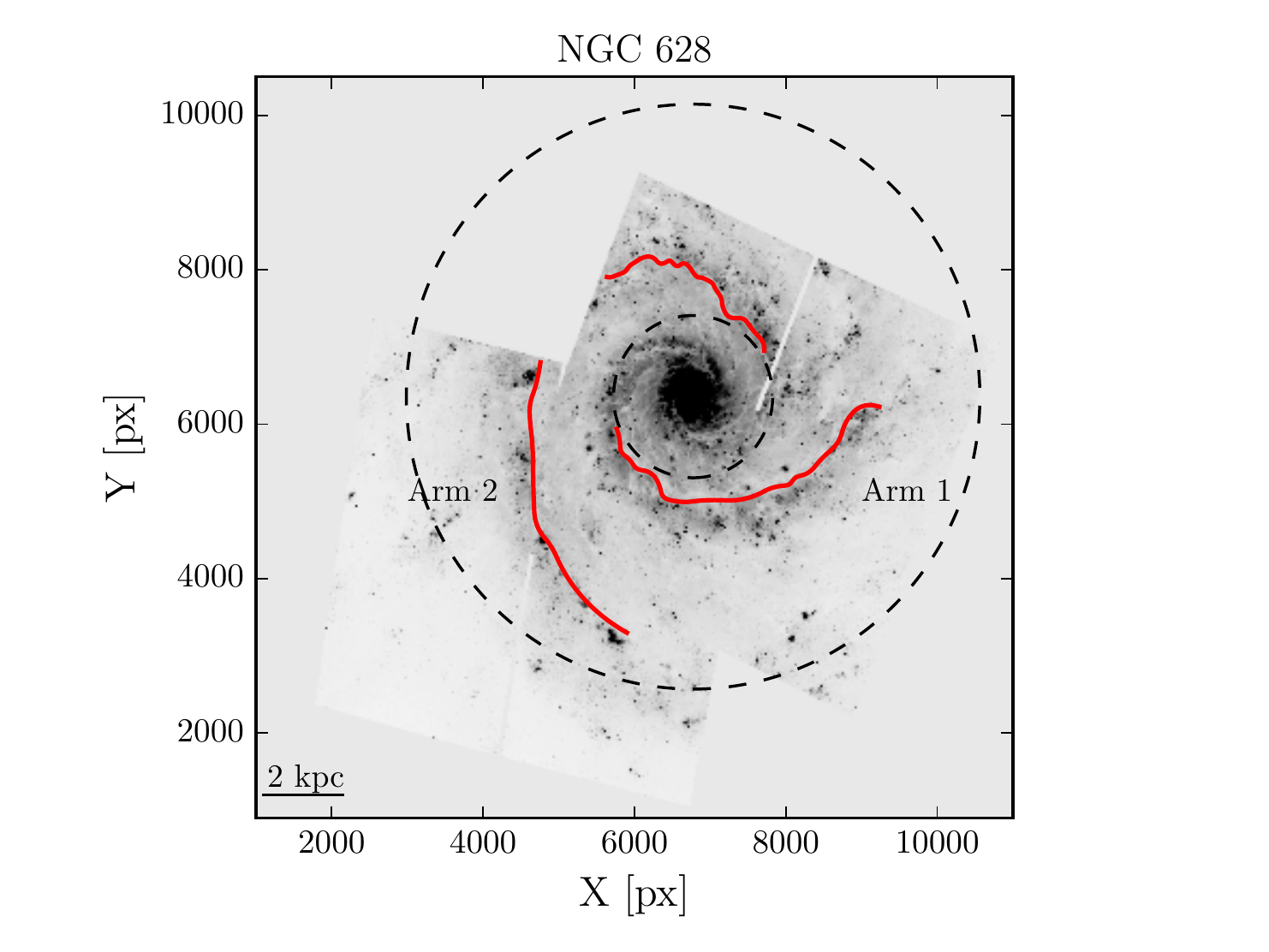}
\end{subfigure}
\caption{The location of spiral arm ridge lines is shown by red lines overplotted on the smoothed $B$---band images of NGC~1566, M51a, and NGC~628. We refer to the southern arm and northern arm as "Arm~1" and "Arm~2", respectively. The two black dashed circles in each panel mark the onset of the bulge and the location of the co--rotation radius of the galaxies. The horizontal bar in the lower left corner denotes a length scale of 2 kpc. North is up and East to the left.}
\label {fig:arms}
\end{figure}

To better define the average positions of the dust lanes, we used a Gaussian kernel (with a 10 pixels sigma) to smooth the images, reduce the noise, and enhance the spiral structure. In the smoothed images the dust lanes are clearly visible as dark ridges inside the bright spiral arms. We defined these dark spiral arm ridge lines manually. For the remainder of this paper, we refer to the southern arm and northern arm as \enquote{Arm~1} and \enquote{Arm~2}, respectively. Fig.~\ref{fig:arms} presents the defined spiral arm ridge lines (red lines) overplotted on the smoothed $B$--band images of  NGC~1566, M51a, and NGC~628.  

\subsection{Measuring azimuthal offset}
Knowing the position of star clusters and spiral arm ridge lines in our target galaxies allowed us to measure the azimuthal distance of a star cluster from its closest spiral arm, assuming that it rotates on a circular orbit. 

We limited our analysis to the star clusters located in the disk where spiral arms exist. The disk of a galaxy can be defined by its rotation curve. The rotational velocity increases when moving outwards from the central bulge--dominated part and becomes flat in the disk--dominated part of the galaxy. We derived a radius of 2~kpc for the bulge--dominated part of our galaxies using the rotation curves of \cite{k2000} for NGC~1566, \cite{sofi2, sofi1} for M51a, and \cite{combes} for NGC~628. Furthermore, we limited our analysis to star clusters located inside the corotation radius. If stationary density waves are the dominant mechanism driving star formation in spiral galaxies we expect to find an age gradient from younger to older clusters inside the corotation radius. 
The bulge--dominated region and co--rotation radius of each galaxy are shown in Fig.~\ref{fig:arms}. The adopted corotation radii of the galaxies are listed in Tab.~\ref{tab:properties of galaxies}. 

\begin{figure*}
\centering
\begin{subfigure}[t]{0.49\textwidth}
   \includegraphics[width=1\linewidth]{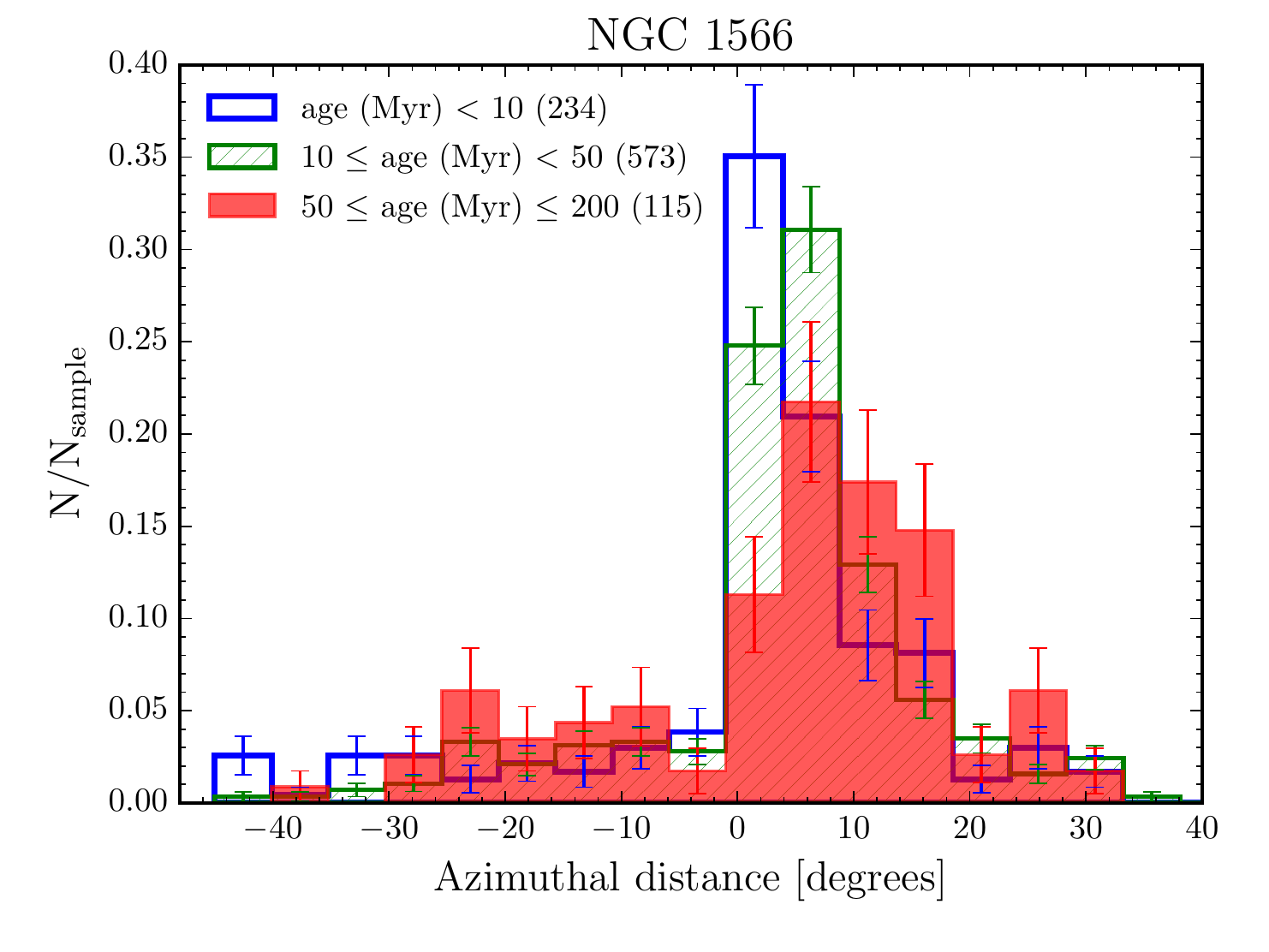}
\end{subfigure}%
~
~ \hspace{-0.5cm}
  \begin{subfigure}[t]{0.49\textwidth}
   \includegraphics[width=1\linewidth]{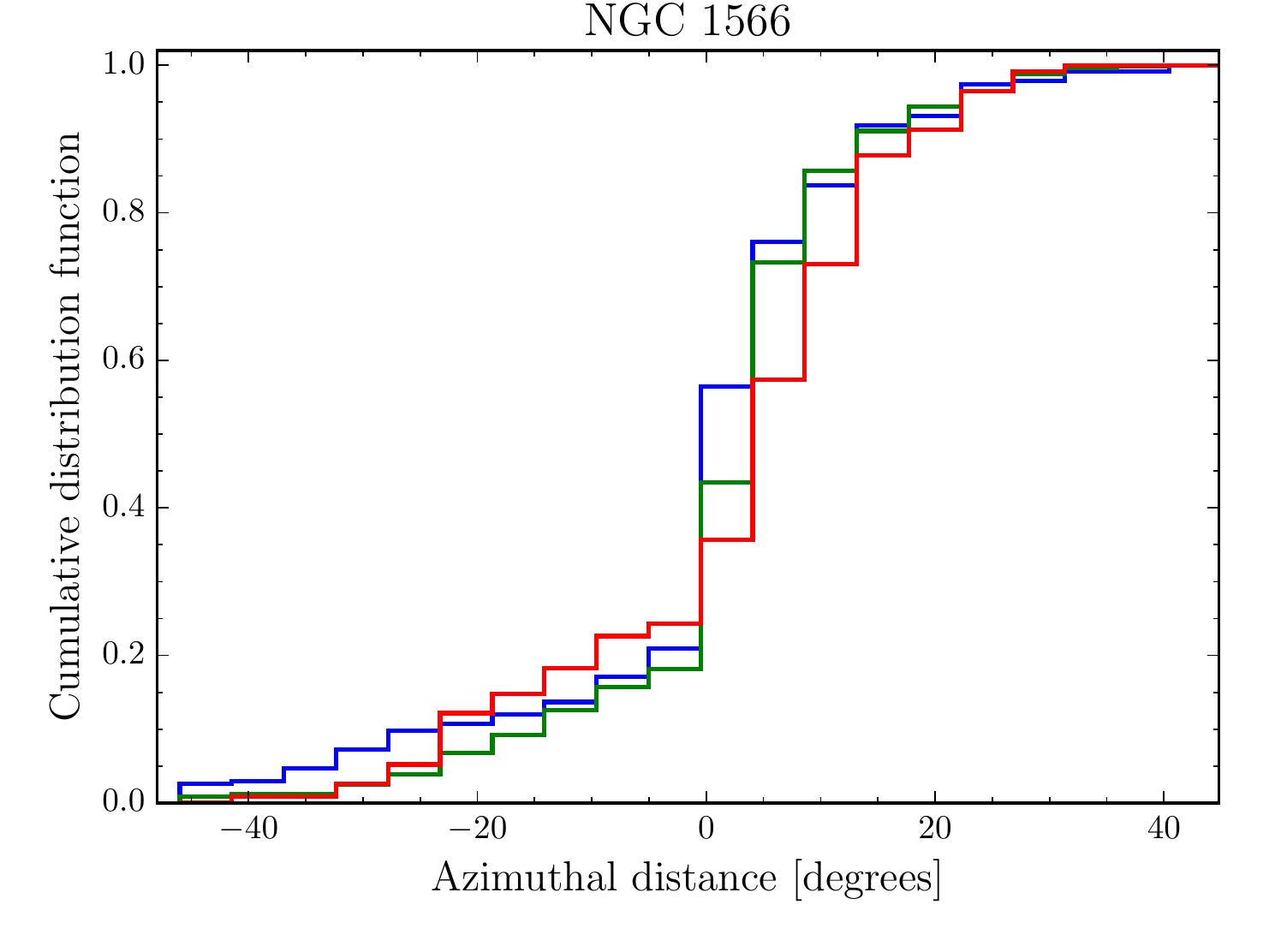}
\end{subfigure}

\begin{subfigure}[t]{0.49\textwidth}
   \includegraphics[width=1\linewidth]{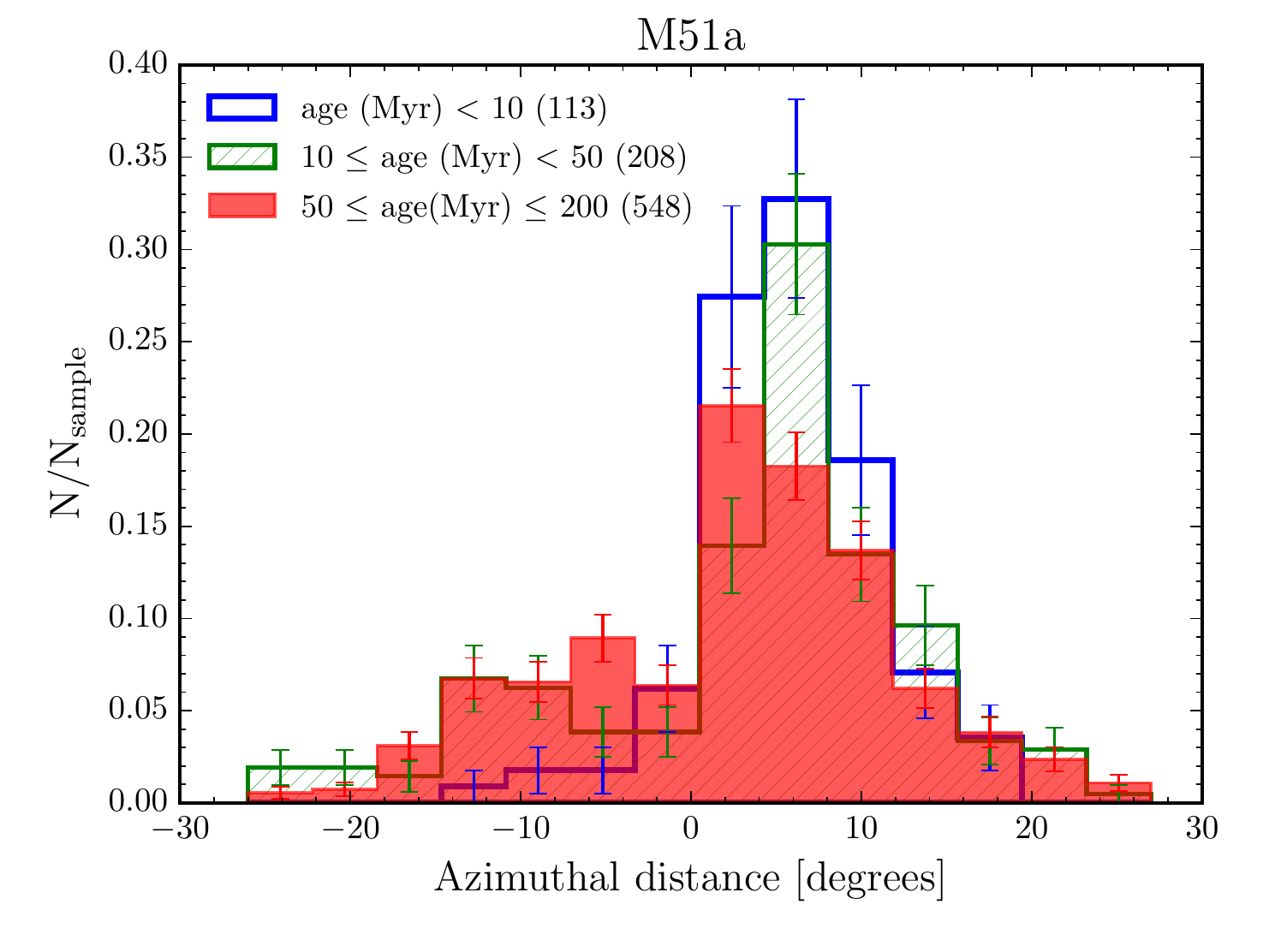}
\end{subfigure}%
~
~ \hspace{-0.5cm}
  \begin{subfigure}[t]{0.49\textwidth}
   \includegraphics[width=1\linewidth]{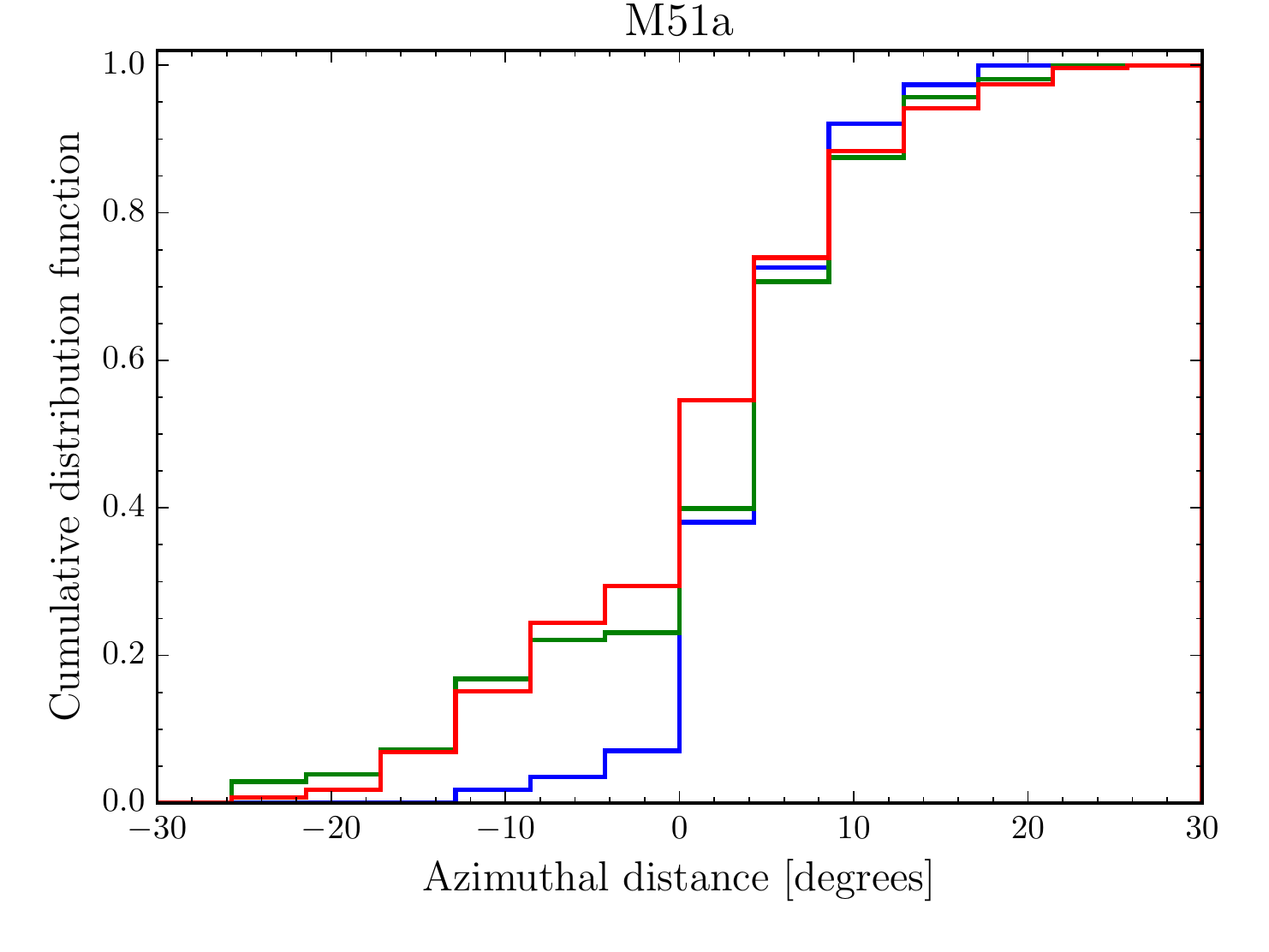}
\end{subfigure}

\begin{subfigure}[t]{0.49\textwidth}
   \includegraphics[width=1\linewidth]{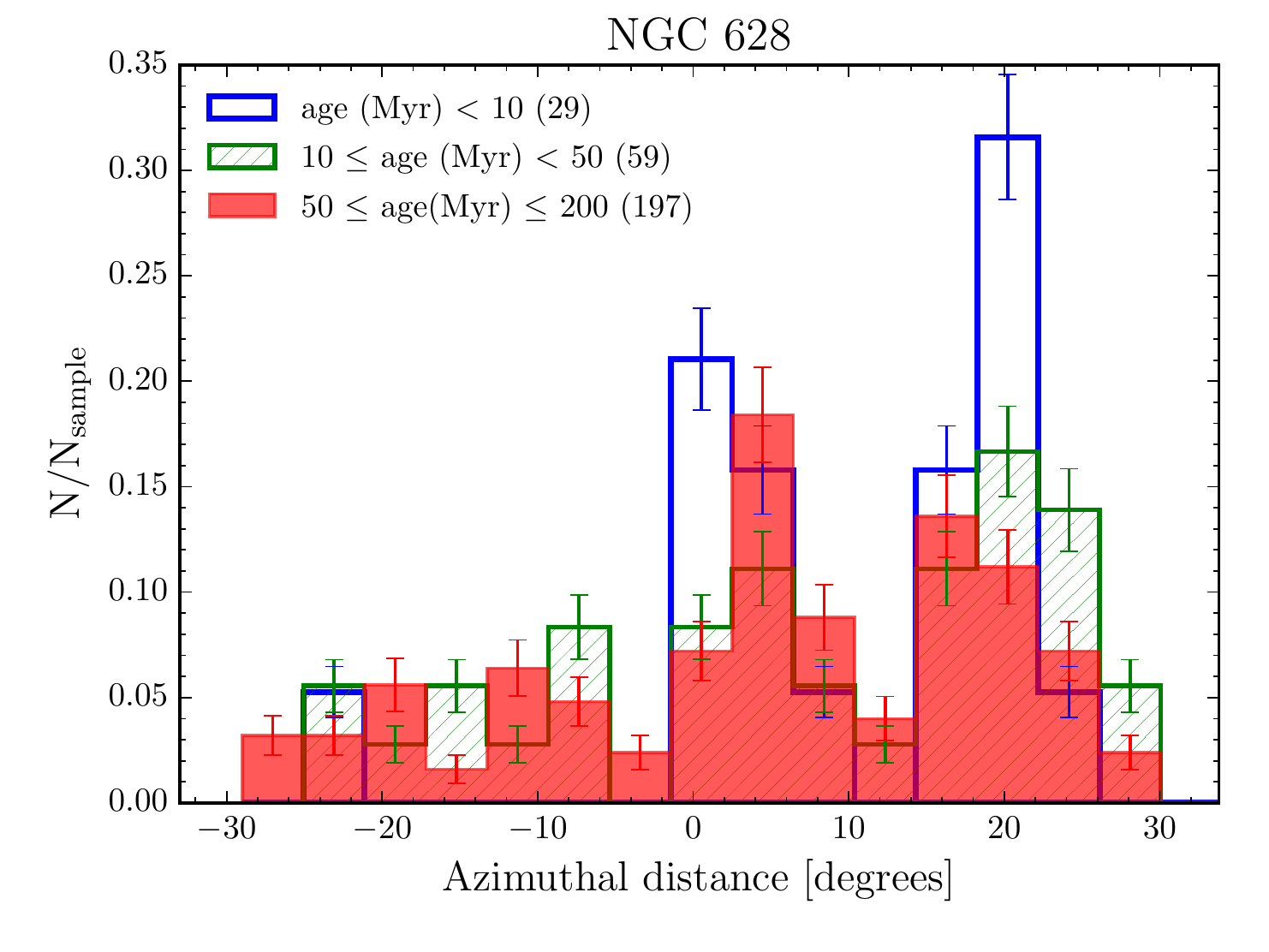}
\end{subfigure}%
~
~ \hspace{-0.5cm}
  \begin{subfigure}[t]{0.49\textwidth}
   \includegraphics[width=1\linewidth]{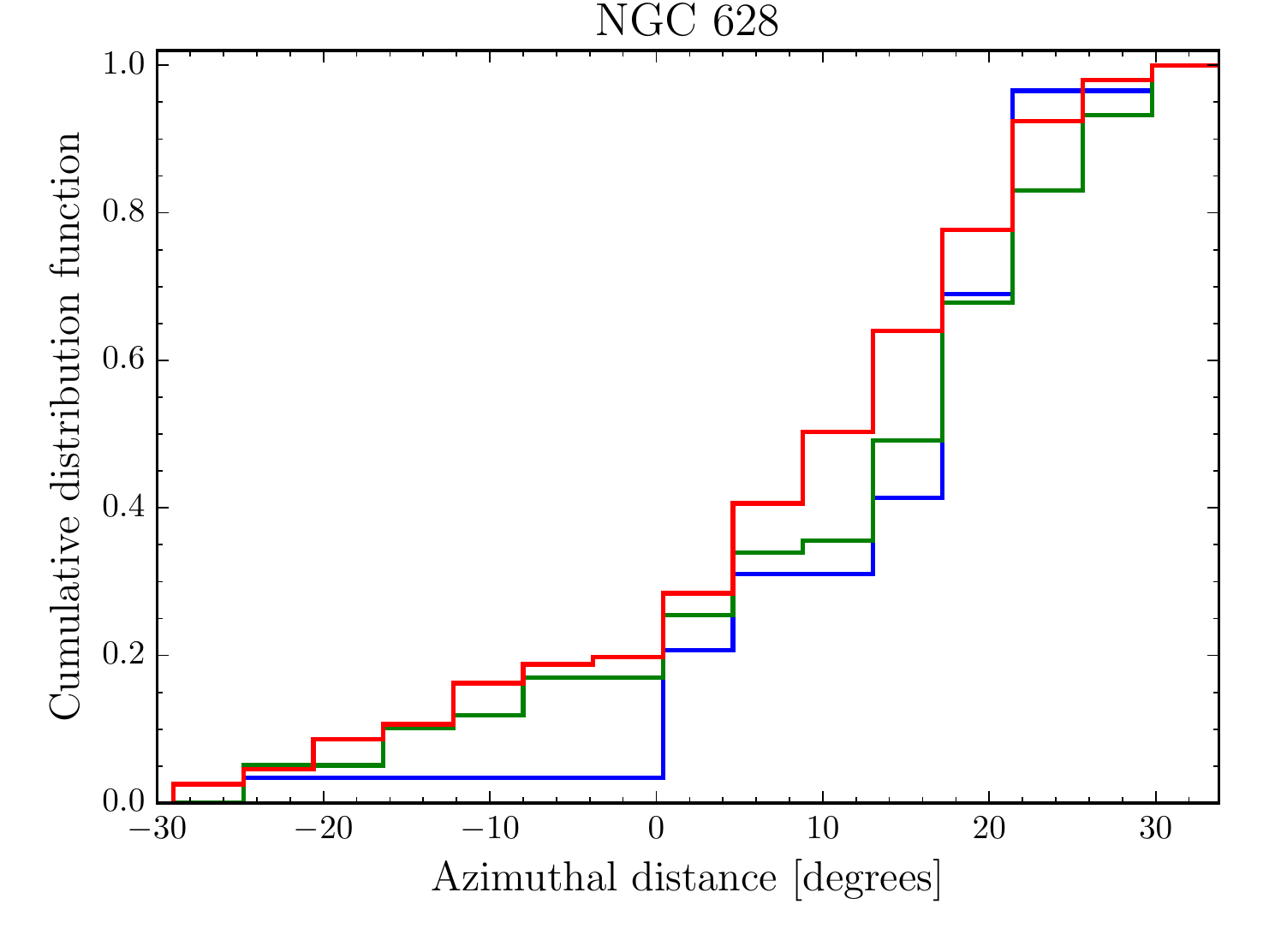}
\end{subfigure}
\caption{The normalized distribution of azimuthal distance (in degrees) of the star cluster samples from their closest spiral arm (left panels) and the cumulative distribution function of star clusters of different ages as a function of the azimuthal distances (in degrees)  in NGC~1566, M51a, and NGC~628. The young (< 10~Myr),  intermediate--age (10--50~Myr), and old star cluster samples (50--200~Myr) are shown in blue, green, and red, respectively. The number of star clusters located in the disk and inside the corotation radius of each galaxy is listed in parantheses.}
\label {fig:hist}
\end{figure*}

Fig.~\ref{fig:hist} (left panels) shows the normalized distribution of the azimuthal distance of star clusters in the three age bins from their closest spiral arm ridge line in NGC~1566, M51a, and NGC~628. The error bars in each sample were calculated by dividing the square root of the number of clusters in each bin by the total number of clusters. We note that an azimuthal distance of zero degrees shows the location of the spiral arm ridge lines and not the center of the arms. Positive (negative) azimuthal distributions indicate that a cluster is located in front of (behind) the spiral arm ridge lines. Blue, green, and red colours represent the young, intermediate--age, and old star cluster samples, respectively.

Fig.~\ref{fig:hist} (right panels) shows the cumulative distribution function of star clusters as a function of the azimuthal distance. In order to test whether the samples come from the same distribution, we used a two--sample Kolmogorov--Smirnov test (hereafter K--S test). Since we aim at finding the age gradient in front of the spiral arms, the K--S test was only calculated for star clusters with positive azimuthal distances. The probability that two samples are drawn from the same distribution (p--values) and the maximum difference between pairs of cumulative distributions (D) are listed in Tab~\ref{tab3}.

In the case of NGC~1566 (Fig.~\ref{fig:hist}, top), we see that the young and intermediate--age star cluster samples are peaking close to the location of the spiral arm ridge lines (azimuthal distance of 0--5 degrees) while the old sample peaks further away from the ridge lines (azimuthal distances of 5--10 degrees). The derived p--values are lower than the test's significance level (0.05) of the null hypothesis, i.e., that the two samples are drawn from the same distribution. As a consequence, our three star cluster samples are unlikely to be drawn from the same population. A clear age gradient across the spiral arms can be observed in NGC~1566, which is in agreement with the expectation from stationary density wave theory. The existence of such a pattern supports the picture of an age sequence in the model of a grand--design spiral galaxy and a barred galaxy suggested by \cite{DP10, dimit17}. 

No obvious age gradient from younger to older is seen in the azimuthal distributions of the star cluster samples in  M51a (Fig.~\ref{fig:hist}, middle). What is remarkable here is that the older star clusters are located closer to the spiral arm ridge lines than the young and intermediate--age star clusters. The K--S test indicates that the probability that the young star cluster sample is drawn from the same distribution as the intemediate-age and old star cluster samples is more than 10\%. The derived p--value for the intermediate--age and old cluster samples is lower than the significance level of the K--S test and rejects the null hypothesis that the two samples are drawn from the same distribution. The lack of an age pattern is consistent with the observed age trend for an interacting galaxy, modeled based on M51a, suggested by \cite{DP10}. Our result is compatible with a number of observational studies have found no indication for the expected spatial offset from the stationary density wave theory in M51a \citep{S09, k10, Foyle, S17}. 

There is no evident trend in the azimuthal distribution of star clusters in NGC~628 (Fig.~\ref{fig:hist}, bottom). The majority of the young star clusters tends to be located further away from the ridge lines (azimuthal distance of 20--25 degrees). The calculated p--values from the K--S test are larger than 0.05, which suggests weak evidence against the null hypothesis. As a result, the three young, intermediate--age, and old star cluster samples are drawn from the same distribution. The absence of an age gradient across the spiral arms in NGC~628 is consistent with a simulated multiple arm spiral galaxy by \cite{grand}.

\section{The origin of two spiral arms}
\label{2arms}

An observational study by \cite{Egusa17}, based on measuring azimuthal offsets between the stellar mass (from optical and near--infrared data) and gas mass distributions (from CO and HI data) in two spiral arms of M51a, suggest that the origin of these spiral arms differs. One spiral arm obeys the stationary density wave theory while the other does not. 

In another recent study of M51a, \cite{chandar17} quantified the spatial distribution of star clusters with different ages relative to different segments of the two spiral arms of M51a traced in the 3.6~$\mu$m image. They observed a similar trend for the western and eastern arms: the youngest star clusters (< 6~Myr) are found near the spiral arm segments, and the older clusters (100--400~Myr) show an extended distribution.

In this section, we test whether measuring the azimuthal offset of star cluster samples from each spiral arm individually leads to different results. We assume that a star cluster whose distance from Arm~1 is smaller than its distance from Arm~2 belongs to Arm~1 and vice versa.

\begin{figure}
\centering
\begin{subfigure}[t]{0.50\textwidth}
   \includegraphics[width=1\linewidth]{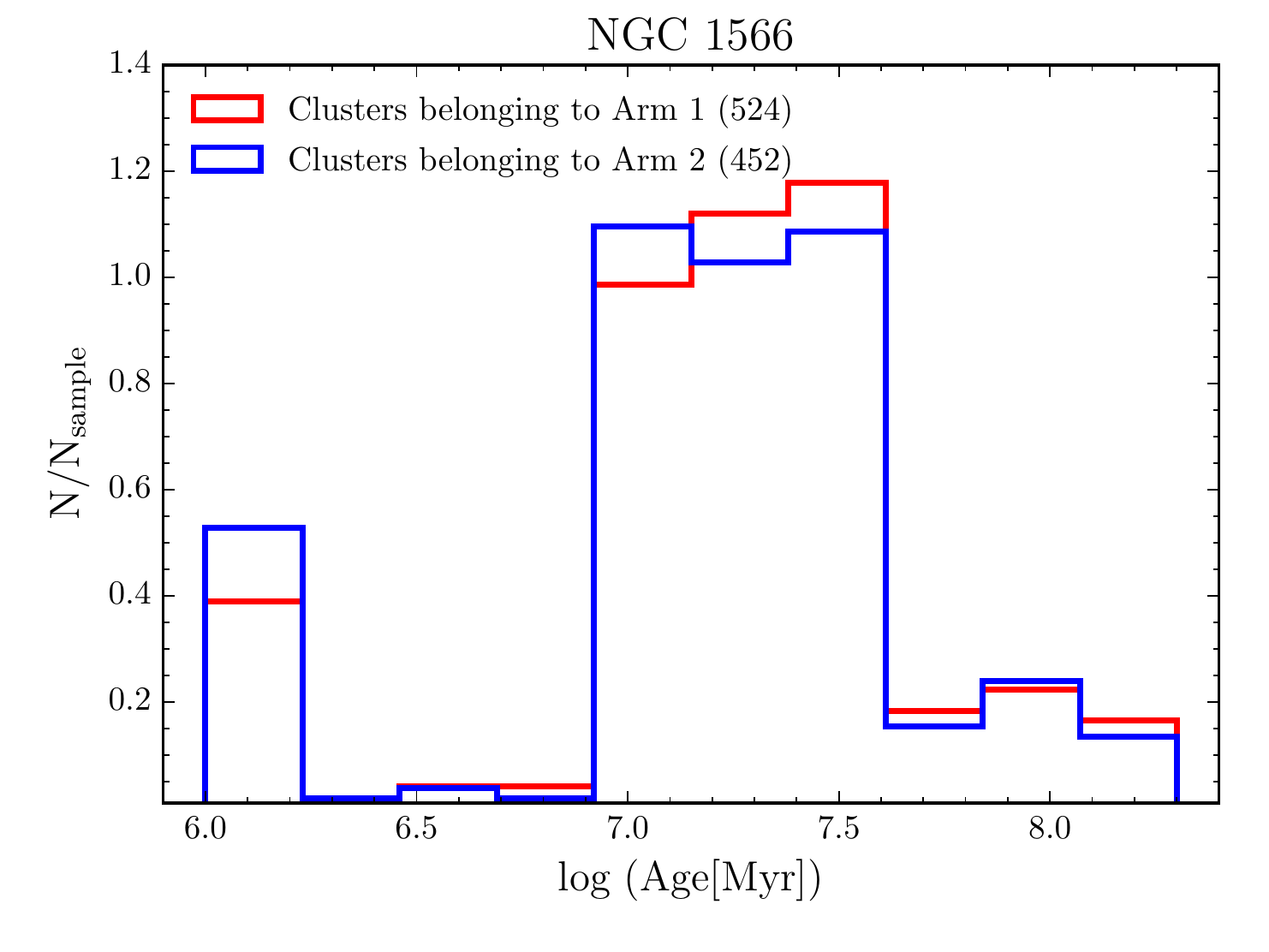}
\end{subfigure}

  \begin{subfigure}[t]{0.50\textwidth}
   \includegraphics[width=1\linewidth]{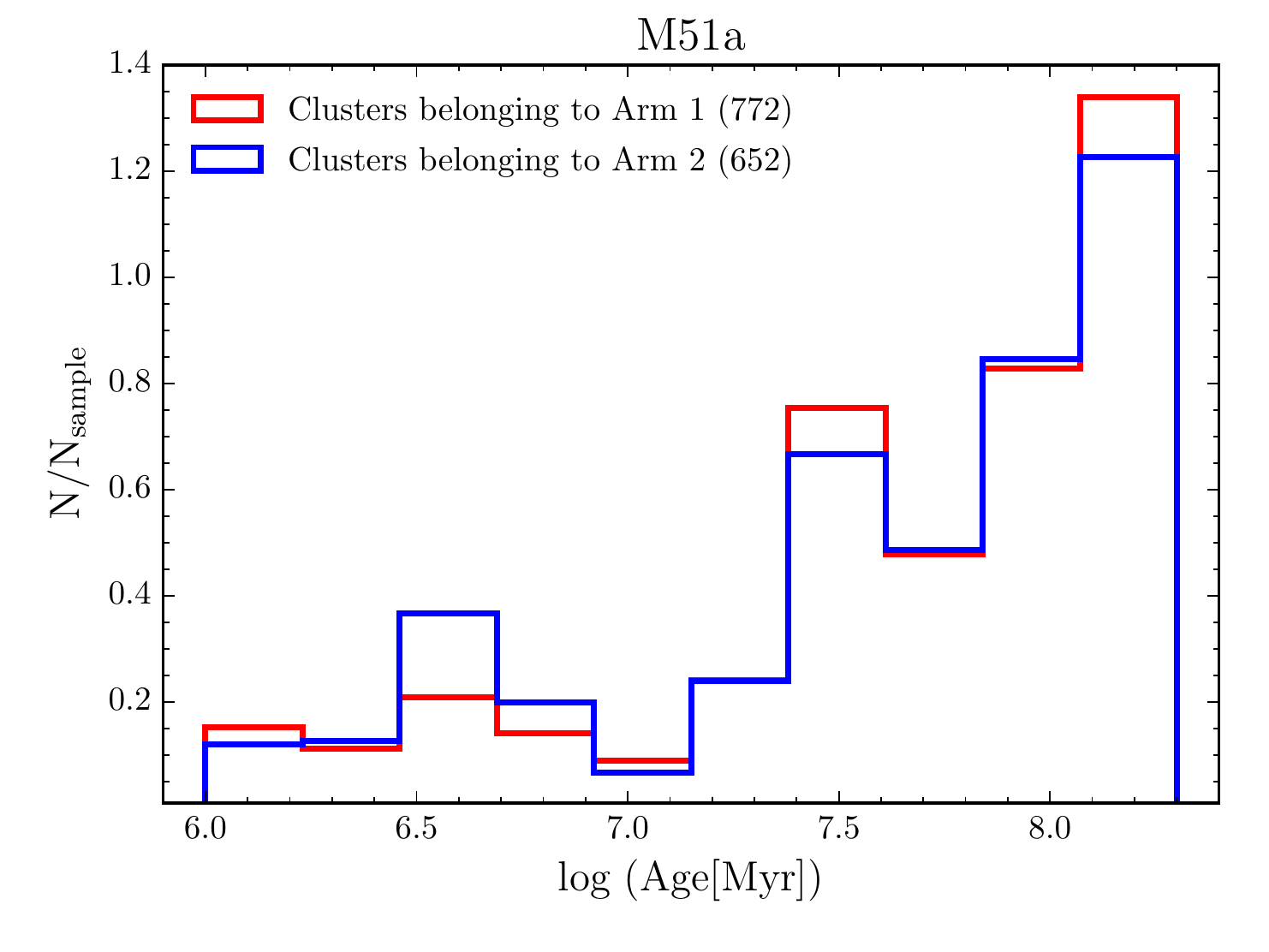}
\end{subfigure}

  \begin{subfigure}[t]{0.50\textwidth}
   \includegraphics[width=1\linewidth]{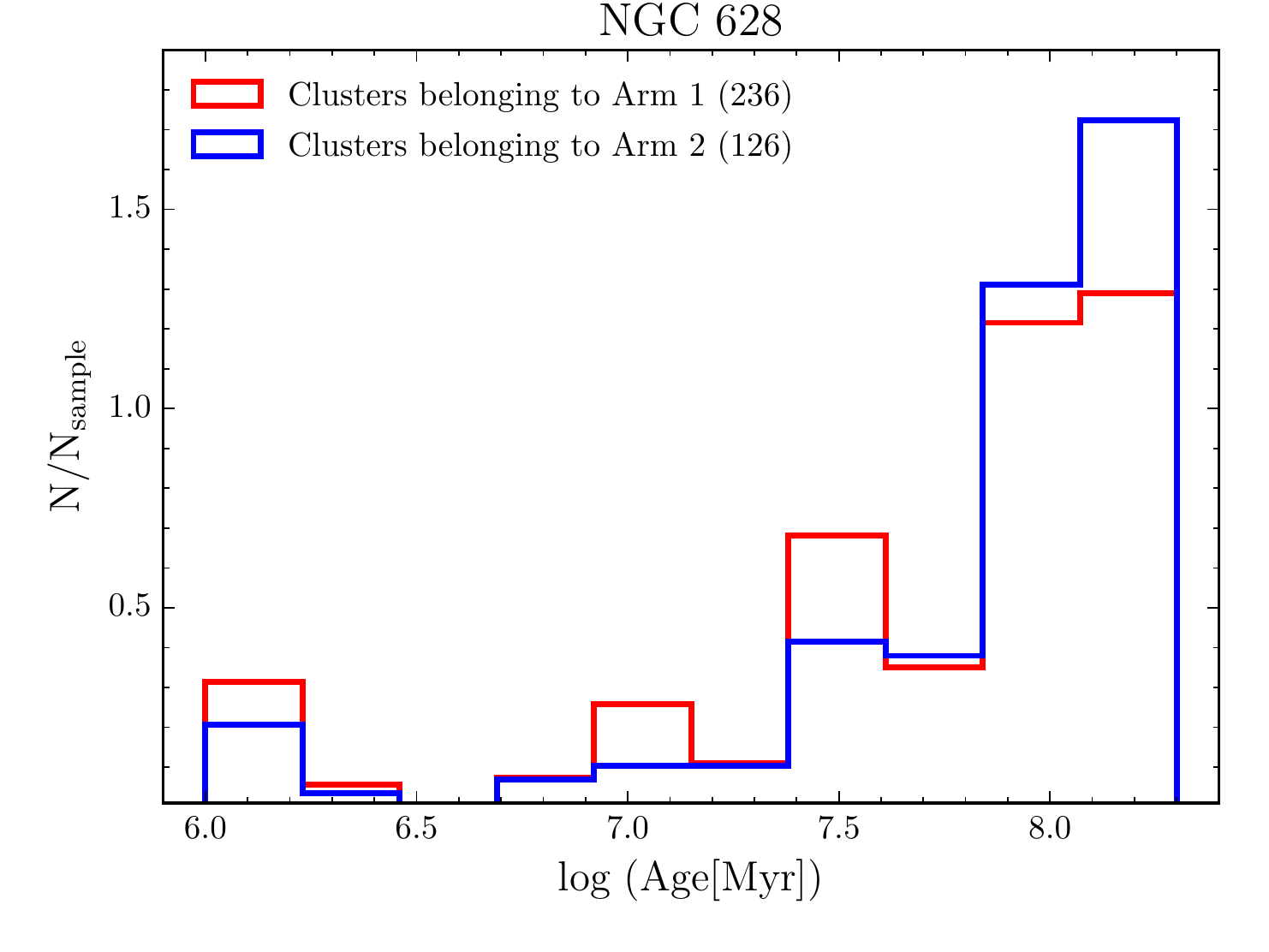}
\end{subfigure}
\caption{The distribution of the age of star clusters associated with Arm~1 (red) and Arm~2 (blue) in NGC~1566, M51a, and NGC~628. The number of star clusters relative to the  Arm~1 and Arm~2 is listed in parantheses.}
\label {fig:age_hist}
\end{figure}

Fig.~\ref{fig:age_hist} shows the normalized distribution of ages of star clusters associated with Arm~1 (shown in red) and Arm~2 (shown in blue) in each of the galaxies. No significant differences between the age distribution of star clusters belonging to the two spiral arms in our target galaxies can be observed. Also, the K--S test indicates that the age distributions of star clusters relative to Arm~1 and Arm~2 in each galaxy are drawn from the same population.

\begin{figure*}
\centering
\begin{subfigure}[t]{0.49\textwidth}
   \includegraphics[width=1\linewidth]{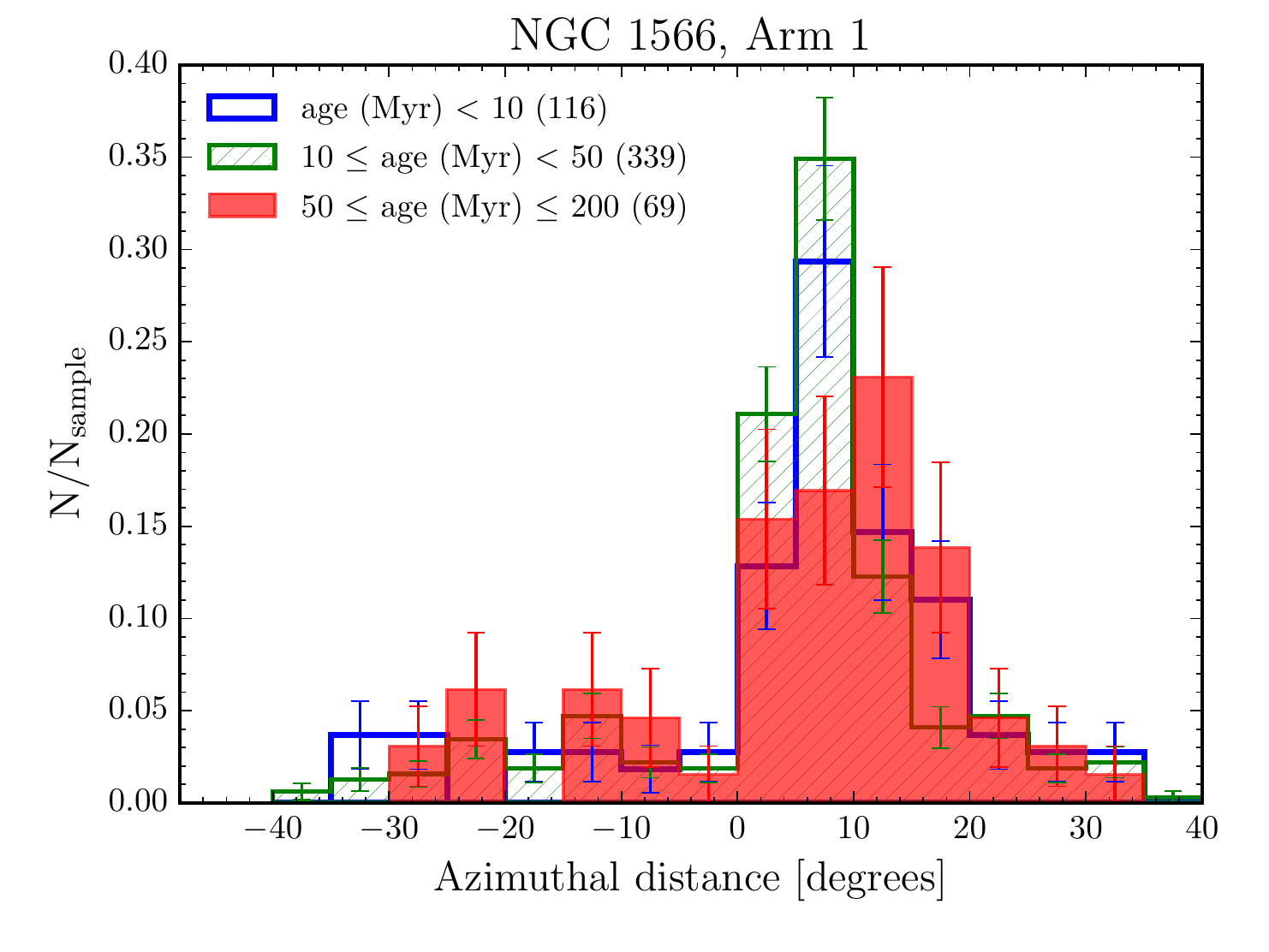}
\end{subfigure}%
~
~ \hspace{-0.5cm}
  \begin{subfigure}[t]{0.49\textwidth}
   \includegraphics[width=1\linewidth]{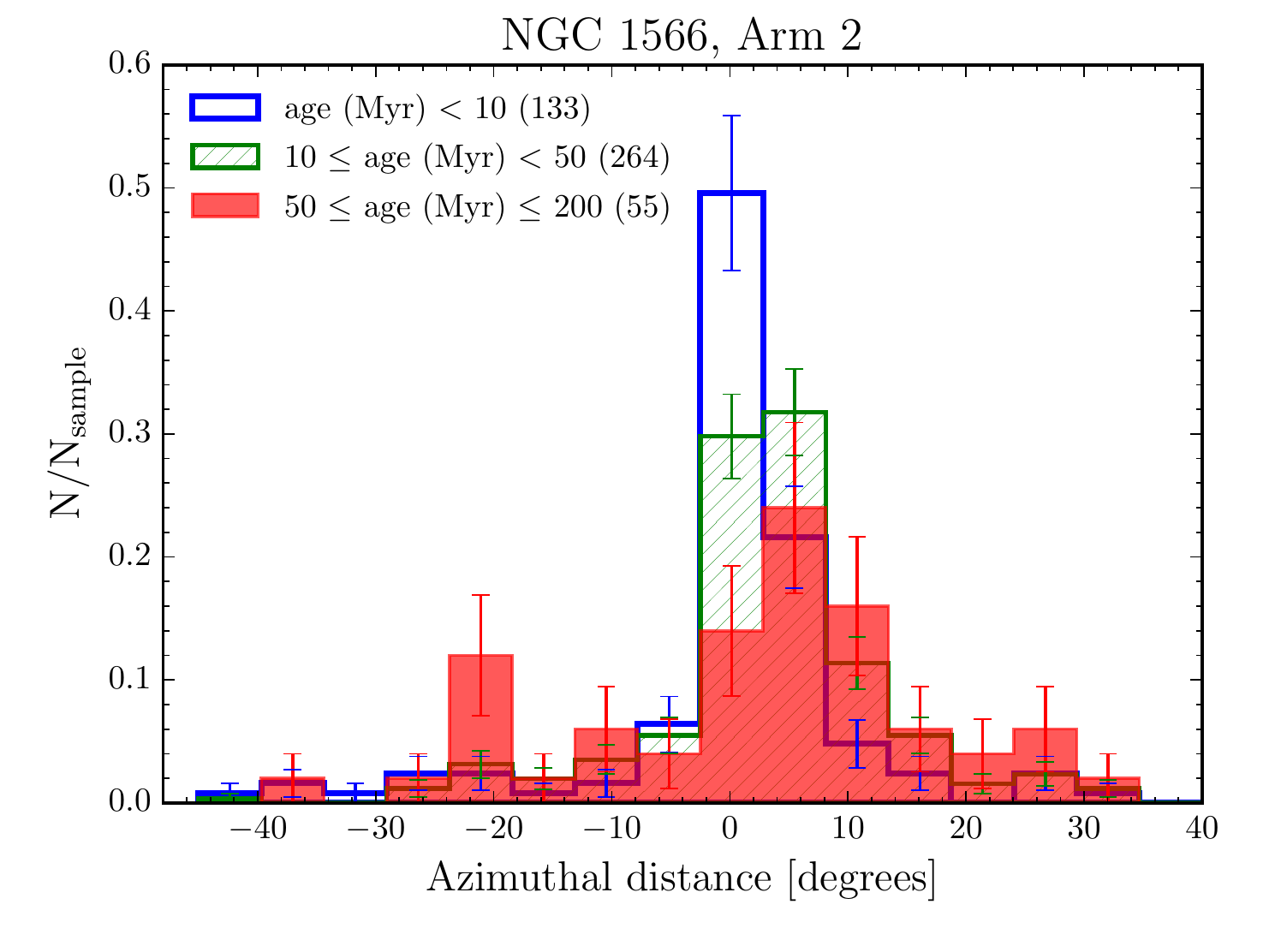}
\end{subfigure}

\begin{subfigure}[t]{0.49\textwidth}
   \includegraphics[width=1\linewidth]{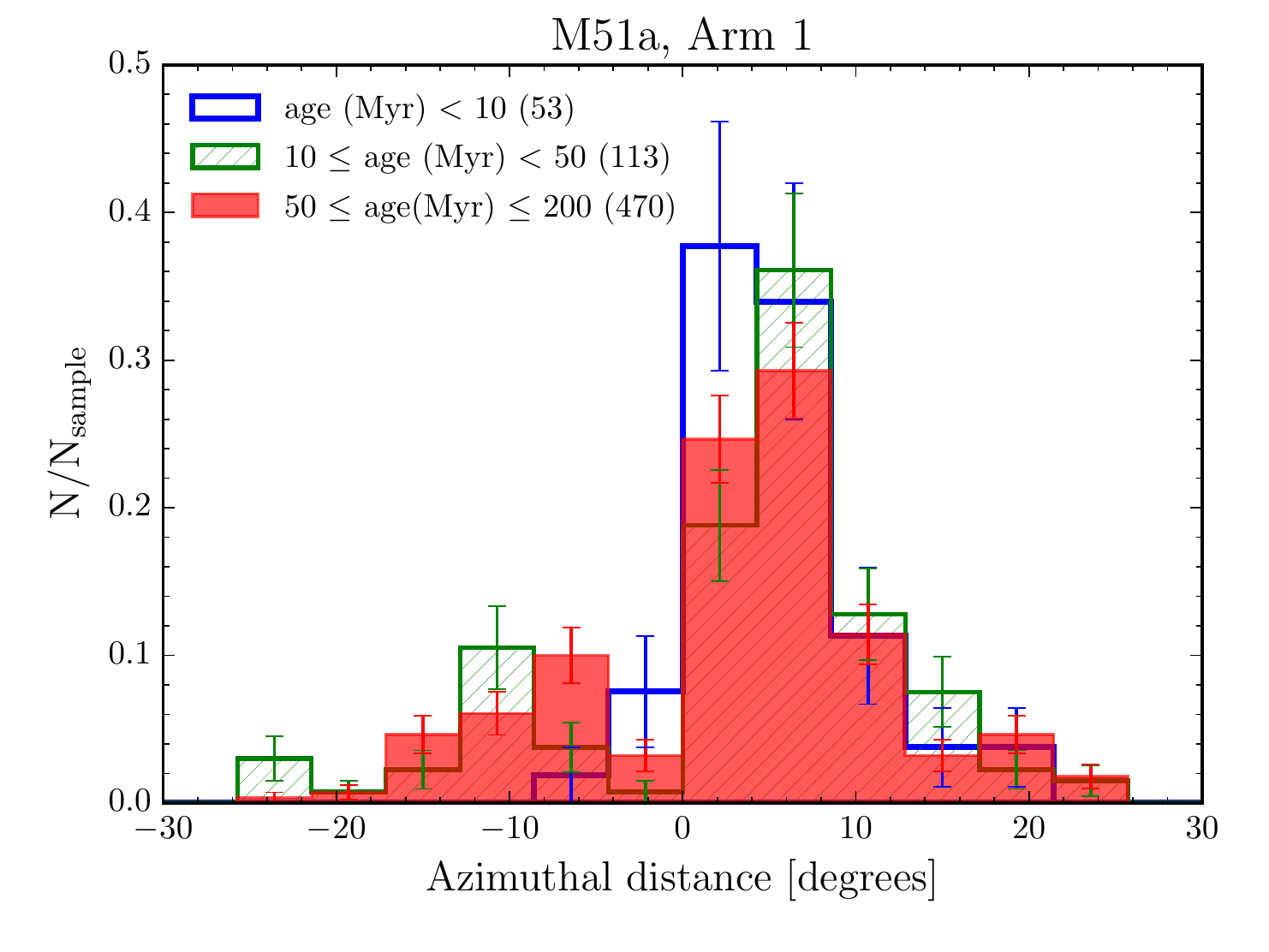}
\end{subfigure}%
~
~ \hspace{-0.5cm}
  \begin{subfigure}[t]{0.49\textwidth}
   \includegraphics[width=1\linewidth]{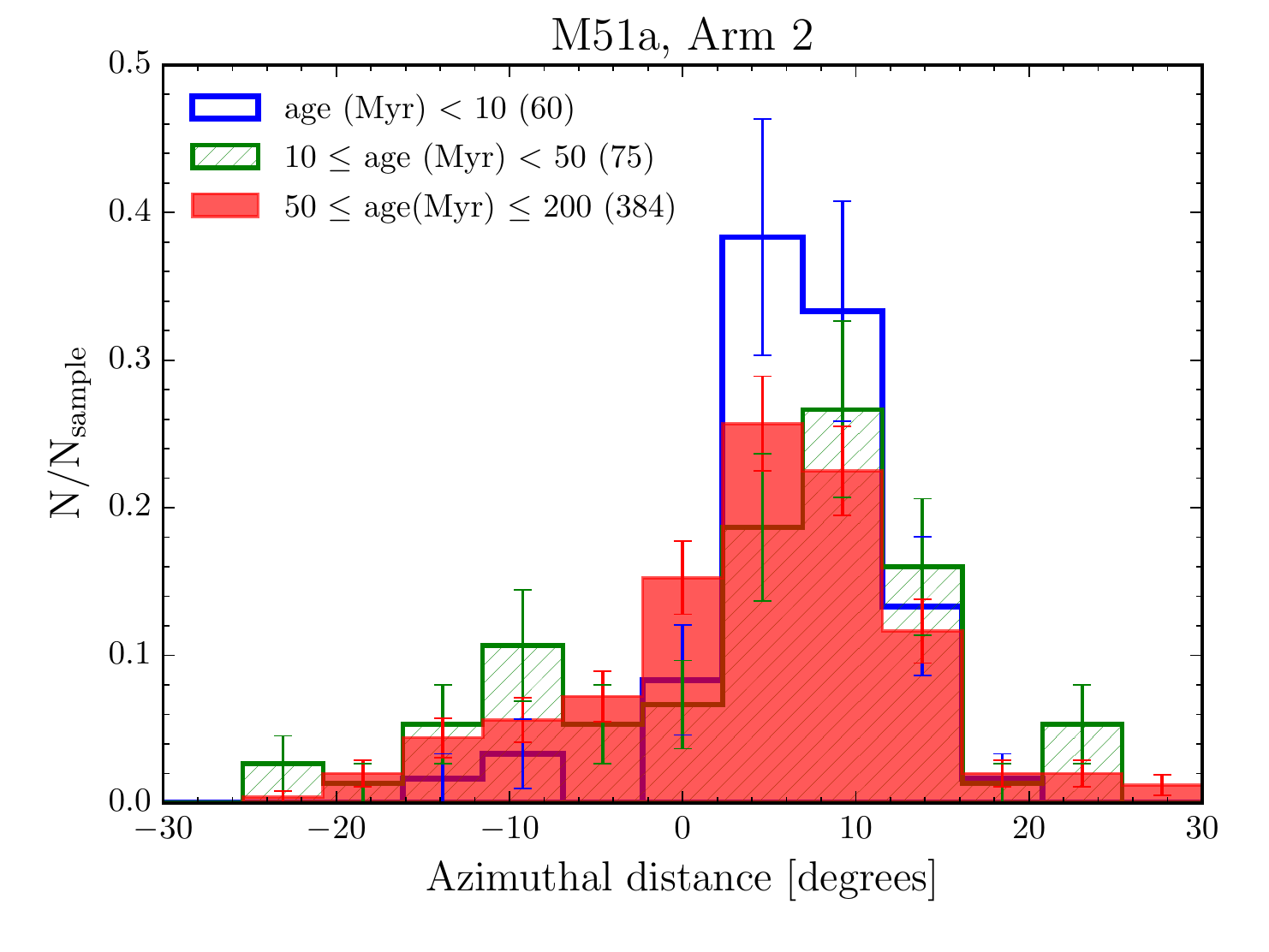}
\end{subfigure}

\begin{subfigure}[t]{0.49\textwidth}
   \includegraphics[width=1\linewidth]{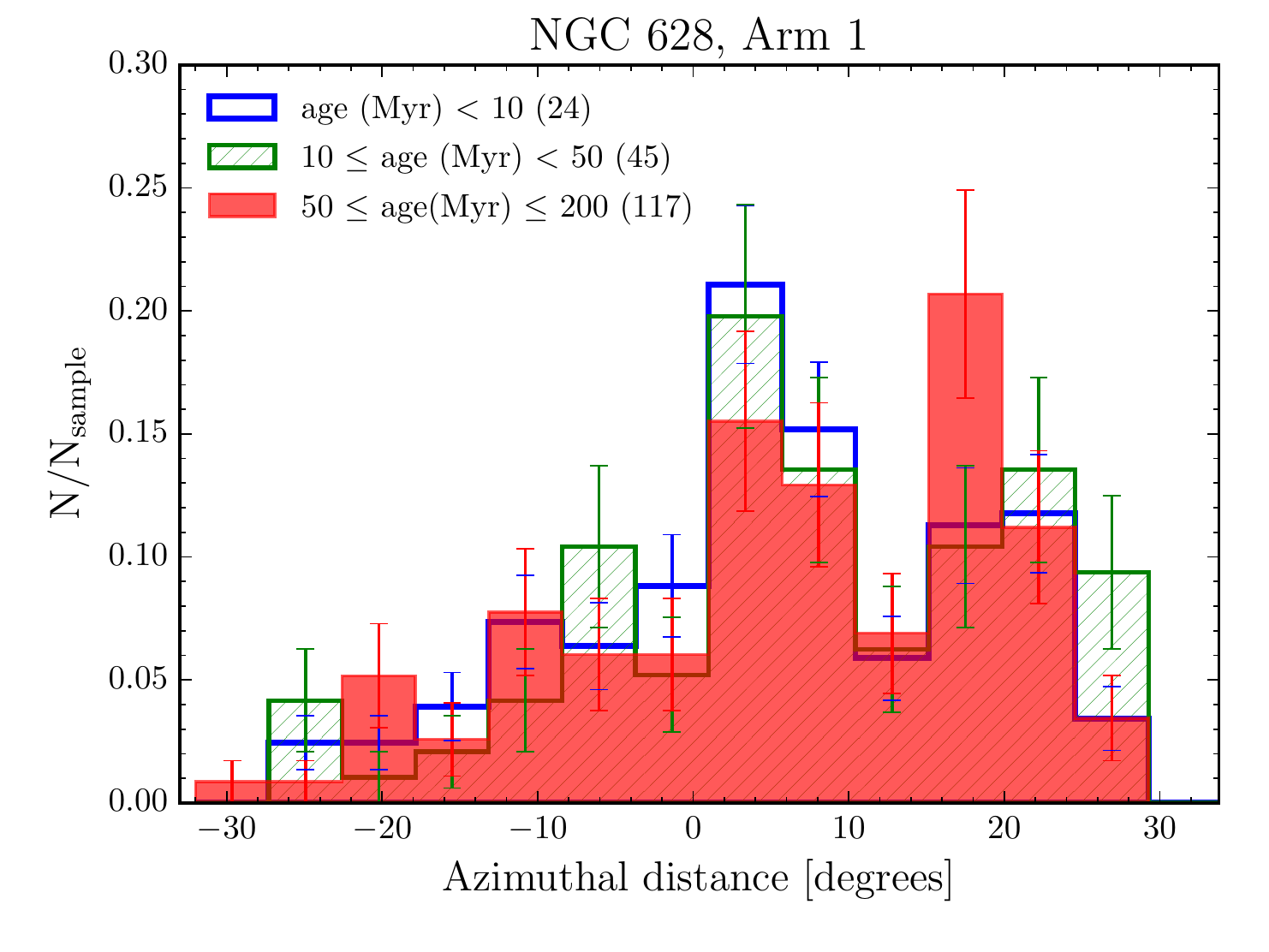}
\end{subfigure}%
~
~ \hspace{-0.5cm}
  \begin{subfigure}[t]{0.49\textwidth}
   \includegraphics[width=1\linewidth]{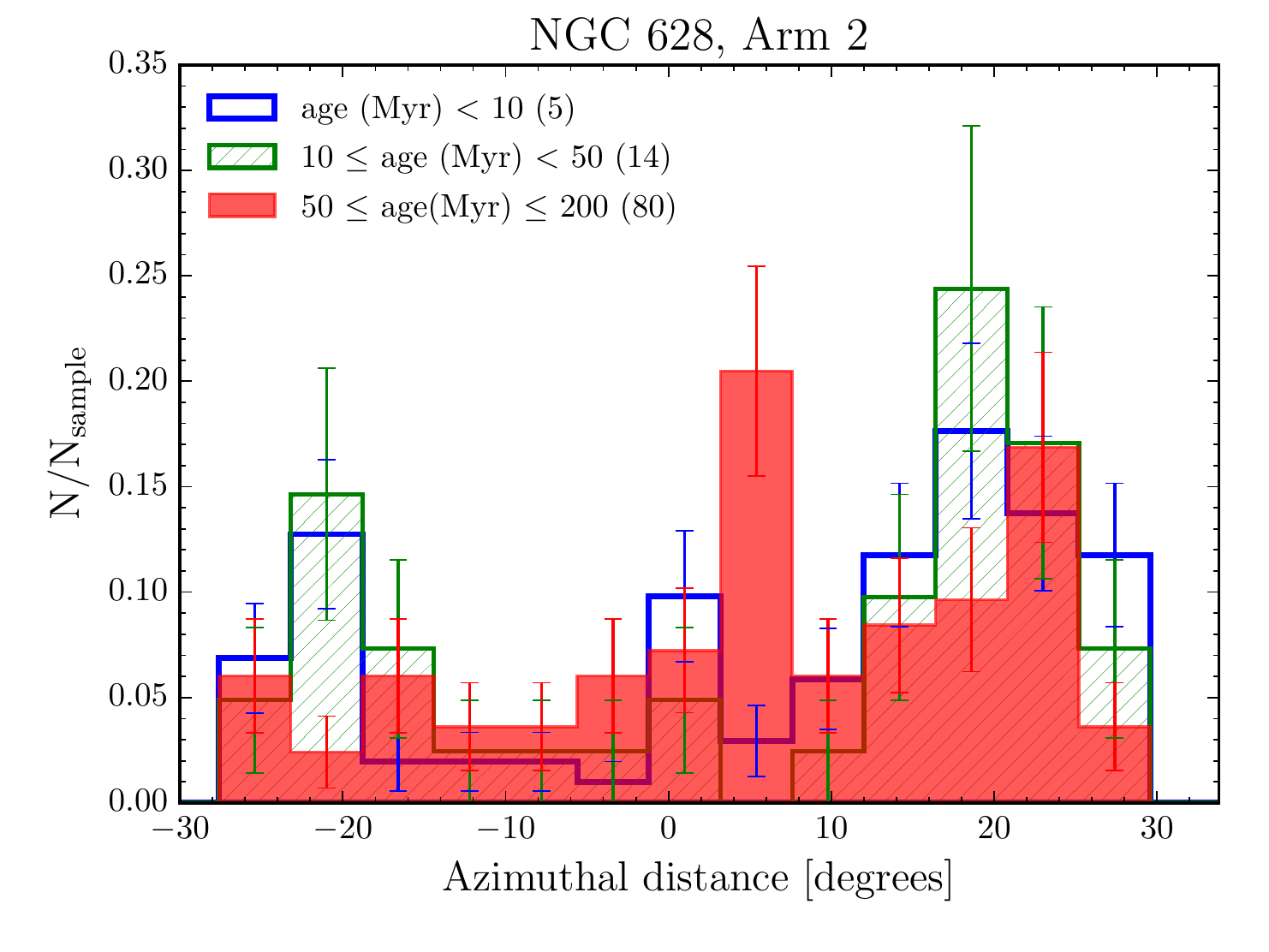}
\end{subfigure}

\caption{The normalized distribution of azimuthal distance (in degrees) of the star cluster samples belonging to Arm~1 (left panels) and Arm~2 (right panels) in NGC~1566, M51a, and NGC~628. Blue, green, and red colours present the young (< 10 Myr ),  intermediate--age (10--50 Myr), and old star cluster samples (50--200 Myr), respectively. The number of star clusters corresponding to Arm~1 and Arm~2 is listed in parantheses. The error bars in each sample were calculated by dividing the square root of the number of clusters in each bin by the total number of clusters.}
\label {fig:hist-arms}
\end{figure*}

In Fig.~\ref{fig:hist-arms} we compare the normalized azimuthal distribution of the three young, intermediate--age, and old star cluster samples relative to Arm~1 (left panels) and Arm~2 (right panels) in our target galaxies. As before, our analysis was limited to the star clusters positioned in the disk and inside the corotation radius of our target galaxies. 

The upper panels of Fig.~\ref{fig:hist-arms} exhibit a noticeable age gradient across both spiral arms of NGC~1566. The young star clusters are highly concentrated towards the location of Arm~1 and Arm~2 while the older ones are peaking further away from the two spiral arms. 

The second row panels of Fig.~\ref{fig:hist-arms} show the azimuthal distance of star cluster samples across the two arms of M51a. This galaxy displays an offset in the location of young and old star clusters across Arm~1. The young star clusters culminate close to Arm~1 (at azimuthal distances of 2--6 degrees) while the old ones are positioned further away (at azimuthal distances of 6--10 degrees). Even though M51a shows an age gradient across the Arm~1 at first glance, the K--S test does not imply significant differences between the young and old star cluster samples (all derived p--values are larger than the test's significance level). We do not observe any shift in the azimuthal distribution of the star cluster samples across Arm~2 in M51a. 

In the case of NGC~628, no obvious age gradient across Arm~1 and Arm~2 is observed (the lower panels of Fig.~\ref{fig:hist-arms}). It is important to note that our results are inconclusive for the young star clusters associated with Arm~2 due to the small number statistics.  Hence, we also explored the change in the azimuthal distribution of the star clusters by including clusters with masses < 5000 $\rm M_{\sun}$ and ages > 200 Myr. The observed differences are not significant and the general trend is the same as before. 

Thus, measuring the azimuthal distance of the star clusters from the two individual spiral arms in each galaxy suggests that the two spiral arms of our target galaxies may have the same physical origin.

\section{Comparison with the non--LEGUS cluster catalogue of M51}
\label{chandra}
\begin{table*}
\centering
\caption{The maximum difference between pairs of cumulative distributions (D) of azimuthal distance of star clusters and the probability that two samples are drawn from the same distribution (p--values) of the two sample K--S test in NGC~1566, M51a, and NGC~628.}
\label{tab3}
\begin{tabular}{llccccc}
\hline \hline
\multirow{2}{*}{Galaxy} & \multicolumn{2}{c}{Young \& Intermediate--age}     & \multicolumn{2}{c}{Young \& Old}                     & \multicolumn{2}{c}{Intermediate--age \& Old}          \\ \cline{2-7} 
                        & \multicolumn{1}{c}{D} & p--value                   & D                         & p--value                   & D                         & p--value                   \\ \hline
NGC~1566           & 0.15                 & $ \rm 3.78\times 10^{-3}$ & 0.31                      & $\rm 2.88 \times 10^{-5}$ & 0.26                   & $\rm 6.19 \times 10^{-5}$ \\ 
M51a           & 0.15                  & 0.10 & 0.13                      & 0.10 & 0.17                      & $\rm 2.4 \times 10^{-3}$ \\ 
NGC~628         & 0.21                  & 0.49                     & \multicolumn{1}{c}{0.17} & 0.47                      & \multicolumn{1}{c}{0.19} & 0.10 \\ \hline \hline
\end{tabular}
\end{table*}
In this section, we use the \cite{chandar16} catalogue (hereafter CH16 catalogue) to measure the azimuthal offsets of star clusters with different ages in M51a and to compare the results with our analysis based on the LEGUS catalogue. We caution that the south--eastern region of M51a is not covered by the LEGUS observations. We also investigated whether our results are biased due to the absence of star clusters from that region. 

\cite{chandar16} provided a catalogue of 3816  star clusters in M51a based on HST ACS/WFC2 images obtained the equivalents of $UBVI$ and H$\rm \alpha$ filters. \cite{messa} compared the age distributions of star clusters in common between the LEGUS and CH16 catalogue. They observed that  
a large number of young star clusters (age < 10~Myr) in \cite{chandar16} have a broad age range (age: 1--100~Myr) in the LEGUS catalogue. They argued that the discrepancies in the estimated ages are due to the use of different filter combinations. 

\begin{figure}
\centering
\includegraphics[width=0.49\textwidth]{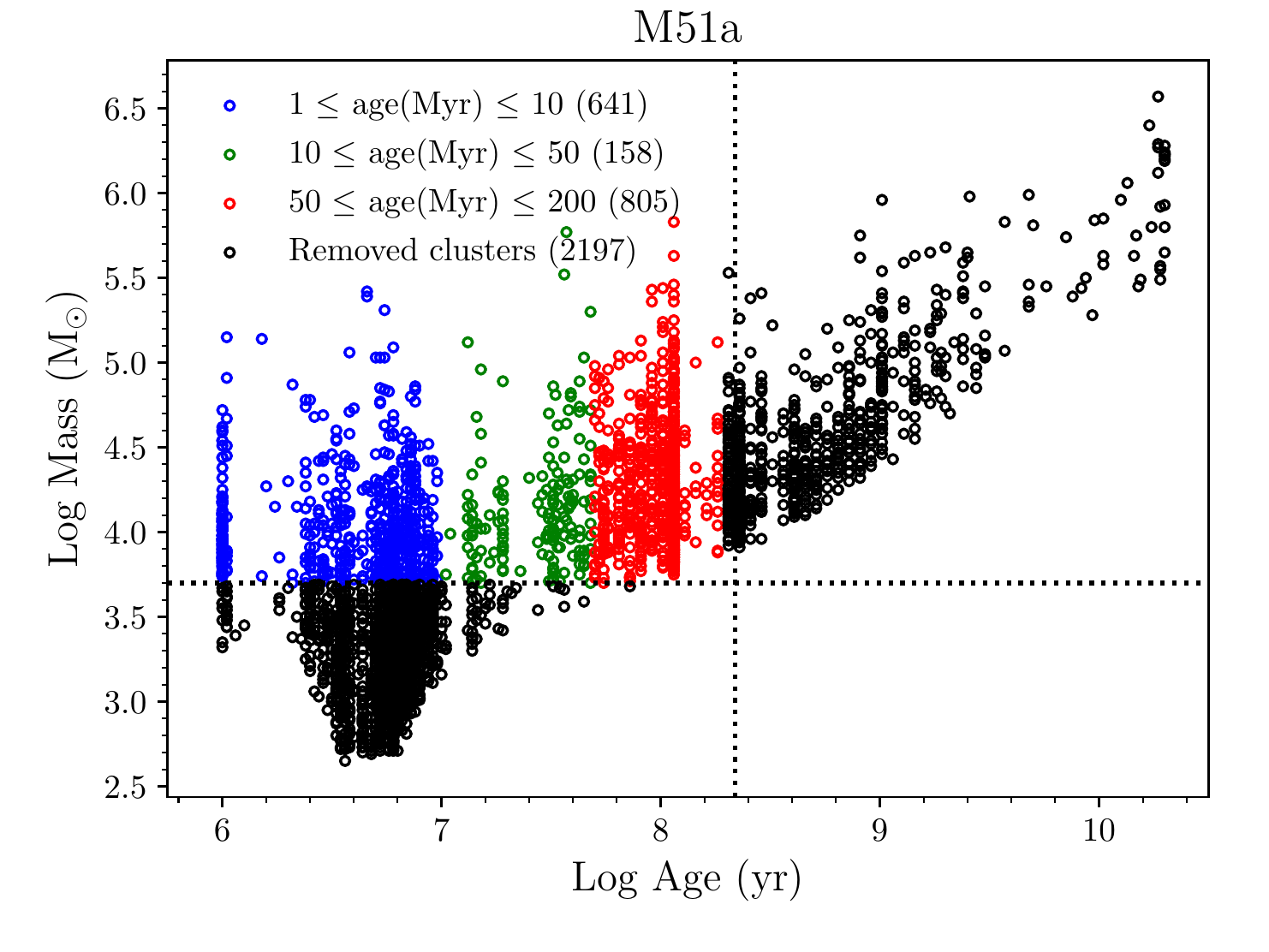}
\caption{The distribution of ages and masses of the 3816 star clusters in M51a, based on the CH16 catalogue. The young (<10~Myr ), intermediate--age (10--50~Myr), and old (50--200~Myr) star clusters are shown in blue, green, and red, respectively. The black points indicate excluded star clusters due to the applied mass cut and the imposed completeness limit. The number of clusters in each sample is listed in parentheses. The horizontal and vertical dotted lines show the applied mass cut of 5000 $\rm M_{\sun}$ and the corresponding detection completeness limit of 200~Myr, respectively. }
\label {fig:age_mass_chandar}
\end{figure}

In Fig.~\ref{fig:age_mass_chandar}, we show the distribution of ages and masses of star clusters in M51a from the CH16 catalogue. In order to be able to compare our results, we considered a mass--limited sample with masses > 5000 $\rm M_{\sun}$ and ages < 200~Myr and selected the same age bins as before: The young (< 10~Myr), intermediate--age (10--50~Myr), and old star cluster samples (50--200~Myr). 

\begin{figure*}
\centering

     \begin{subfigure}[t]{0.40\textwidth}
   \includegraphics[width=1\linewidth]{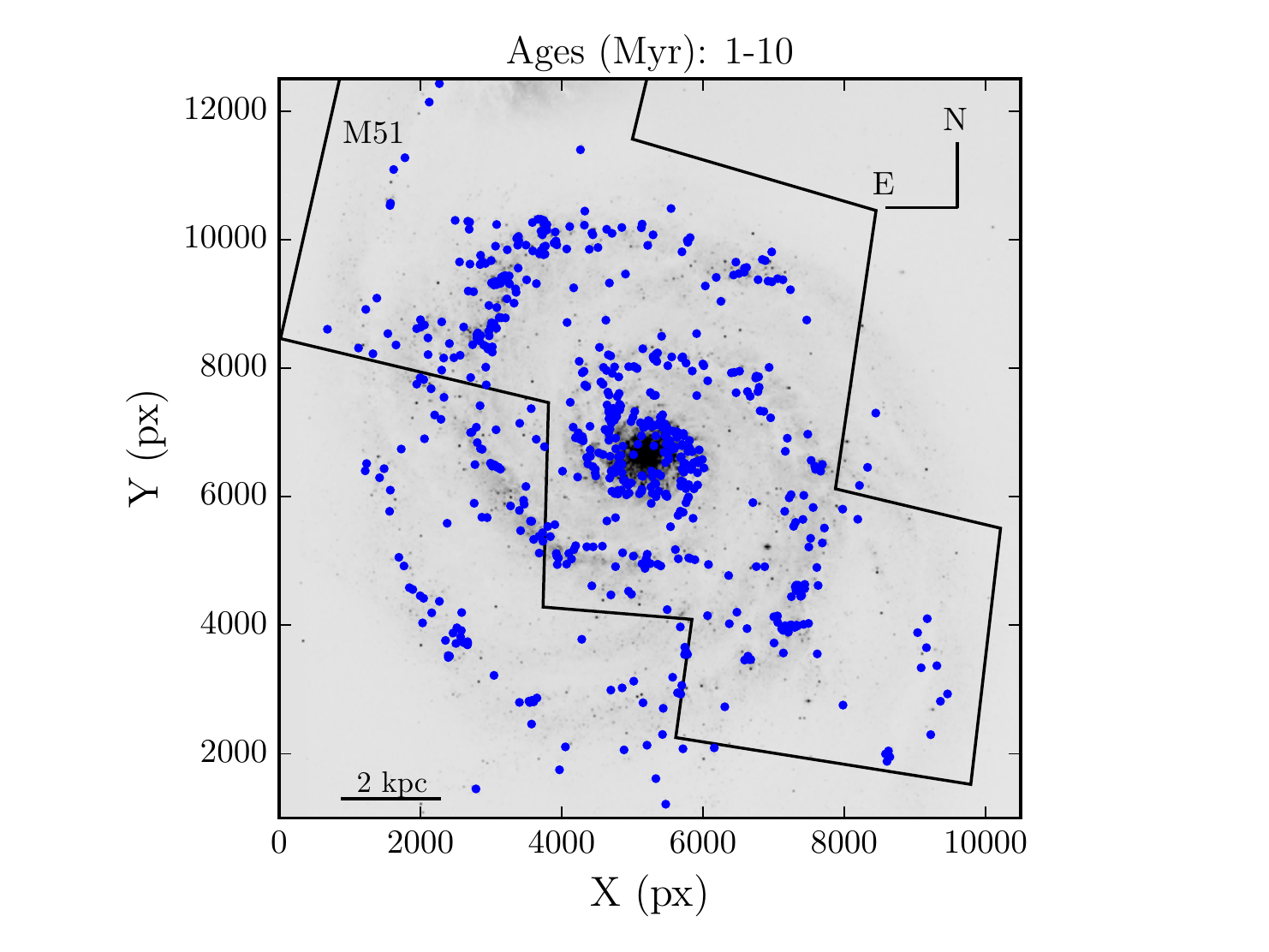}
    \end{subfigure}%
    ~
     ~ \hspace{-1.54cm}
  \begin{subfigure}[t]{0.40\textwidth}
   \includegraphics[width=1\linewidth]{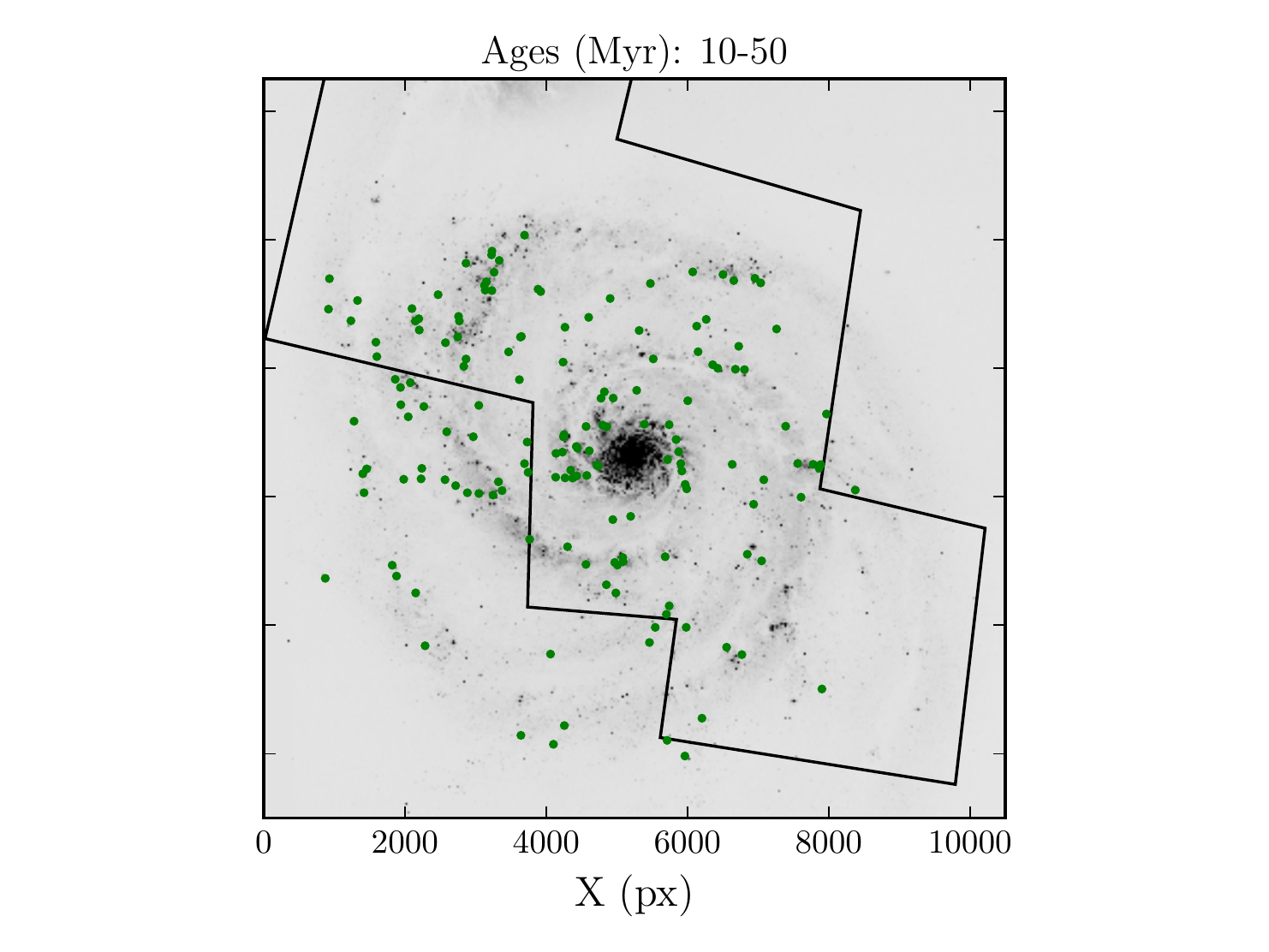}
    \end{subfigure}%
    ~
    ~ \hspace{-1.54cm}
    \begin{subfigure}[t]{0.40\textwidth}
   \includegraphics[width=1\linewidth]{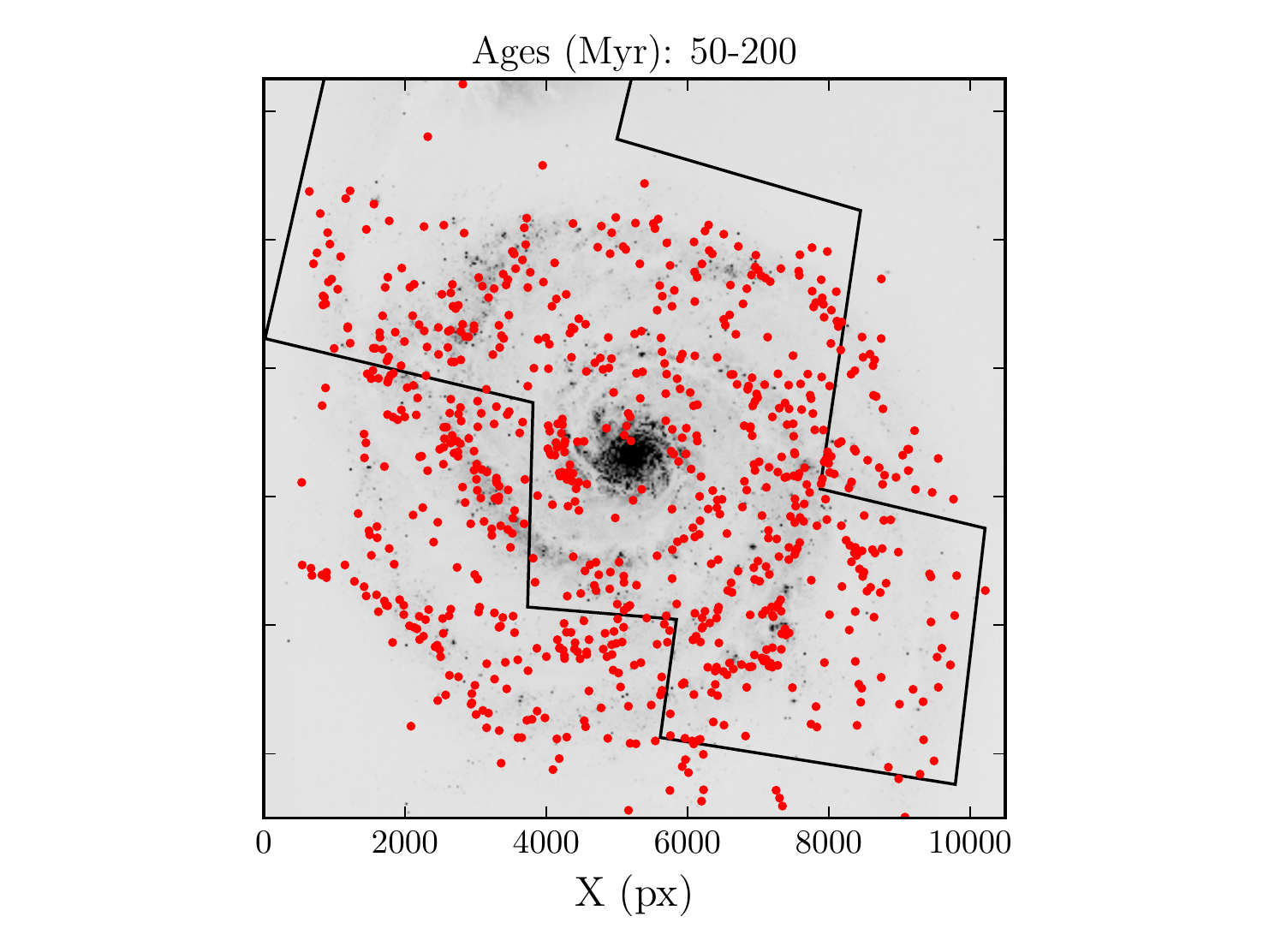}
    \end{subfigure}%

\caption{The spatial distribution of the young (blue), intermediate--age (green), and old (red) star clusters in M51a taken from the CH16 catalogue. The black outlines show the area covered  by the LEGUS observations. The horizontal bar indicates the length of 2~kpc, corresponding to $\rm 54^ {\arcsec}$. North is up and East to the left.}
\label {fig:clusters-ch16}
\end{figure*}

In Fig.~\ref{fig:clusters-ch16}, we plot the spatial distribution of the young, intermediate--age, and old star clusters based on the CH16 catalogue in M51a. As we can see, M51a displays a very clear and strong spiral pattern in the young star clusters. The intermediate--age star clusters tend to be located along the spiral arms while the old ones are more scattered and populate the inter--arm regions. Recently, \cite{chandar17} using the CH16 catalogue found that the youngest star clusters (< 6~Myr) are concentrated in the spiral arms (defined based on 3.6~$\mu$m observations). The older star clusters (6--100~Myr) are also found close to the spiral arms but they are more dispersed, and the spiral structure is not clearly recognisable in older star clusters (> 400 Myr).

\begin{figure*}
\centering
\begin{subfigure}[t]{0.50\textwidth}
   \includegraphics[width=1\linewidth]{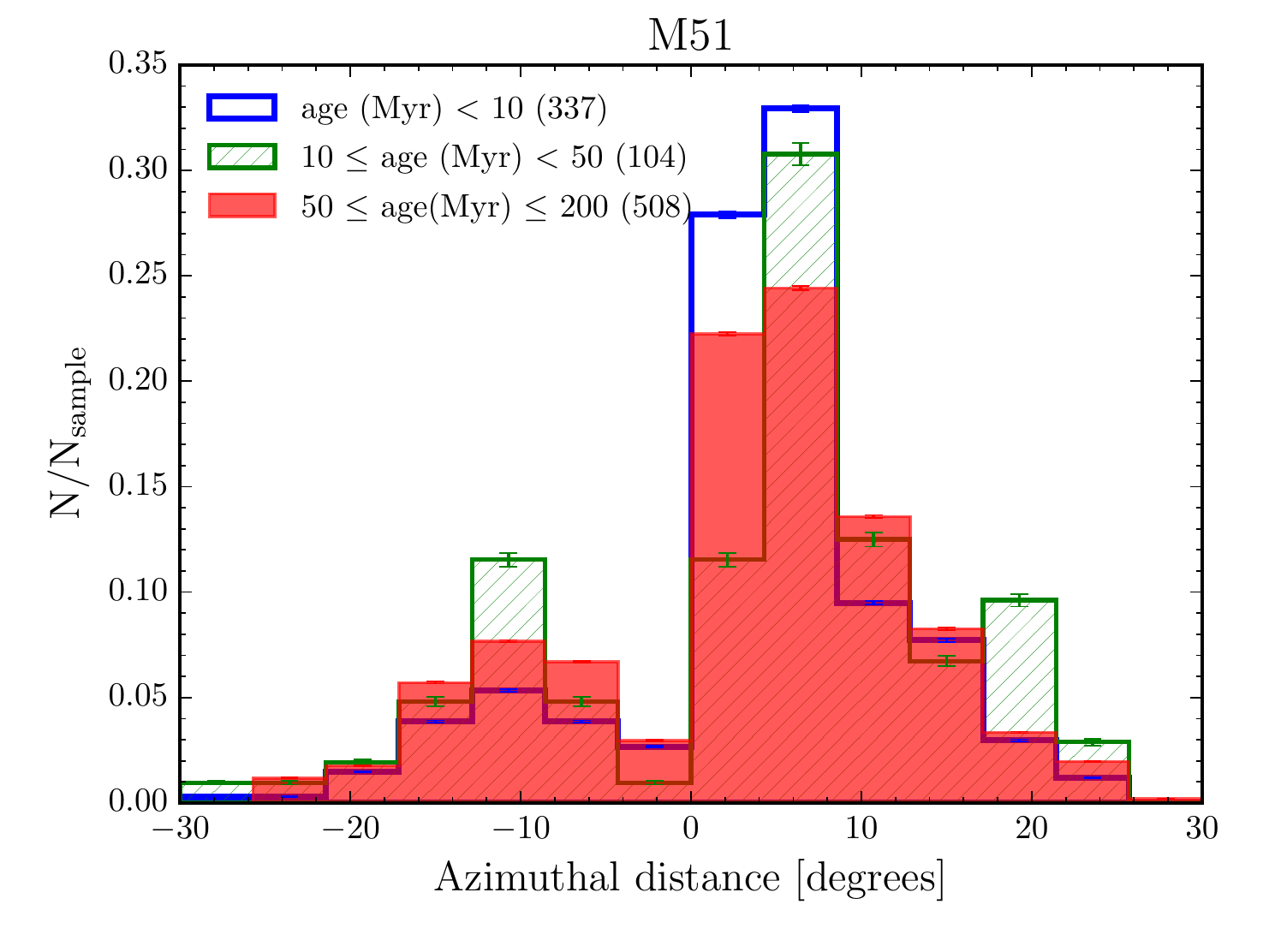}
\end{subfigure}%
~
~ \hspace{0 cm}
  \begin{subfigure}[t]{0.50\textwidth}
   \includegraphics[width=1\linewidth]{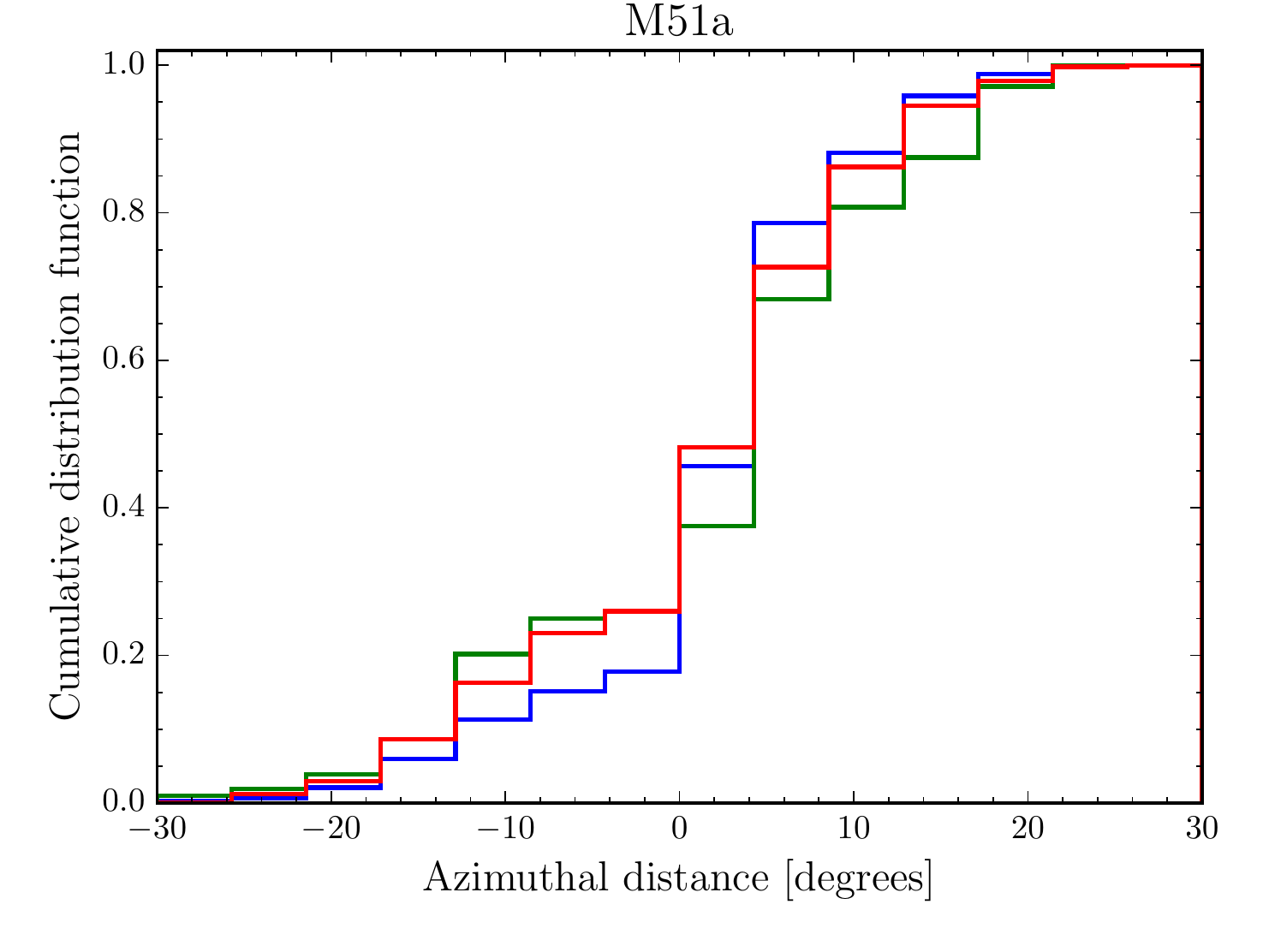}
\end{subfigure}
\caption{The normalized azimuthal distribution (left) and the cumulative distribution function as a function of the azimuthal distance (right) of three star cluster samples in M51a, based on the CH16 catalogue. The young, intermediate--age, and old star clusters are shown in blue, green, and red, respectively. The number of star clusters in each age bin (located in the disk and inside the co--rotation radius of M51a) is listed in parentheses. The error bars in each sample were calculated by dividing the square root of the number of clusters in each bin by the total number of clusters.}
\label {fig:azi_ch}
\end{figure*}

In order to quantify the possible spatial offset in the location of the three young, intermediate--age, and old star cluster samples from the CH16 catalogue across the spiral arms, we computed the normalized azimuthal distribution and corresponding cumulative distribution function of the star cluster samples in Fig.~\ref{fig:azi_ch}. We applied our analysis only to the star clusters positioned in the disk and inside the co--rotation radius of M51a (2.0--5.5~kpc). Our result demonstrates that the three young, intermediate--age, and old star cluster samples peak at an azimuthal distance of 6 degrees from the location of the spiral arms. We observe no obvious offsets between the azimuthal distances of the three star cluster age samples in M51a. \cite{chandar17}, using the same cluster catalogue, quantified the azimuthal offset of molecular gas (from PAWS and HERACLES) and young (<10~Myr) and intermediate--age (100--400~Myr) star clusters in the inner (2--2.5~kpc) and outer (5--5.5~kpc) annuli of the spiral arms. They found that in the inner annuli the young star clusters show an offset of 1~kpc from the molecular gas while there is no offset between the molecular gas and young and old star clusters in the outer portion of the spiral arms.  

Adopting the CH16 catalogue, we found that there is no noticeable age gradient across the spiral arms of M51a, which is in agreement with our finding based on the LEGUS star cluster catalogue.

\section{DISCUSSION AND CONCLUSIONS}
\label{Summary}
The stationary density wave theory predicts that the age of star clusters increases with increasing distance away from the spiral arms. Therefore, a simple picture of the stationary density wave theory leads to a clear age gradient across the spiral arms. In this study, we are testing the theory that spiral arms are static features with constant pattern speed. For this purpose, we use the age and position of star clusters relative to the spiral arms.

We use high--resolution imaging observations obtained by the LEGUS survey \citep{C15} for three face--on LEGUS spiral galaxies, NGC~1566, M51a, and NGC~628. 
We have measured the azimuthal distance of the LEGUS star clusters from their closest spiral arm to quantify the possible spatial offset in the location of star clusters of different ages (< 10~Myr, 10--50~Myr, and 50--200~Myr) across the spiral arms. We found that the nature of spiral arms in our target galaxies is not unique. The main results are summarized as follows:

 \begin{itemize}
\item Our detailed analysis of the azimuthal distribution of star clusters indicates that there is an age sequence across spiral arms in NGC~1566.   NGC~1566 shows a strong bar and bisymmetric arms typical of a massive self--gravitating disk \citep{Elena15}. We speculate that when disks are very self--gravitating the bar and the two--armed features dominate a large part of the galaxy, producing an almost constant pattern speed. The observed trend is also in agreement with what was found by \cite{DP10} in simulations of a grand design and a barred spiral galaxy. 
  
\item We find no age gradient across the spiral arms of M51a. This galaxy shows less strong arms and a weaker bar and hence a less 
self--gravitating disk. The absence of an age sequence in M51a indicates that the grand--design structures of this galaxy are not the result of a steady--state density wave, with a fixed pattern speed and shape, as in the early analytical models. More likely, the spiral is a density wave that is still changing its shape and amplitude with time in reaction to the recent tidal perturbations. A possible mechanism to explain the formation and presence of grand--design structures in spiral galaxies is an interaction with a nearby companion \citep{Toomre 72, k, B3}. Since such an interaction is obviously occurring in M51a, tidal interactions could be the dominant mechanism for driving its spiral patterns. \cite{DP10} simulated M51a with an interacting companion (M51b), and observed no age gradient across the tidally induced grand--design spirals arms.
Our findings are consistent with the results of several other observational studies, which did not find age gradients as expected from the spiral density wave theory in M51a \citep{S09, k10, Foyle, S17}.

 \item NGC~628 is a multiple--arm spiral galaxy with weak spiral arms consistent with a pattern speed decreasing with radius and multiple corotation radii. In this case we find no significant offset among the azimuthal distributions of star clusters with different ages, which is consistent with the swing amplification theory. The lack of such an age offset is in agreement with an earlier analysis of NGC~628 \citep{Foyle}, and consistent with the spatial distribution of star clusters with different ages in the simulated multiple--arm spiral galaxy by \cite{grand}. 
 \end{itemize}

\section*{Acknowledgements}

This work is based on observations made with the NASA/ESA Hubble Space Telescope, obtained at the Space Telescope Science Institute, which is operated by the Association of Universities for Research in Astronomy, Inc., under NASA contract NAS 5--26555. These observations are associated with program 13364. Support for Program 13364 was provided by NASA through a grant from the Space Telescope Science Institute.

This research has made use of the NASA/IPAC Extragalactic Database (NED), which is operated by the Jet Propulsion Laboratory, California Institute of Technology, under contract with the National Aeronautics and Space Administration.

A.A. acknowledges the support of the Swedish Research Council (Vetenskapsr{\aa}det) and the Swedish National Space Board (SNSB). D.A.G kindly acknowledges financial support by the German Research Foundation (DFG) through programme GO 1659/3--2.










\bsp	
\label{lastpage}
\end{document}